\documentclass[12pt]{article}

\usepackage[utf8]{inputenc}

\usepackage{fullpage}
\usepackage[margin=1in]{geometry} 
\usepackage{caption}
\usepackage{times} 
\usepackage{subcaption}
\usepackage{graphicx}
\usepackage{natbib,har2nat}  
\usepackage{amssymb} 
\usepackage{bm}
\usepackage[tbtags]{amsmath}
\usepackage{calc}            
\usepackage{setspace}
\usepackage[bottom]{footmisc} 
\usepackage{xcolor}
\usepackage{caption}
\usepackage{booktabs}
\usepackage{rotating}
\usepackage{tabularx}
\usepackage{tabularx,ragged2e}
\usepackage{pdflscape}
\usepackage{graphicx}
\usepackage{graphicx,caption}
\usepackage{graphics}

\usepackage{array}
\newcolumntype{L}[1]{>{\raggedright\let\newline\\\arraybackslash\hspace{0pt}}m{#1}}
\newcolumntype{C}[1]{>{\centering\let\newline\\\arraybackslash\hspace{0pt}}m{#1}}

\usepackage[capposition=top]{floatrow}
\usepackage{threeparttable}

\usepackage{appendix}    
\usepackage[final]{pdfpages} 
\usepackage{verbatim}     
\usepackage{setspace}     
\PassOptionsToPackage{hyphens}{url}
\usepackage{url}
\usepackage{hyperref}
\hypersetup{
colorlinks = true,
linkcolor = blue!60!black,
anchorcolor = blue!60!black,
citecolor = blue!60!black,
filecolor = blue!60!black,
pagecolor = blue!60!black,
urlcolor = blue!60!black
}

\usepackage{multibib}
\newcites{ltex}{Appendix References}

\begin{document}

\singlespacing 
\title{Pay Transparency and Gender Equality\thanks{{\scriptsize This paper combines two studies that previously circulated separately: ``Pay Transparency and Gender Equality" by E. Duchini, S. Simion, and A. Turrell, and ``Wage responses to gender pay gap reporting requirements'' by J. Blundell. We are grateful to Ghazala Azmat, Manuel Bagues, Nick Bloom, Mirko Draca, Gabrielle Fack, James Fenske, Caroline Hoxby, Libertad Gonzalez, Victor Lavy, Luigi Pistaferri, and Giuseppe Pratobevera for their valuable advice. We further thank Leonardo Bursztyn, Rebecca Diamond, Mitch Hoffman, Matt Gentzkow, Maia G\"{u}ell, Guido Imbens, Ed Lazear, Joseph Mullins, Muriel Niederle, Bobby Pakzad-Hurson, Paul Oyer, Ricardo Perez-Truglia, Barbara Petrongolo, Joahnna Rickne, Paul Robinson, Daniel Schaefer, Sebastian Seitz, Kathryn Shaw, Hans H. Sievertsen, Carl Singleton, Isaac Sorkin, and Simone Traini for their useful comments. A special thanks goes to Mihir Chandraker, Sam Pickering, George Taylor, Marios Tsoukis, and Giulia Vattuone for excellent research assistance. We also acknowledge participants in Warwick internal seminars, Brunel research seminar, the Productivity Insight Network Workshop, Bank of England research seminar, the 2020 EALE-SOLE-AASLE conference, 19th IZA/SOLE Transatlantic Meeting for Labor Economists, 2020 World Congress, 2020 EEA conference, 2020 ESCoE Conference on Economic Measurement, MILLS Seminar, seminars at Padua, Milano Cattolica, Bristol, Royal Holloway, Essex, 2021 EEA Conference, ONS Women's Network seminar, Helsinki GSE for their constructive suggestions. We finally thank Miranda Kyte for facilitating our access to YouGov data. Blundell thanks Stanford Institute for Economic Policy Research for support, in particular the Dixon and Carol Doll Graduate Fellowship and the George P. Shultz Graduate Fellowship. Duchini gratefully acknowledges financial support from the Productivity Insight Network (PIN), and the Centre for Competitive Advantage in the Global Economy (CAGE). The views expressed are those of the authors and may not reflect the views of the Office for National Statistics and the wider UK government. All mistakes are our own.}}}
\author{\large Emma Duchini\thanks{{\scriptsize\ University of Essex, Department of Economics, United Kingdom. \textit{Email:} e.duchini@essex.ac.uk.}} \qquad \c{S}tefania Simion\thanks{{\scriptsize\ University of Bristol, School of Economics, United Kingdom. \textit{Email:} stefania.simion@bristol.ac.uk.}} \qquad Arthur Turrell\thanks{{\scriptsize\ Office for National Statistics, United Kingdom. \textit{Email:} arthur.turrell@ons.gov.uk.}} \qquad Jack Blundell\thanks{{\scriptsize\ Centre for Economic Performance, London School of Economics, United Kingdom. \textit{Email:} j.r.blundell@lse.ac.uk.}}}

\date{November 2022}
\maketitle

\thispagestyle{empty}
\vspace{-2.2em}
\begin{abstract}
\singlespacing
\noindent Since 2018 UK firms with at least 250 employees have been mandated to publicly disclose gender equality indicators. Exploiting variations in this mandate across firm size and time we show that pay transparency closes 18 percent of the gender pay gap by reducing men’s wage growth. The public availability of the equality indicators seems to influence employers’ response as worse performing firms and industries more exposed to public scrutiny reduce their gender pay gap the most. Employers are also 9 percent more likely to post wages in job vacancies, potentially in an effort to improve gender equality at entry level. 

%

\vspace{0.3cm}

\noindent\textbf{JEL codes}: J08, J16, J24. \\
\textbf{Keywords}: pay transparency; gender equality; public disclosure. 
\end{abstract}

\doublespacing

\setcounter{page}{1}

\newpage

\section{Introduction}

In recent years, many governments have adopted pay transparency policies with the aim of improving gender equality.\footnote{Following the recommendations of the European Commission, Austria, Denmark, Italy, and Germany introduced transparency laws (\citealt{aumayr2018pay}). Though pay transparency requirements are less common in the United States, many states have prohibited employers from imposing pay secrecy clauses on their employees (\citealt{siniscalco2017state}).} By increasing the salience of gender gaps in the labor market, transparency measures
are meant to act as an information shock that may alter the relative bargaining power of male and female employees vis-\`{a}-vis the firm. Coupled with the potential negative effects of unequal pay on firms’ reputation, transparency policies have the potential to improve women’s relative pay and career outcomes. The magnitude of these effects is also likely to depend on how strong and salient the information shock is.

This paper studies the impact of pay transparency in a context where firms' gender equality performance is publicly disclosed. Each year since 2018, UK firms with at least 250 employees have been required to publish a series of gender equality indicators both on their own websites and on a dedicated government website. These indicators include percentage gaps in mean and median hourly pay, and the percentage of women in each quartile of the firm's wage distribution.

We begin our analysis by studying the impact of this policy on the gender gap in hourly pay using the Annual Survey of Hours and Earnings (ASHE), the UK matched employer-employee data set, from 2013 to 2019. To identify causal effects, we exploit the variation across firm size and time in the application of the transparency policy. To avoid capturing any potential impact of this policy on firm size, we define the treatment status based on firms’ number of employees prior to the introduction of the mandate. To enhance comparability between the treatment and control groups, we restrict the sample to firms with +/-50 employees from the 250-employee threshold in our main specification.

Our results show that the UK pay transparency policy leads to an 18 percent reduction in the gender pay gap, off a base of \textsterling 2.60. Importantly, and consistent with the hypothesis that pay transparency reduces the relative bargaining power of better-paid employees (\citealt{cullen2019equilibrium}), we find that this effect is driven by a significant 2.6 percent decrease in men's real hourly pay in treated firms relative to control ones. Further evidence suggests that this slowdown of men's pay growth is the result of fewer promotions accompanied by nominal cuts in highly paid occupations.

Event-study exercises show that these results are not driven by different pre-policy trends in the outcomes of interest between treatment and control groups. A battery of placebo regressions further exclude the possibility that our estimates capture the impact of time shocks affecting firms above and below the 250-employee threshold differently. Also, our estimates are not sensitive to choices made in the main specification, such as the bandwidth around the 250-employee cutoff.

As for the mechanisms behind these results, we provide several pieces of evidence that point to the importance of the public availability of the equality indicators to increase firms' accountability. First, we find descriptive evidence for a behavioural response whereby worse performing firms in 2018 -- employers reporting a larger gender pay gap in 2018 -- decrease their gender pay gap the most between 2018 and 2019. Second, we use two YouGov surveys that, since 2018, measure firms' reputation using representative samples of, respectively, British women and British employees, to show that, each year, firms publishing a larger gender pay gap obtain worse placements in both the Women's Rankings and the Workforce Rankings. Third, we show that the drop in men’s real pay is larger in the two sectors, ``Distribution and Hospitality'' and ``Banking and Finance'', that are potentially the most exposed to public scrutiny, as measured by their presence in the YouGov surveys. In sum, by enhancing public scrutiny and enabling comparisons across firms, the public disclosure of the equality indicators may have magnified the disciplinary effects of the policy (\citealt{perez2015shaming}, \citealt{luca2018digital}, \citealt{johnson2020regulation}).

To complement these results, we study whether and how the policy affects firms' hiring practices by combining the difference-in-differences strategy with a text analysis of online job vacancies collected by Burning Glass Technologies (BGT hereafter). One of the most remarkable findings of our analysis is that the policy does not directly improve women's pay or career outcomes, despite this being an explicit goal of the legislation.\footnote{As explained in Section \ref{setting}, the UK pay transparency policy is an amendment to the 2010 \textit{Equality Act}.} One possibility is that it simply takes time for these improvements to materialize, while the policy could have the more immediate effect of increasing firms’ efforts to improve gender equality at the entry level. A growing number of papers document that a factor contributing to the persistence of the gender pay gap is the so-called gender ask gap, whereby women shy away from wage bargaining or propose a lower ask salary when stating how much they want to make in their next job (\citealt{babcock2003nice}, \citealt{hall2012evidence}, \citealt{leibbrandt2015women}, \citealt{card2016bargaining}, \citealt{roussille2020central}, \citealt{biasi2022flexible}). Upfront wage information in the recruitment process may help address this gap by reducing the room for wage bargaining. Consistent with this hypothesis, we first show that, while less than 50 percent of BGT listings provide salary details, firms that are more likely to post wage information in their vacancies also tend to have a larger percentage of women at the top of the firm wage distribution and a lower gender pay gap. More importantly, while these are only correlations, we bring causal evidence that, after the introduction of the policy, the probability that targeted firms post wages in their job ads increases by 9 percent compared to the control group. In other words, the policy induces employers to provide upfront wage information, potentially in an effort to improve gender equality at the entry level.\footnote{In light of the effects on men's pay, we also study the impact of the policy on firms' profits, using the Annual Business Survey (ABS). Despite the significant reduction in labor costs, we find no effect on companies' profits. Consistent with previous results in this literature (\citealt{card2012inequality}, \citealt{breza2018morale}, \citealt{dube2019fairness}, \citealt{bennedsen2020firms}, \citealt{cullen2018much}), this finding points to a negative effect of the policy on labor productivity, which does not hurt profits as it is compensated by the reduction of the wage bill.}

Our paper contributes to different strands of literature. To begin with, we make several contributions to the growing number of studies analyzing the impact of pay transparency on the gender pay gap and wage inequality more broadly (\citealt{card2012inequality}, \citealt{mas2017does}, \citealt{breza2018morale}, \citealt{cullen2018salary}, \citealt{burn2019more}, \citealt{dube2019fairness}, \citealt{bennedsen2020firms}, \citealt{cullen2019equilibrium}, \citealt{cullen2018much},  \citealt{baker2019pay}, \citealt{gulyas2020payreporting}). The closest studies to ours are \citet{bennedsen2020firms}, \citet{baker2019pay}, and \citet{gulyas2020payreporting}. \citet{baker2019pay} study the effect on the gender pay gap of a Canadian law requiring public sector organizations to publish employees’ salaries above a certain pay threshold, while \citet{bennedsen2020firms} and \citet{gulyas2020payreporting} analyze the effect on the gender pay gap of, respectively, a 2006 Danish law and a 2011 Austrian law, mandating that private firms provide their employees with pay data by gender and occupation. Both
\citet{bennedsen2020firms} and \citet{baker2019pay} find that transparency leads to pay compression by slowing down men's wage growth. In contrast, \citet{gulyas2020payreporting} find no impact on individuals' wages or the gender pay gap, and suggest that the fact that, in Austria, the pay information is not disclosed publicly may contribute to explain these null results. Relative to these studies, the UK legislation has two unique features that could help improve our understanding of the effects of pay transparency. First, the information disclosed focuses on the gender pay gap rather than pay levels by gender. While in the latter case, workers could react both to cross-gender comparisons and to comparisons with their own gender, in the UK this second channel is shut down. Second, in the UK setting, the information is disclosed publicly rather than provided exclusively to employees' representatives, which allows us to study the role of performance comparisons and enhanced public scrutiny in influencing firms' responses to the policy. 

Our paper also contributes to the increasing number of studies that use job advert data to analyze different dimensions of the labor market (\citealt{deming2018skill}, \citealt{adams2020flex}, \citealt{azar2020labor}, \citealt{azar2020concentration}). Our analysis of firms' wage posting decision specifically adds to the literature on the gender ask gap, by considering potential employers' responses to address this gap (\citealt{roussille2020central}, \citealt{flinn2021firms}, \citealt{biasi2022flexible}). Moreover, this analysis complements contemporary work on the impact of pay history inquiry bans on recruitment practices (\citealt{sran2020employer}), by showing that the content of job vacancies is influenced by broader management practices, such as pay transparency.

The paper proceeds as follows. Section \ref{setting} describes the institutional setting and the UK transparency policy. Section \ref{impact on gpg} presents the identification strategy, data, and impact of the policy on the gender pay gap. Section \ref{robust} illustrates the robustness checks. Section \ref{mechanisms} studies the role of the publicly availability of the gender equality indicators in influencing firms' response. Section \ref{firm-level outcomes} presents results on firms' hiring practices. Section \ref{conclusion} concludes.

\section{Institutional setting}\label{setting}

In 2015, the UK government launched a process of consultations with employers to enhance pay transparency. At that time, the average gender pay gap for all employees in the UK stood at 19.1 percent. Moreover, women made up only 34 percent of managers, directors, and senior officials (\citealt{geo2015}). According to the government's view, ``greater transparency will encourage employers and employees to consider what more can be done to close any pay gaps. Moreover, employers with a positive story to tell will attract the best talent'' (\citealt{geo2015}). 

In February 2017, this process resulted in the passing of the \textit{Equality Act 2010 (Gender Pay Gap Information) Regulations 2017}. This mandate requires all firms registered in Great Britain that have at least 250 employees to publish gender equality indicators both on their own website and on a dedicated website managed by the Government Equalities Office (GEO hereafter).\footnote{The mandate does not apply in Northern Ireland, while in England, Wales, and Scotland, it applies to both private and public sectors. Note also that the public sector in these countries was already subject to some transparency measures. Further details on this are provided on the Equality and Human Right Commission's website: https://www.equalityhumanrights.com/en/advice-and-guidance/public-sector-equality-duty.} Also, organizations that are part of a group must report individually. In sum, around 10,500 firms are subject to this mandate each year, representing only 0.4 percent of all UK firms but accounting for 40 percent of employment and 48 percent of turnover (Business Structure Database).\footnote{The Business Structure Database (BSD) provides information on firm output, employment, and turnover for almost 99 percent of business organizations registered in the UK. The data come from the Inter-Departmental Business Register (IDBR), a live register of firms collected by the tax authorities via VAT and employee tax records. Office for National Statistics. (2021). Business Structure Database, 1997-2021: Secure Access. [data collection]. 14th Edition. UK Data Service. SN: 6697, DOI: 10.5255/UKDA-SN-6697-14.} To the best of our knowledge, no other substantial law exclusively targeted firms in this size band when the transparency mandate was introduced.\footnote{Since 2010, employees working in firms with at least 250 employees have the right to request time off for training. Note that, even if this policy affected employees' outcomes differently below and above the 250-employee cutoff, the difference-in-differences strategy would take care of these effects, unless they interacted with the transparency policy. Also, since 2020, publicly listed firms with at least 250 employees have been required to publish pay gaps between the CEO and the median employee. However, note that only 1 percent of businesses with at least 250 employees are publicly listed.}


The timing of the publication of the equality indicators works as follows. Each year, if a firm has at least 250 employees by April 5th (the end of the financial year in the UK), it has to calculate the gender equality indicators as of that date, and publish them by the end of the following financial year. Firms themselves must calculate their number of employees using guidelines provided by the government. Importantly, they have to adopt an extended definition of an employee that includes agency workers. Partners of firms are also included in the definition of an employee but should not be included in the calculation of the indicators. Finally, part-time workers have the same weight as full-time workers in the calculations.

The indicators that firms have to report include: the gender gap in the median (mean) hourly pay, expressed relative to men's pay; the gender gap in the median (mean) bonus pay, expressed relative to men's bonus pay; the proportion of male and female employees who receive any bonus pay; and the percentage of female employees in each quartile of a company's wage distribution. Table \ref{public_indicators} provides sample means of these indicators for 2017/18 and 2018/19. Note that in mid-March 2020 the transparency mandate was temporarily paused due to the Coronavirus outbreak, and firms were only asked to start publishing the equality indicators again in October 2021. Considering this, our main analysis focuses on the effects of the policy in the first two years since its introduction, though Appendix Table \ref{including 2020-21} shows that the results do not change when adding data for the financial years 2019/20 and 2020/21.

The first row of Table \ref{public_indicators} shows that the gender gap in median pay is just below 12 percent in 2017/18, and increases slightly in the following year. The gap in mean pay is around 14 percent in 2017/18, and instead decreases a little the year after. Both gaps in median and mean bonuses tend to be smaller than pay gaps but it is worth noting from the large standard deviation that some firms mistakenly reported their level gap rather than a percentage, making it difficult to interpret these bonus gaps.\footnote{When excluding the bottom and top 1 percent, the median (mean) bonus gap stands at 13.14 (23.56) percent in 2017/18 and 12.35 (23.46) percent in the second year.} The proportion of women receiving bonus pay is smaller than for men in each year, and both increase slightly in the second year. The gender ratio along the wage distribution is balanced at the bottom, but less than 40 percent of employees in the upper part of the wage distribution are women. Also, the proportion of women in each quartile of the wage distribution increases slightly over the two years. 

Clearly, the magnitude of these raw firm-level indicators depends both on compositional and observable factors, such as gender differences in educational choices, occupation held and experience, and unobservable factors such as employers' unconscious biases and subtle discrimination on the workplace (\citealt{azmat2020gender}, \citealt{bertrand2020gender}). As they are, these aggregate measures do not allow to distinguish the importance of each underlying factor, and statistics broken down by occupation, or even better hierarchy position, would be more informative in this respect. Yet, the firm-level indicators may reflect a compromise between the government's will to disclose these statistics publicly, and firms' privacy concerns, and it is a matter of empirical analysis to understand how effective they are at pushing firms to tackle the underlying causes of gender inequality. From now on, we will refer to these data as the GEO data.

Three other features of this policy are important to understand the UK context. First, the policy does not impose sanctions on firms that do not improve their gender equality indicators over time. However, the Equality and Human Rights Commission, the enforcement body responsible for this regulation, can issue court orders and unlimited fines for firms that do not publish these indicators. As of 2020, all firms targeted by the law were deemed to have complied. Panel A of Figure \ref{institutional_setting} reports the distribution of submission dates for the first two years that the mandate was in place. While some firms do not meet the deadline, the majority publish their data in the last month before it. Note also that only 235 firms with fewer than 250 employees published gender equality indicators in 2018. These represent less than 0.1 percent of active UK firms with fewer than 250 employees in 2018 (Business Structure Database). This tiny percentage is consistent with the hypothesis that firms are reluctant to disclose information on employees' pay if they are not forced to do so (\citealt{siniscalco2017state}). It is also important to take into account this figure when thinking about the potential general equilibrium effects of this policy.  

Second, according to a survey conducted on behalf of GEO prior to the introduction of this policy, out of 855 private and non-profit firms with at least 150 employees, only one third of firms had ever computed their gender pay gap, and just 3 percent had made these figures publicly available. Moreover, up to 13 percent declared that staff were discouraged from talking about their pay with colleagues and 3 percent reported that their contracts included a clause on pay secrecy (\citealt{geo2015iff}). These figures suggest that the transparency policy is likely to represent an information shock both inside and outside the firm. 

Finally, this policy is salient. Not only are the figures publicly available via both a dedicated government website and companies' own website, but they also receive extensive media attention each year when they are published (\textit{BBC} \citeyear*{bbc2018}, \textit{The Guardian} \citeyear*{guardian2018}, \textit{Financial Times} \citeyear*{ft2018}, \textit{Financial Times} \citeyear*{ft2019}), and firms are not spared from ``naming and shaming'' articles.\footnote{For example, the Independent, a prominent daily newspaper ran a story titled “Gender pay gap: worst offenders in each sector revealed as reporting deadline passes”(\textit{Independent} \citeyear*{independent2018}). In this article a championship football club and an airline were revealed as having among the greatest gender pay gaps in the country.} Importantly, Panel B of Figure \ref{institutional_setting} shows that Google searches for the term ``gender pay gap'' spike around each year's deadline, indicating that this policy attracts significant public interest. And while there is no direct evidence that employees of targeted firms consult the gender equality indicators, the law imposes companies to publish this information on their website ``in a manner that is accessible to all its employees and to the public; and for a period of at least three years beginning with the date of publication'',\footnote{The full text of the law is available at https://www.legislation.gov.uk/ukdsi/2017/9780111152010.} which makes it unlikely that employees are completely unaware of it. 

\section{Impact on the gender pay gap}\label{impact on gpg}

\subsection{Identification strategy}\label{identification strategy}

Our primary goal is to identify the impact of the UK pay transparency policy on the gender gap in hourly pay and unpack this into the effect on women's and men's pay. For this, we exploit the variation in the implementation of the policy across firm size and over time, and compare the evolution of the outcomes of interest in firms whose size is slightly above (treatment group) or below (control) the 250-employee cutoff. As firm size can be endogenously determined, we define treatment status based on firm size in 2015, prior to the start of the consultation process to implement the mandate.\footnote{Appendix Figure \ref{density} shows the distribution of firms around the 250-employee cutoff in each year since the introduction of the mandate. Data are drawn from the Business Structure Database, covering 99 percent of UK firms. While a McCrary test performed separately for each year does not reject the null that there is no jump at the cutoff, it seems cautious to define treatment status based on pre-policy firm size.} Moreover, to enhance comparability between treatment and control group, we consider firms with $+/-50$ employees from the 250 threshold in the main specification. As both choices could be considered to be arbitrary, we show in the next section that our results are robust both to the use of a different year to define the treatment status and to changes in the bandwidth used to construct the estimation sample. Based on these choices, we estimate the following triple-differences regression model that aims to estimate the relative impact of the policy on men's and women's outcomes:
\begin{eqnarray}\label{tripledid}
Y_{ijt}&=&\alpha_{ij}+\theta^{M}_{t}+\theta^{F}_{t}+\beta\;(TreatedFirm_{j}*Post_{t}) \nonumber\\ 
&&+\gamma\;(TreatedFirm_{j}*Post_{t}*Fem_{i}) \nonumber\\ 
&&+X_{ijt}'\pi+u_{ijt},
\end{eqnarray} 

\noindent where $i$ is an employee working in firm $j$, which has between 200 and 300 employees, in year $t$, running between 2013 and 2019,\footnote{As explained in Section \ref{Ashe data}, we choose this time window because it is the maximum number of years over which we observe all outcomes of interest.} and $M$ and $F$ stand respectively for men and women. The outcome $Y_{ijt}$ is either a pay or career outcome, as defined in the next Section. As for the regressors,  $\alpha_{ij}$ are individual-firm fixed effects that capture the impact of individual-firm-specific time-invariant characteristics such as the quality of the match between the employee and the employer; $\theta^{M}_{t}$ and $\theta^{F}_{t}$ are gender-year fixed effects that control for time shocks common to all firms but gender-specific such as the introduction of family-friendly policies; $Fem_{i}$ is a dummy variable that is equal to one if $i$ is a woman; $TreatedFirm_{j}$ is a dummy variable equal to one if a firm has at least 250 employees in 2015; $Post_{t}$ is a dummy variable equal to one from 2018 onward. Note that, in our data, we observe outcomes over fiscal years, which is also the time span over which firms have to publish equality indicators. Hence, for instance, 2018 refers to the fiscal year 2017/18 that starts in April 2017 and ends in March 2018. The vector $X_{ijt}$ includes time-varying individual and firm controls. In our main specification, we only include region-specific time shocks to control for shocks to the local labor market where the firm operates and the individual works, though we present robustness checks in Section \ref{robust} where we alternatively include industry-specific time shocks, or individual controls such as age and age squared. Standard errors are clustered at the firm level.

Our main coefficient of interest is $\gamma$ which, conditional on the validity of this identification strategy, should capture any deviation from a parallel evolution in the outcome's gender gap between the treatment and the control group due to the introduction of the mandate. Put differently,  $\gamma$ should identify the differential effect of the policy on women compared to men. Equally important are $\beta$ and $\beta +\gamma$, which identify, respectively, the effect of the policy on male and female employees. Thus, at the bottom of results' tables we also report the p-value of the t-test on women's effect. 

The validity of our identification strategy depends on three assumptions. First, it has to satisfy the parallel-trend assumption, that is, prior to the introduction of the policy, the evolution of the outcomes of interest must be comparable in treated and control firms. Second, our estimates should not capture the effect of other time shocks that coincide with the introduction of pay transparency and affect firms on either side of the 250-employee cutoff differently. Third, the results should not depend on the size of the bandwidth considered around the policy cutoff, nor should they depend on the year chosen to define the treatment status. 

To support the validity of the parallel-trend assumption and study the dynamic impact of the pay transparency policy, we will open the discussion of our main findings by illustrating the results of the following event-study exercise:
\begin{eqnarray}\label{event_study_reg}
Y_{ijt}&=&\alpha_{ij}+\theta^{M}_{t}+\theta^{F}_{t}+X_{ijt}'\pi \nonumber\\ 
&&+\sum_{k=2013}^{2019}\beta{_k}(TreatedFirm_{j}* \bm{1}[t=k]) \nonumber\\ 
&&+\sum_{k=2013}^{2019}\gamma{_k}(Fem_{i}*TreatedFirm_{j}* \bm{1}[t=k])+\nu_{ijt},
\end{eqnarray}

\noindent where $\bm{1}[t=k]$ is an indicator variable that takes value 1 when $t = k$ and 0 otherwise. In what follows, we take 2017, the year prior to the introduction of pay transparency, as the reference year.

Next, in Section \ref{robust}, we will provide evidence that the other two conditions necessary for the validity of the identification are also met.

\subsection{Data}\label{Ashe data}

To study the impact of the policy on the gender pay gap, we use the Annual Survey of Hours and Earnings (ASHE). ASHE is an employer survey covering 1 percent of the UK workforce that is conducted every year and is designed to be representative of the employee population.\footnote{Office for National Statistics (2022). Annual Survey of Hours and Earnings, 1997-2021: Secure Access. [data collection]. 20th Edition. UK Data Service. SN: 6689, DOI: 10.5255/UKDA-SN-6689-19.} The ASHE sample is drawn from National Insurance records for working individuals, and the selected workers' employers are required by law to complete the survey. Specifically, ASHE asks employers to report data on gender, pay, hours, and tenure for the selected employees, using a snapshot in April each year. Information on age, occupation, and industrial classification is also available. Once workers enter the survey, they are followed even when changing employer, though individuals are not observed when unemployed or out of the labor force. In practice, ASHE is an unbalanced panel data set at the employee level. 

The main limitation of ASHE is its small sample size and the fact that we do not observe all the employees of a firm, which does not allow us to compute a firm-level measure of the gender pay gap. However, this is the only data set available in the UK that provides both a large range of employees' outcomes, including salary components, and information on the total number of employees in a firm and year, which allows us to define the treatment status in our identification strategy.\footnote{When none of the employees of a firm is interviewed in ASHE in the year used to define the treatment status, we recover the information on firm size from BSD. This concerns 28 percent of firms in our estimation sample. In Section \ref{robust} we show that our results are not affected if these firms are excluded from the estimation sample.}
From ASHE, we create the following variables: 

{\bf {\textit{Pay measures}}}. Our main variable of interest is employees' hourly pay, including bonuses but excluding overtime pay; we also separately consider the basic hourly pay and bonus payments, as well as weekly pay and hours worked. To study the impact of the policy on pay variables, we take log transformations, while for bonuses we use the inverse hyperbolic sine transformation to take into account the fact that 80 percent of workers do not receive these payments.\footnote{While it would be interesting to explore separately the impact of the policy on the probability of receiving bonuses and on the amount received, the fact that the latter outcome is only observed for 20 percent of the sample strongly limits our ability to do this.} All monetary values are deflated using the ONS' 2015 CPI Index. To complement the analysis on pay outcomes, we also study the impact of the policy on employees' promotion prospects. For this, we follow the ONS' definition of a promotion and construct a dummy variable that is equal to one if, within the same firm, an employee has experienced at least a 30 percent increase in his/her hourly pay since the previous year, and/or has acquired managerial responsibility since then (ONS \citeyear*{ONS2020promotions}). 

{\bf {\textit{Career outcomes}}}. To get a full understanding of the impact of the policy on employees, in our analysis we also consider its effects on job mobility. First, we use months of tenure in a firm to study mobility into that firm; this variable is missing for around 2 percent of the estimation sample. Second, we construct a dummy variable that is equal to one if the employee leaves the firm in $t+1$, either by moving to a different firm, or by leaving the labor market. Finally, we construct a dummy variable equal to one if a worker is employed in a managerial occupation (1-digit SOC 1).\footnote{Though it is important to stress that by including firm times individual fixed effects in our regressions, only individuals who change 1-digit occupation within employers contribute to identify any effect at this margin.} 

In the main specification, we use data over the period 2013--2019. We start from 2013, as we can observe all outcomes since then, and stop in 2019, as we prefer not to include years affected by the pandemic. However, note that we use information from 2012 and 2020 to construct, respectively, the promotion dummy and the indicator for leaving a firm in $t+1$. In terms of sample restrictions, we drop individuals with missing id or missing firm id (0.5 percent of the sample); we drop secondary jobs (3 percent); we drop individuals who work at least once more than 100 hours per week and those with an hourly pay greater or equal to \textsterling 1000 (0.2 percent). Our resulting sample is formed of 5,135 men and 4,451 women, for a total of 15,769 individual-year observations for men and 13,457 for women. We observe men across 2,929 firms and women across 2,640 firms.

Table \ref{summary_stats} provides summary statistics for the main outcomes measured in the pre-policy period, 2013--2017. Several features are worth noting. First, the profile of workers in treated and control firms is remarkably similar. Second, focusing on the treatment group (columns 1 and 3), the unconditional gap in hourly pay amounts to \textsterling 2.57, or 16 percent of men's pay. This gap reaches 29 percent when looking at weekly pay, as there is also a 16 percent gap in hours worked. As for bonuses, there is a large  gender gap in the probability of receiving bonuses (34 percent), and a very large gap in the amount received (61 percent). And while there is a 33 percent promotion gap in favor of women, there is a 30 percent gender gap in favor of men in the probability of working in managerial occupations. This drops to 5 percent if we consider technical, professional, and managerial occupations, that is the three highest-paid occupations based on the pre-policy median hourly pay at 1-digit SOC level. Men are also more likely to stay longer in a firm than women, and to work in the private sector -- though this share is already around 80 percent for women, which prevents us from studying heterogeneous effects between public and private-sector employees. Finally, note that among both men and women, only one third of workers are covered by a collective agreement, which similarly limits our ability to study heterogeneous effects between unionized and non-unionized workers.

\subsection{Results}\label{main results}

\begin{figure}[H]
\caption{Event studies - log hourly pay}\label{event studies} 
\centering

    \vspace{0.5cm}
 \begin{subfigure}{.5\textwidth}
  \centering
  \includegraphics[scale=0.6]{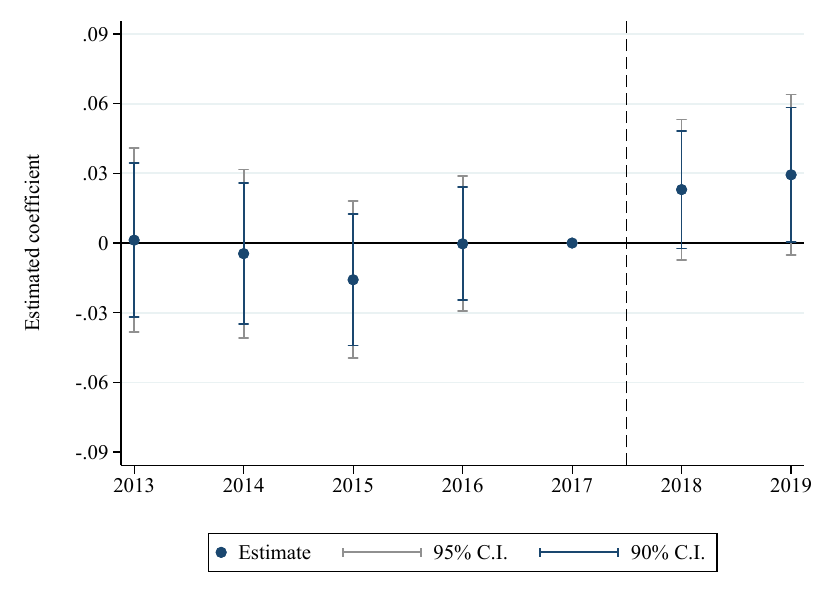}
  \caption{Gender pay gap}
\end{subfigure}\\
\vspace{0.5cm}
\begin{subfigure}{.5\textwidth}
  \centering
 \includegraphics[scale=0.6]{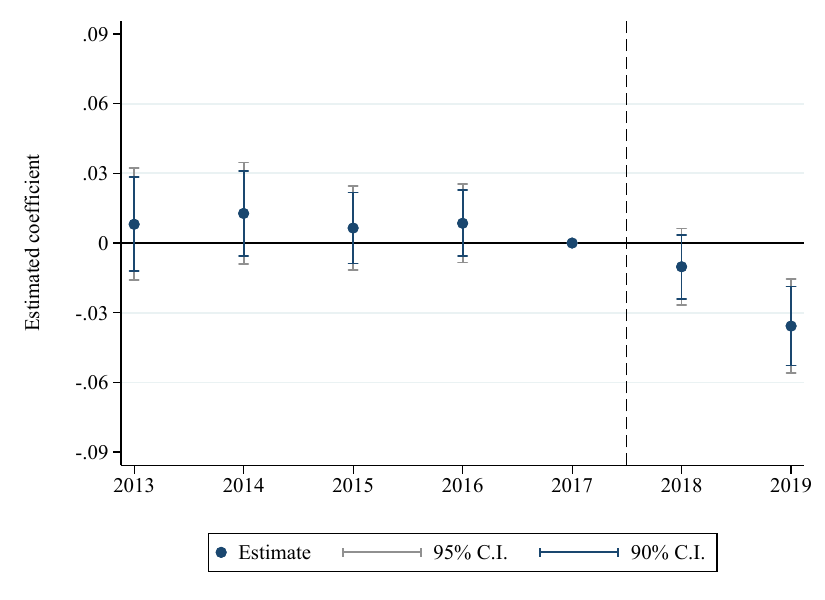}
  \caption{Men}
\end{subfigure}%
\begin{subfigure}{.5\textwidth}
  \centering
 \includegraphics[scale=0.6]{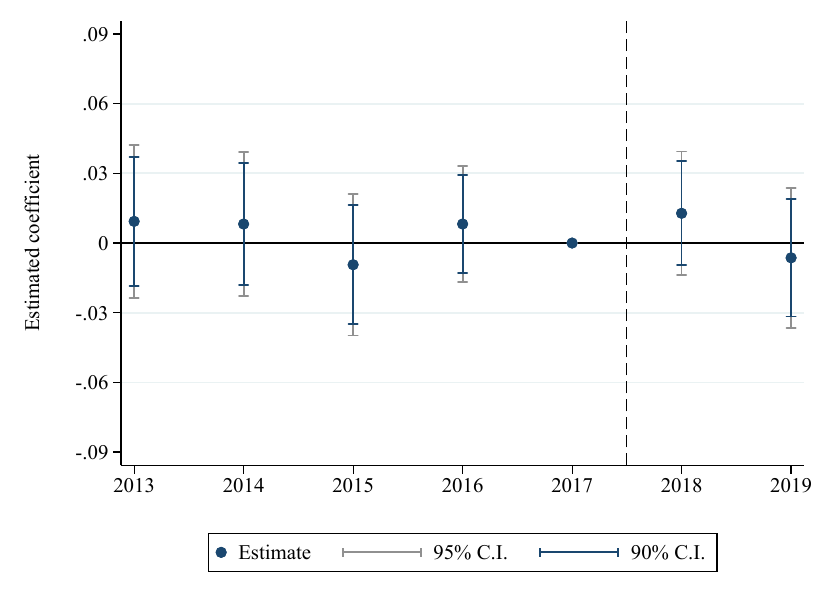}
  \caption{Women}
\end{subfigure}\\
    \vspace{0.5cm}
\begin{tablenotes}
\item \textit{Source:} ASHE, 2013--2019.
\item\textit{Notes:} These graphs present the estimates of the leads and lags of the policy on the gender pay gap (Panel A), and  men's and women's pay (Panel B and C, respectively). These results are obtained from the estimation of regression \ref{event_study_reg}. In each graph, the estimation sample includes workers employed in firms with 200 to 300 employees. The graphs also report 90 and 95 percent confidence intervals associated with firm-level clustered standard errors. The dash vertical line indicates the month when the mandate is approved, i.e., February 2017.
\end{tablenotes}
\end{figure}

This section presents the impact of the pay transparency policy on employees’ outcomes. Figure \ref{event studies}  shows the event studies for the gender gap in hourly pay, in Panel A, and the log hourly pay, separately for men and women, in Panels B and C. From these figures, we observe, first, that the evolution of the outcomes in the pre-policy period seems to be comparable across treatment and control groups, both for what concerns the gender pay gap, and separately for male and female employees' pay. Second, Panel A shows that women's pay increases relatively to men's pay after the introduction of the policy, with this effect being significant at 10 percent in 2019. Third, from Panel B, we see that this dynamic is driven by a decrease in men’s hourly pay in treated firms relative to control firms after the introduction of the mandate, with this effect becoming significant at the 5 percent level in 2019. In contrast, the policy does not appear to have any visible impact on women’s pay (Panel C).

Table \ref{pay outcomes} reports the estimates of the corresponding average effects of the policy. Each column shows a different outcome. At the bottom of the table, we report the p-value of the t-test on the effect on women and the pre-policy mean for the treatment group calculated over the period 2013–2017. Consistent with the dynamics seen in the event studies, Column 1 shows that the policy leads to a significant 2.9 percentage-point increase in women's hourly pay compared to men's pay. Relative to the pre-policy value of 16 percent, this corresponds to an 18 percent decrease in the gender gap in hourly pay. Importantly, the coefficient on \textit{TreatedFirm*Post} confirms that this effect is driven by a 2.6 percent significant reduction in men's real hourly pay, while on average, the policy has no impact on women's pay.

While these results are consistent with the estimated effects of pay transparency in other settings (\citealt{bennedsen2020firms}, \citealt{baker2019pay}), they definitely deserve further explanations. On the one hand, we need to explain whether the fall in men's real hourly pay is driven by bonuses' cuts or a slowdown in the contractual part of pay. In turn, if it was due to the latter, it is important to understand to what extent treated firms have slowed down men's promotions or even cut their nominal pay. On the other hand, the null effect on women's pay is also remarkable, and we want to understand to what extent it could be driven by general equilibrium effects.

To open this discussion, note that our specification includes worker times firm fixed effects, which implies that the results are not driven by compositional effects, such as high-paying men leaving treated firms or inexperienced women joining them after the introduction of the policy.\footnote{In Appendix Table \ref{job mobility}, we also estimate directly the impact of the policy on job mobility and find no significant effects on either men's or women's tenure in the firm, probability of separation, or probability of working in a managerial occupation. These results are interesting per se and help understand the impact on employees' pay, by ruling out major compositional effects. Appendix Table \ref{hours vs. pay} further shows that the effect on men's hourly pay comes from a negative impact on weekly pay (Column 2) rather than an increase in hours worked (Column 3). Note also that the coefficients in the second row point to a negative effect on women's hours. This could suggest that employers are more willing to accept women's requests to work part-time in an effort to retain them, though we do not want to speculate extensively on insignificant results.}

Columns 2 to 4 of Table \ref{pay outcomes} then unpack the wage effects into the impact on the different pay components. This exercise reveals two aspects of the negative impact on men's pay. First, this effect is mostly driven by the contractual part of pay, while the impact on bonuses is negative but not significant, nor significantly different from the effect for women. Second, while on average the policy has no effect on men's probability of promotion, the dynamic specification depicted in Panel C of Appendix Figure \ref{event studies other pay outcomes} shows a significant 10 percent drop in 2019, when men experience most of the negative pay effect. 

To complement these results, Appendix Figure \ref{trends men basic pay} compares the dynamics of log real basic pay to the trends in log nominal basic pay. While many macroeconomic models still adopt the assumption of downward nominal wage rigidity since Keynes first proposed it, there is growing evidence from different countries, summoned by \citet{elsby2019prevalent}, that at least 20 percent of job stayers experience nominal wage cuts each year, with an even larger prevalence in the upper half of the within-firm wage distribution. Curiously, newspapers reported cuts in the salaries of high-paid male employees following the introduction of pay transparency.\footnote{According to the New York Times, when the pay transparency policy was introduced in the UK, Johan Lundgren, easyJet’s chief executive, took a 4.6 percent pay cut to match the salary of his female predecessor (\textit{New York Times} \citeyear*{nyt2019}). Similarly, in January 2018, The Guardian reported that "six high-profile male presenters have already agreed to pay cuts, including John Humphrys, Jeremy Vine and Nick Robinson'' (\textit{The Guardian} \citeyear*{guardianjan2018}).} Consistent with this anecdotal evidence, Panel A of Appendix Figure \ref{trends men basic pay} shows that both real and nominal pay of male employees decrease after the introduction of the pay transparency policy in treated firms compared to control firms. Importantly, Panel B of Appendix Figure \ref{trends men basic pay} further shows that the drop in men's nominal pay is almost entirely driven by workers in better-paid occupations, i.e., managerial, professional, and technical occupations. In particular, while treated workers in lower-paid professions only experience a 1 percent drop in their nominal pay from 2018 to 2019, treated workers in better-paid occupations experience a 5 percent drop over this period. At the same time, as shown in Appendix Table \ref{results by occupation}, when comparing the impact of the policy across the two subgroups on both men's real pay and the probability of promotion, the effect is not statistically different across the two occupational groups. In other words, the slowdown of men’s pay growth seems to be the result of fewer promotions across the occupational distribution, accompanied by nominal cuts in high-paid occupations.

As for the null effect on women’s pay, in principle, this could be due to the fact that both treated and control firms have raised women’s pay as they compete for the same workers. Yet,  Panel A of Appendix Figure \ref{trends women basic pay} shows  that women's pay increase in both treated and control firms over the period considered, but we do not see any sharp increase after the introduction of pay transparency in either of the two groups. In other words, it seems implausible that general equilibrium effects could completely explain the null effect on women’s pay. Thus, at least in the short run, the main effect of pay transparency is to generate pay compression through a reduction of men's real pay.

The next section is dedicated to showing that these results are not driven by time shocks that affect treated and control firms differently, and that they are robust to the use of different models and changes in the regression specification. Following this, Section \ref{mechanisms} will explore the contribution of different channels in explaining these results.

\section{Robustness checks}\label{robust}

This section presents two sets of robustness checks. First, we show that our results are unlikely to be driven by contemporaneous shocks to the policy that have heterogeneous effects across treated and control firms. Second, we show that our results do not depend on the choices made in the main specification, in particular in terms of the size of the bandwidth around the policy cutoff, and the year used to define the treatment status. To summarize all these results, we visually represent them in Figure \ref{robustness_checks}, and report detailed regression tables in  Appendix Section \ref{detailed robustness checks}. We then conclude this section by discussing the external validity of our results.

\paragraph{Contemporaneous shocks.} To make sure that our estimates do not capture the effect of other events occurring at the same time as the introduction of pay transparency requirements that could affect treated and control firms differently, we run a series of placebo tests pretending that the mandate binds at different firm size thresholds. The estimated effects of these placebo reforms on the gender pay gap, together with 95\% and 90\% confidence intervals, are displayed in Panel A of Figure \ref{robustness_checks}.  In each regression, the estimation sample includes firms with $+/-50$ employees from the threshold indicated on the vertical axis. Reassuringly, none of the placebo mandates has a significant impact on the gender pay gap. This exercise helps exclude the possibility that our estimates capture the impact of time shocks that happen at the same time as the mandate and affect larger firms differently to smaller firms.\footnote{Note that the regressions corresponding to placebo cutoff values "300" to "450" include all treated firms. The fact that the point estimates are positive may simply point to heterogeneous effects of the policy across firm size, consistent with the idea that larger firms are more exposed to public scrutiny.} 

\paragraph{Specification.} Our second set of robustness checks aims to verify that our results are robust to the choice of the bandwidth around the 250-employee cutoff, do not depend on the year we use to define the treatment status, and are not sensitive to the other choices made in the main specification. Panel B of Figure \ref{robustness_checks} shows that the estimates of $\gamma$ from equation \ref{tripledid} change very little when restricting or enlarging the bandwidth around the 250-employee cutoff. Appendix Table \ref{changing bandwidth} displays the corresponding detailed regression results. Specifically, the estimated effect of the policy on the gender pay gap only becomes marginally insignificant when using a bandwidth of $+/-100$, where treated and control firms may start to be less comparable. Importantly, Appendix Table \ref{changing bandwidth} shows that the estimated negative effect on men's pay is always significant and comparable in magnitude across the different regressions. 

Next, Panel C of Figure \ref{robustness_checks}  and Appendix Table \ref{treatment year} compare the impact of the policy on the gender pay gap when changing the year used to define the treatment status. Note that, to avoid capturing any impact of the policy on firm size, we only consider years before the announcement of the employee-cutoff, which took place in the fall of 2015. While the estimates on the gender pay gap become insignificant when using the firm size in 2014 to define the treatment status, Appendix Table \ref{treatment year} shows that the negative impact on men's pay is always significant and comparable in magnitude across the different regressions. 

Finally, Panel D of Figure \ref{robustness_checks}  and Appendix Table \ref{other robustness checks} further show that the estimated impact on the gender pay gap changes little when: including industry-specific time shocks in place of region-specific time shocks; adding age controls; using Labour Force Survey weights; restricting the sample to either workers aged 16-65 or those aged 25+; considering only full-time employees or those working in the private sector; or restricting the sample to firms for which we can use only ASHE-based information on the number of employees to define the treatment status.

To sum up, our estimates are very stable across different specifications and samples, which strongly supports the validity of our identification strategy.

\paragraph{External validity.} One of the limitations of identification designs that exploit the variation in the application of a policy around a specific cutoff is that the estimated effects are by construction local. Thus, to provide more insights regarding the external validity of our estimates, Appendix Figure \ref{comp_samples} compares the occupational and industry distribution of men and women in the estimation sample to that of the entire ASHE population, over the period studied. The occupational distribution displayed in Panel A is remarkably similar in the two samples both for men and women, with only some under-representation of sales occupations in women's estimation sample. As for the industry distribution, Panel B of Figure \ref{comp_samples} shows that, with the exception of the manufacturing sector being over-represented in men's estimation sample, the distribution matches well across the two samples. Taken together, these figures suggest that, in the absence of large equilibrium effects, the estimated effects can hold across the firm size distribution.

\section{Mechanisms}\label{mechanisms}

Our results show that pay transparency reduces the gender pay gap through a slowdown of men's pay growth. This finding is remarkably consistent with the evidence produced by contemporaneous studies (\citealt{bennedsen2020firms}, \citealt{baker2019pay}). As an increasing number of countries introduce pay transparency policies, it is especially important to understand in what circumstances these laws are effective at reducing gender inequality.\footnote{In the conclusion, we will return to the failure of these policies to increase the salaries of low-paid workers.} 

One of the most innovative features of the UK transparency policy as compared to the mandates introduced in other countries is that firms must make their equality indicators publicly available. By enhancing public scrutiny and enabling comparisons across firms, the public availability of this information has the potential to magnify the disciplinary effects of transparency policies (\citealt{perez2015shaming}, \citealt{luca2018digital}, \citealt{johnson2020regulation}). Notably, the evidence on the impact of pay transparency is mixed in contexts where the information on firms’ gender equality performance is only revealed internally, to employees' representatives (\citealt{bennedsen2020firms}, \citealt{gulyas2020payreporting}). In contrast, our findings, and those of \citet{baker2019pay}, show that pay transparency enhances gender equality in contexts where this information is publicly available.

In this section, we provide three complementary pieces of evidence that point to the importance of the public availability of gender equality indicators to increase firms’ accountability. First, we find descriptive evidence for a behavioural response whereby worse performing firms in 2018 -- employers reporting a larger gender pay gap in 2018 -- decrease their gender pay gap the most between 2018 and 2019. Second, we use two YouGov surveys that, since 2018, measure firms' reputation using representative samples of, respectively, British women and British employees, to show that, each year, firms publishing a larger gender pay gap obtain worse placements in both the Women's Rankings and the Workforce Rankings. Third, we show that the drop in men’s real pay is larger in the two sectors, ``Distribution and Hospitality'' and ``Banking and Finance'', that are potentially the most exposed to public scrutiny, as measured by their presence in the YouGov surveys. 

\paragraph{Performance comparisons.} The behavioural economics literature provides evidence that when individuals receive information on their relative performance, those performing worst improve the most afterwards (\citealt{allcott2019welfare}). The same may be true of firms comparing their relative performance in terms of gender equality. Unfortunately, we cannot use the difference-in-differences design to study whether firms react in this way as we cannot compute the firm-level gender pay gap pre-policy in ASHE.\footnote{ASHE does not provide information on all employees in a firm.} However, we explore this mechanism descriptively by exploiting the publicly available data on the gender equality indicators in conjunction with ASHE. Column 1 of Appendix Table \ref{behavioral response table} correlates changes in firms' gender pay gap with their 2018 gender pay gap. To net out sectoral characteristics, we control for 5-digit SIC fixed effects. Standard errors are also clustered at the level of 5-digit SIC. Consistent with the predictions of the behavioural literature, we find a negative correlation: within each sector, a one percentage point increase in the 2018 gender pay gap leads to a 1.7 percent lower growth in the gender pay gap between 2018 and 2019. Moreover, by merging the publicly available data with ASHE,\footnote{We find 6,917 firms targeted by the policy in ASHE, or two thirds of all businesses that have to publish gender equality indicators.} we find in Columns 2 and 3 that, within each sector, worse performing firms exhibit a relatively lower (higher) growth in men's (women's) hourly pay. In particular, a one percentage point increase in the 2018 gender pay gap leads to a 0.6 (0.5) percent lower (higher) growth in men's (women's) hourly pay between 2018 and 2019. While these results are mostly descriptive, they are consistent with the hypothesis that the public availability of gender equality indicators allows firms to compare themselves to other firms and prompts the worst performing employers to improve gender equality the most.

\paragraph{Firms' reputation.} In 2018 and 2019, YouGov compiled two distinct rankings of 1,342 firms (self-selected) called, respectively, YouGov Women's Rankings and YouGov Workforce Rankings, by interviewing a representative sample of 50 to 100 people per day between January and December of each year. Women's Rankings are based on women’s answers to the question: ``Overall, of which of the following brands do you have a positive/negative impression?''. The Workforce Rankings are instead obtained by asking both men and women: ``Which of the following brands would you be either proud or embarrassed to work for?'' The resulting ``impression score'' in the case of Women's Rankings, and ``reputation score'' in the case of the Workforce Rankings, are constructed as the percentage difference between all the positive and negative answers relative to all the answers received in the survey; the higher the score that a firm receives in a survey, the better its placement in the corresponding ranking.

Ideally, we would like to compare the evolution of firms' placement in the two rankings before and after the introduction of the policy, but unfortunately YouGov only started these surveys in 2018. However, we can study descriptively how a firm's placement correlates with its gender equality performance. For this, we manually matched YouGov data with firms' gender equality indicators. Taking into consideration that more than one YouGov firm is associated with the same GEO parent company, we match 943 YouGov companies, or 70 percent of the YouGov sample, to 540 companies disclosing their equality indicators in 2018 and 527 firms publishing them in 2019, or around 5 percent of GEO companies each year.\footnote{The YouGov companies that we cannot link with the GEO data are mostly below the 250-employee cutoff or not registered in the UK. Note also that the impression scores of women are not available for three companies. Finally, while the list of firms included in YouGov surveys does not change over time, the pool of GEO employers varies from one year to another as it only includes firms with at least 250 employees as of that year.} Note that while firms volunteer to be part of the YouGov surveys, Appendix Table \ref{selection_gpg_yougov} shows that GEO firms included in the YouGov list have a slightly larger gender pay gap than the other GEO companies, especially in 2018. 

Before proceeding to explore the patterns of correlation between firms' placement in YouGov Rankings and their gender equality performance, it is important to consider the timing of the two data sets. GEO firms publish their gender equality indicators by April each year, and YouGov surveys are run from January to December. This implies that, each year, at least two thirds of people interviewed by YouGov have access to the information on firms' gender equality performance for the year when the interview takes place. Given this timing, we explore within-year correlations between firms' equality indicators and their placements in YouGov Rankings.\footnote{It would also be interesting to study how firms' placement changes from one year to the next depending on their gender equality performance the first year. However, the fact that, each year, the majority of YouGov interviews take place after the equality indicators for that year become available makes it difficult to isolate the influence that their disclosure has on the evolution of firms' reputation. Similarly, it would be interesting to study whether firms that perform worse in the 2018 YouGov Rankings reduce their gender pay gap the most by the following year, but, unfortunately, it is unlikely that firms publishing gender equality indicators by April 2019 already have the information on their performance in 2018 YouGov Surveys, as these are run until December 2018.} Importantly, the availability of two years of data allows us to compute these correlations conditional on firm and year fixed effects. We also cluster standard errors at the level of the GEO company. Lastly, because a large number in the ranking means a worse placement, we invert the ranking for ease of interpretation. Appendix Table \ref{Corr YouGov GPG} shows that a one percentage point increase in a firm's gender pay gap is associated with a loss of almost one position in both YouGov Women's Rankings and YouGov Workforce Rankings. While these dynamics could be influenced by other factors in addition to year and firm fixed effects, they are consistent with the hypothesis that the public availability of the equality indicators increases the attention of the public audience. This further motivates us to study firms' response to increased public scrutiny.

\paragraph{Firms' response to public scrutiny.} These two pieces of descriptive evidence support the hypothesis that comparisons across firms and increased public scrutiny play an important role in shaping firms' response to the pay transparency policy. Yet one may argue that these factors may be less relevant for smaller firms, such as the ones included in ASHE estimation sample. Indeed, as shown in Appendix Figure \ref{yougov distributions}, Panel A, firms participating in YouGov surveys tend to be among the largest that publish gender equality indicators. One way to address this concern is to compare the effect of the policy in the ASHE sample across those industries that may be more or less exposed to the public scrutiny. In this respect, it seems plausible that the main sectors present in YouGov's rankings are also the ones most concerned about their public image. As shown in Appendix Figure \ref{yougov distributions}, Panel B, ``Distribution and Hospitality'' and ``Banking and Finance'' represent more than 50 percent of firms included in YouGov Rankings, with the former being especially over-represented compared to the GEO sample.\footnote{According to the UK Labour Force Survey, ``Distribution and Hospitality'' includes UK SIC sections G and I, that is ``Wholesale and retail trade'' and ``Accommodation and food service activities''. ``Banking and Finance'' includes UK SIC sections K, L, M, N, that is ``Financial and insurance activities'', ``Real estate activities'', `` Professional, scientific, technical activities'', and ``Administrative and support service activities''. Note also that, as shown in Appendix Figure \ref{yougov distributions},  ``Distribution and Hospitality'' and ``Banking and Finance'' represent at most 40 percent of GEO firms.} Interestingly, Table \ref{results by industry} shows that the negative effect on men's pay in these sectors is 5 times larger than the impact in other industries and statistically different from it. While these two sectors could share other characteristics that influence firms' reactions, these results are consistent with the hypothesis that reputation concerns play an important role.

In sum, these findings suggest that the public availability of firms' gender equality indicators spurs comparisons across employers, prompting the worst performing firms to reduce their gender pay gap the most, and magnifies the response of those sectors that are potentially the most exposed to the public scrutiny.

\section{Impact on firm-level outcomes}\label{firm-level outcomes}

To evaluate the effectiveness of transparency policies, it is important to consider all of their implications for workers and firms. Thus, we conclude our analysis by studying the impact of the UK policy on firm-level outcomes. 

First, in light of the negative impact of the policy on men's pay, it seems natural to ask whether this reduction of the wage bill translates into higher profits. We explore this question in Appendix Section \ref{impact on profits} using the Annual Business Survey (ABS), which provides annual data on profits and labor costs for the major sectors of the UK economy (production, construction, distribution, and service industries). Despite the significant reduction in labor costs, we find no effect on companies' profits. Consistent with previous results in this literature (\citealt{card2012inequality}, \citealt{breza2018morale}, \citealt{dube2019fairness}, \citealt{bennedsen2020firms}, \citealt{cullen2018much}), this finding points to a negative effect of the policy on labor productivity, which does not hurt profits as it is compensated by the reduction of the wage bill.

Together with the negative effect on men's pay, our analysis has identified another remarkable result so far: the policy does not directly improve women’s pay or career outcomes, despite this being an explicit goal of the legislator. One possibility is that it simply takes time for these improvements to materialize, while the policy could have the more immediate effect of increasing firms' efforts to improve gender equality at the entry level. To investigate this hypothesis, in the next section we study the impact of the policy on firms' hiring practices, by combining the difference-in-differences strategy with a text-analysis of online job listings from Burning Glass Technologies (BGT).

\subsection{Impact on hiring practices}\label{hiring practices}

Many studies document that women have a lower propensity to negotiate than men, and that this so-called gender ask gap helps explain the persistence of the gender pay gap (\citealt{card2016bargaining}). For example, women are less likely to ask for wage increases (\citealt{babcock2003nice}, \citealt{bowles2007social}, \citealt{biasi2022flexible}), tend to avoid bargaining for higher wages when they apply for jobs that leave wage negotiation ambiguous (\citealt{hall2012evidence}, \citealt{leibbrandt2015women}), and propose lower salaries when asked how much they want to make in their next job (\citealt{roussille2020central}). 

Potentially, upfront wage information in the recruitment process may help address this gender ask gap by reducing the room for wage bargaining. Interestingly, \citet{flinn2021firms} show that in a labor market with heterogeneous wage settings, where both wage bargaining and wage posting initially coexist, mandating wage posting reduces the gender pay gap by 6 percent.\footnote{Though, importantly, directed search models and related empirical evidence show that wage posting may increase competition for a job (\citealt{banfi2019high}, \citealt{marinescu2020}, \citealt{wright2021directed}, \citealt{belot2022wage}).}  Also consistent with the hypothesis that wage posting may help address the gender ask gap, when combining the GEO data with the information extracted from BGT vacancies, we find a positive correlation between firms' wage-posting decision and gender equality performance.\footnote{Details of BGT data, the procedure to match BGT with other firm-level data, and the definition of wage posting are given in the following section and Appendix Section \ref{appendix BGT}.} Specifically, the bar graph in Appendix Figure \ref{bgt gpg correlations} reports the correlation between GEO firms' average percentage of vacancies posting wage information between 2015 and 2019 and, respectively, the average percentage of women in the top quartile of the firm wage distribution (blue bar), and the average gender pay gap (red bar) between 2018 and 2019. When computing these correlations, we control for firms' 5-digit SIC codes, firms' size bands, and the occupational composition of vacancies; we also cluster the standard errors at the 5-digit SIC level. The graph shows that firms that are more likely to post wage information also tend to have a larger percentage of women at the top of the firm wage distribution and a lower gender pay gap. While these are only correlations, they further motivate us to study the impact of the transparency policy on firms' wage-posting decision. For this, we must first introduce the BGT job advert data, how we match these data with other firm-level data to build the difference-in-differences sample, and how we construct the variables of interest.

\paragraph{BGT data.} BGT scrapes online job ads from company websites and job boards. UK data are available from the 2013 financial year and cover more than 50 million (de-duplicated) individual job vacancies collected from a wide range of online job listing sites. While the data set only includes online advertisements, and hence misses vacancies not posted online (e.g. those advertised informally and internal vacancies), it includes a rich set of information that is especially useful for our analysis. First, each observation includes the text of the job advertisement. Second, more than 95 percent of vacancies have an occupational SOC identifier. Third, around one third of the vacancies, or 17 million observations, include the name of the employer. As this is the only variable that can facilitate the merging of BGT data with other firm-level data, we focus on the restricted sample with non-missing employer names. Importantly, to exclude potential selection issues related to the presence of the firm name, Appendix Figure \ref{bgt vs ons} shows that the industry distribution of the stock of vacancies in BGT and the stock of vacancies in the official ONS Vacancy Survey match well, mitigating concerns regarding the representativeness of BGT. 

\paragraph{Difference-in-differences sample.} To implement the difference-in-differences strategy, we need to identify treatment and control groups in BGT, and for this we need to know firms’ number of employees. To retrieve this information, we merge BGT with FAME, the UK version of Amadeus, which is an European firm-level data set managed by Bureau Van Dijk. Appendix Section \ref{appendix BGT} explains in detail the name-matching strategy that we use to merge the two data sets, and the additional steps that we take to build the sample for the difference-in-differences analysis. In the final sample, we retain only BGT firms matched with FAME that have a firm size between 200 and 300 employees and non-missing firm size in 2015, the year used to define the treatment status. We also restrict the sample to vacancies with non-missing SOC and SIC codes; and we exclude vacancies for part-time jobs, representing around 10 percent of the sample, to take into account that, when an ad for a part-time job posts a wage, we cannot distinguish whether this is the full-time equivalent or the part-time wage. Finally, in the main specification, we only keep vacancies posted from the fiscal year 2015, as BGT expressed concerned over the quality of the data at the beginning of the sample (\citealt{adams2020flex}).\footnote{Though in Appendix Table \ref{bgt outcomes from 2013}, we show that our results change little when including data for the financial years 2013 and 2014.}. The final sample includes 97,467 vacancies from 2,556 firms posted over the fiscal years 2015--2019.

\paragraph{BGT outcomes.} To study firms' wage-posting decision, we extract wages offered from the job-ad text using natural language processing. In particular, to identify wages in the text, we use a series of targeted regular expressions that pick up phrases such as ``30-35k per annum'' and ``20,000/year''. The frequency of the wage offer (annual, weekly, hourly) is similarly inferred from the text, and all values are transformed into annual wages and deflated using the ONS' 2015 CPI Index. When a vacancy posts a wage interval (46 percent of cases) we consider the mid-point of the interval as the value for the posted wage. 

A series of validation exercises conducted by research assistants show that we correctly classify the presence of wages in 98 percent of cases. The residual 2 percent are false negatives, meaning that our code indicates that there is no wage posted when there actually is one. As for the value of posted wages, we correctly identify it in 97 percent of cases. The remaining 3 percent are either the false negatives noted above, or have a typo in the posted wage. Finally, we correctly identify part-time vacancies in 95 percent of cases, with the remaining 5 percent being false positives. 

Our main outcome of interest is a dummy variable equal to one if the vacancy contains wage information, either in the form of a wage interval or a point offer. To complement the analysis on wage posting, we also consider the impact of the policy on the posted wage, on whether the vacancy reports a wage interval, and, in case, on its width.  

Appendix Table \ref{bgt summary_stats} provides summary statistics for the outcomes considered, broken down by treatment status, in the pre-policy period. Note that we define treated firms as in the ASHE analysis, that is, employers with at least 250 employees in 2015. As shown in Table \ref{bgt summary_stats}, the proportion of full-time vacancies is, respectively, 92 and 90 percent for the treatment and control groups. Focusing on full-time vacancies, treated and control firms post, respectively, 61 and 56 percent of their job ads for higher-paid professions, i.e., managerial, professional, and technical occupations. Treated firms are more likely to post wage information, though the share of vacancies providing wage information remains below 50 percent for both groups (46 vs. 45 percent).\footnote{For comparison, according to contemporaneous studies analyzing online job ads data, the share of vacancies posting wage information is, respectively, 20 percent in the United States, 13 percent in Chile, and 17 percent in China (\citealt{kuhn2013gender}, \citealt{banfi2019high}, \citealt{marinescu2020}).} As for the characteristics of posted wages, the treatment group has a higher probability of posting a wage interval (51 vs. 41 percent), though the interval dispersion, that is the ratio between the upper and lower bound posted, is similar across the two groups (1.25 vs. 1.23). Also, on average, treated firms offer a lower annual full-time salary in the pre-policy period (\textsterling 26,000 vs. \textsterling 29,000). Appendix Figure \ref{BGT distributions wage posting} complements these statistics by showing the occupational and industry distribution of wage posting, separately, for treated and control firms. Across both treatment and control firms, wage posting appears to be more frequent in vacancies for caring and leisure occupations and in the public administration, education and health sectors, which is consistent with evidence from the United States (\citealt{hall2012evidence}). Also, with few exceptions, the two distributions are similar across the two groups. 

\paragraph{BGT analysis.} To study the impact of the policy on firms' wage posting decision, we estimate the following difference-in-differences model at the vacancy level:
\begin{equation}\label{did BGT}
Y_{ijt}=\alpha_{j}+\theta_{st}+\beta\left(TreatedFirm_{j}*Post_{t}\right)+u_{ijt}, 
\end{equation} 

\noindent where $Y_{ijt}$ is either a dummy equal to one if vacancy $i$ of firm $j$ in quarter $t$ offers wage information, the log of the posted wage, whether the firm posts a wage interval, and, if so, the interval dispersion. As in the ASHE analysis, $TreatedFirm_{j}$ is a dummy variable equal to one if a firm has at least 250 employees in 2015; $Post_{t}$ is a dummy variable equal to one from the second quarter of 2018 onward; $\alpha_{j}$ are firm fixed effects; and $\theta_{st}$ are 5-digit SIC-specific quarter fixed effects that control for the potential seasonality of labor demand at industry level ($s$ stands for sector). Finally, we cluster standard errors at the firm level.

Table \ref{bgt other outcomes} presents the results of this analysis. First, Column 1 tells us that the policy significantly increases the probability that employers report wage information on their job ads by 4.3 percentage points, or 9 percent compared to the pre-policy sample mean. Panel A of Appendix Figure \ref{event study wage posting} shows the corresponding event study. While the evolution of wage posting seems comparable across treated and control firms before the introduction of the policy, the share of vacancies posting wage information increases in treated firms compared to control firms after its introduction, with this effect being relatively larger and significant at 1 percent in 2019. In the following columns of Table \ref{bgt other outcomes}, we complement this analysis by studying whether the policy affects the characteristics of posted wages, including the mean value of wages, whether the firm posts a wage interval, and, where there is an interval, its dispersion. While the policy has no significant impact on the value of posted wages or the probability of posting a wage interval, it significantly decreases the width of intervals that are posted by 4.6 percentage points, or 4 percent relative to the pre-policy mean.\footnote{In Appendix Table \ref{bgt outcomes with soc fe}, we further show that the magnitude of these results barely changes when controlling for 4-digit SOC fixed effects, though the impact on the wage interval width becomes marginally insignificant. Note that while Appendix Table \ref{bgt vacs} shows that the policy has no significant impact on firms' number vacancies or the occupational distribution of vacancies, it seems cautious not to include SOC fixed effects in the main specification as the policy can in principle affect labor demand for specific occupations.} In sum, these results show that after the introduction of the policy, employers are more willing to provide upfront wage information, potentially in an effort to reduce the gender pay gap at entry level.\footnote{While ASHE small sample size strongly limits our ability to conduct a subgroup analysis by tenure in the firm, in Appendix Table \ref{results by tenure}, we find suggestive evidence that the policy may have increased the pay of recently hired women (those with at most two years of tenure) by up to 5 percent in treated firms compared to control firms, which is consistent with the hypothesis that wage posting may help improve gender equality at entry level. Note also that wage posting may contribute to generate pay compression among incumbent workers insofar as it limits the possibility that, when searching on-the-job, these workers meet firms that are willing to engage in wage bargaining (\citealt{flinn2021firms}). In this respect, the increase in firms' propensity to post wages in job vacancies may also help explain the slowdown in men's pay growth.}


\section{Conclusion}\label{conclusion}

To tackle the persistence of gender inequality in the labor market, many governments are introducing pay transparency policies. Exploiting the variation across firm size and over time in the application of the UK's transparency policy, we show that increased transparency leads to an 18 percent significant reduction in the gender pay gap. Importantly, this effect is the result of a slowdown in men’s wage growth, while the policy has no significant effect on women's pay. In other words, these results
suggest that transparency policies can reduce the gender pay gap with limited costs for firms, but may not be suited to achieve the objective of improving outcomes of lower-paid employees. As an increasing number of studies confirm that transparency policies mainly generate pay compression by pushing down the real wages of better-paid employees (\citealt{mas2017does}, \citealt{bennedsen2020firms}, \citealt{cullen2019equilibrium}, \citealt{baker2019pay}), policy makers should consider whether this is a desirable way to tackle wage inequality. 

To conclude, it is important to stress that our analysis only identifies short-term effects.\footnote{As shown in Appendix Table \ref{including 2020-21}, our main result does not change when extending the sample to the pandemic period. However, we are cautious in interpreting this as a long-term effect, given the disrupting impact of the pandemic on the labor market.} Potentially, pay transparency might only have the effect of stimulating responses from firms in the short run, when it acts as an information shock that attracts strong attention from the public and leads firms to fear for their reputation. If the policy does not produce an actual change in the culture within firms, its effect may fade away over time as the strength of the information shock weakens (\citealt{giuliano2020gender}). In this respect, our analysis of BGT data shows that the UK policy has pushed firms to revise their hiring practices, potentially to reduce the room for wage bargaining. If, over time, this helps firms reduce the gender pay gap at entry level, pay transparency may result in positive long-term effects on gender equality. Considering these ambiguous long-run effects, it is necessary to keep monitoring the impact of these types of policies in order to fully assess their effectiveness.

\clearpage
\newpage

\begin{singlespacing}
\bibliographystyle{aer}
\bibliography{References}
\end{singlespacing}

\clearpage

\section*{Figures and Tables}

\singlespacing


\begin{figure}[H]
\caption{ Institutional Setting}\label{institutional_setting} 
\centering
    \vspace{0.2cm}
\captionsetup[subfigure]{justification=centering}

\begin{subfigure}[a]{\textwidth}
\centerline{
\includegraphics[scale=0.55]{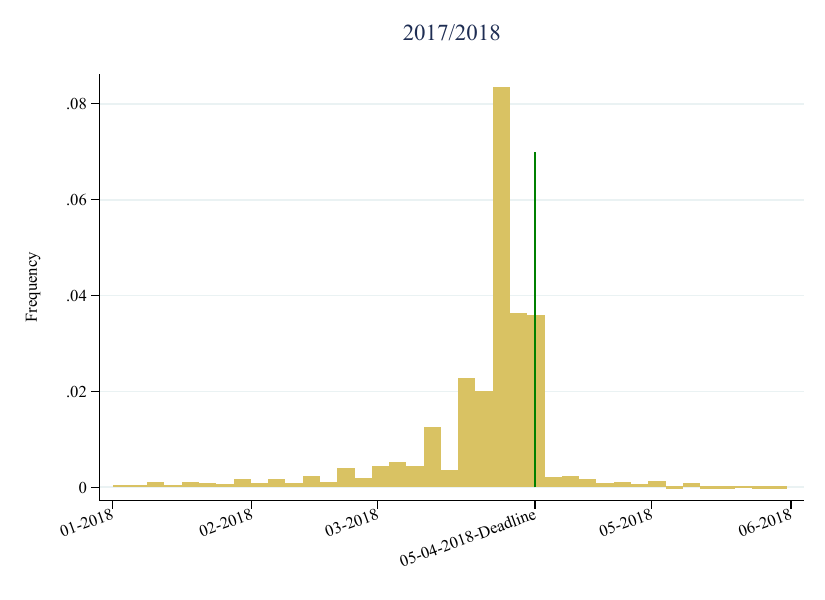}
\includegraphics[scale=0.55]{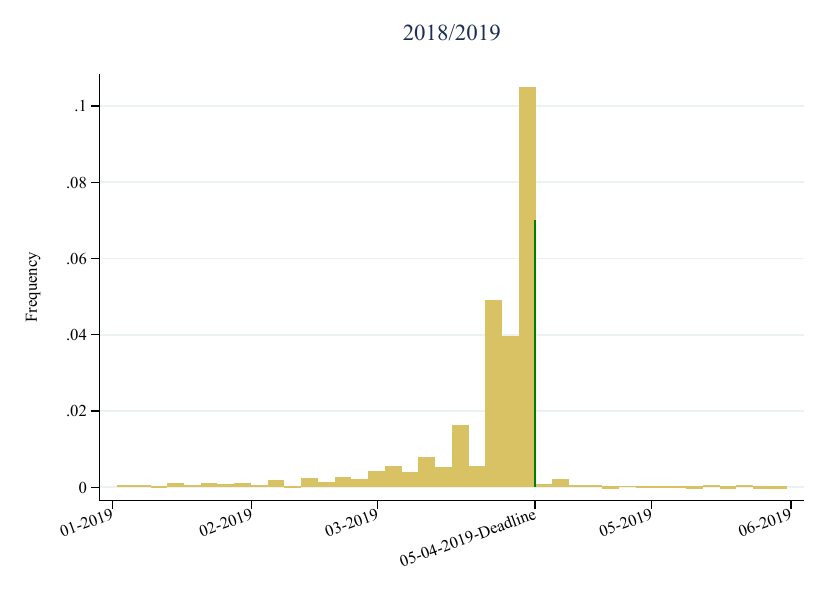}}
\caption{Distribution of submission dates}
\end{subfigure}\\
    \vspace{0.5cm}
\begin{subfigure}[b]{\textwidth}
\centering\includegraphics[scale=0.55]{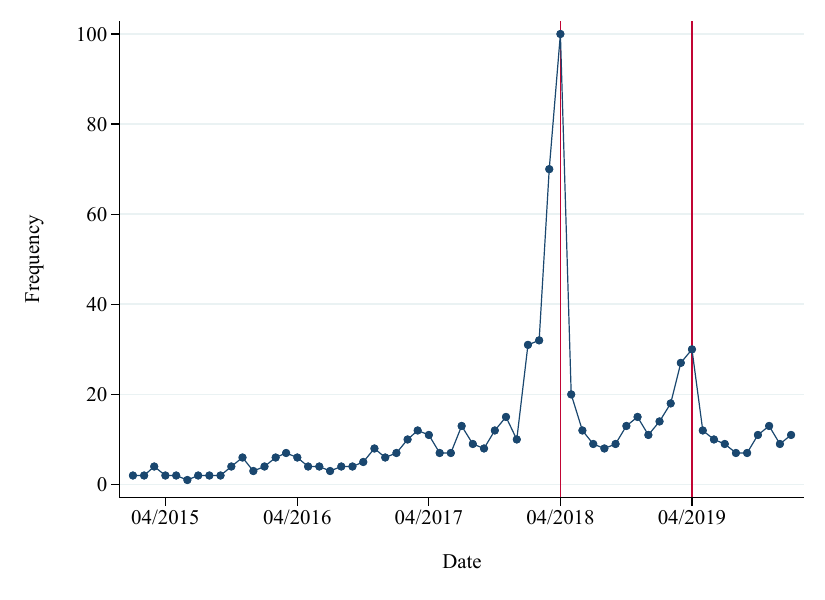}
\caption{Google searches for ``gender pay gap''}
\end{subfigure}\\
    \vspace{0.2cm}
\begin{tablenotes}
\item \textit{Source:} UK Government Equalities Office (GEO); Google, 2015-2019.
\item \textit{Notes:} The figures in Panel A show the distribution of days when firms published their gender equality indicators. The graph on the left refers to the 2017/18 data (10,557 observations), while the one on the right refers to 2018/19 (10,812 observations). Around 5 percent of firms publish before January of the deadline year. The graph in Panel B reports the UK relative search volume for the term ``gender pay gap" between April 2015 and June 2019 using Google's search services. The frequency is indexed to the peak, which occurred in the week commencing 1st April 2018, when firms faced the first deadline to publish gender equality indicators.
\end{tablenotes}
\end{figure}

\begin{figure}[H]
\caption{Robustness checks - gender pay gap}\label{robustness_checks}
\centering
\vspace{0.5cm}
\begin{subfigure}{.5\textwidth}
  \centering
  \includegraphics[scale=0.6]{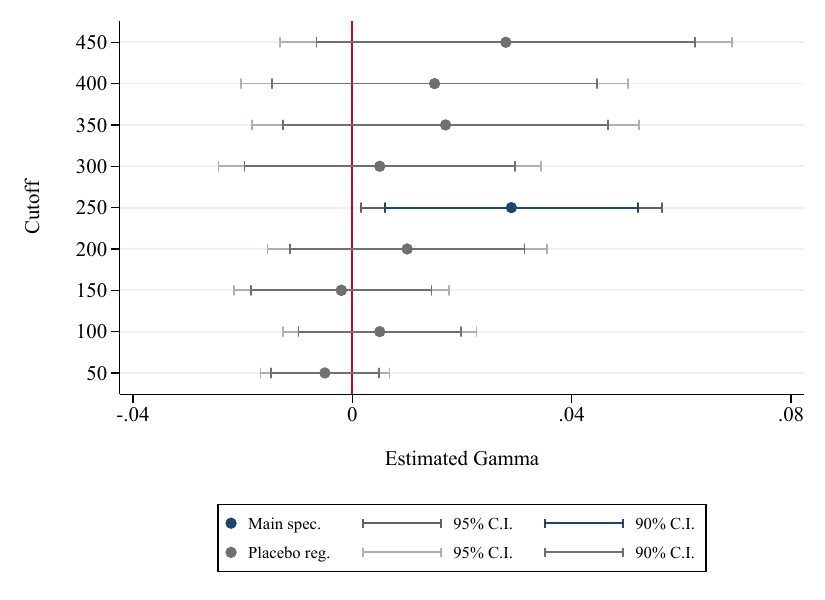}
  \caption{Placebo regressions}
\end{subfigure}%
\begin{subfigure}{.5\textwidth}
  \centering
 \includegraphics[scale=0.6]{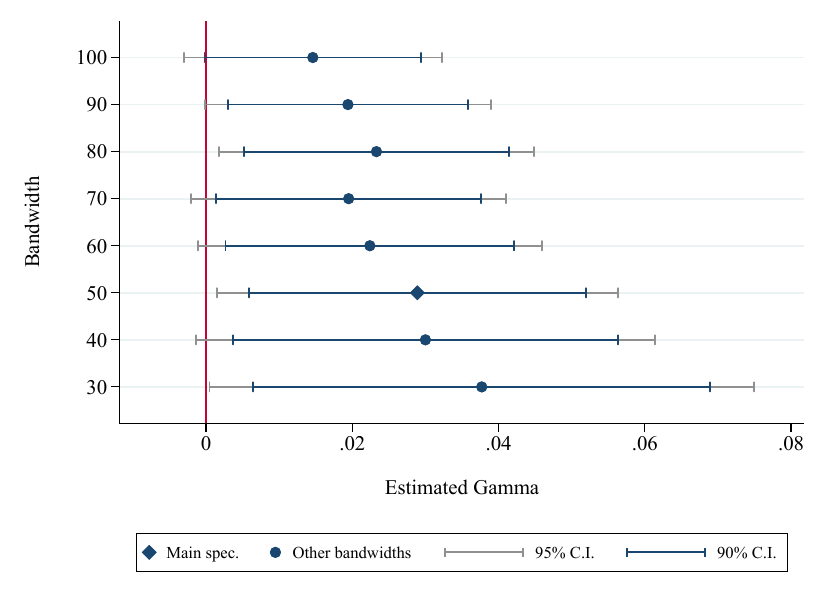}
  \caption{Changing bandwidth}
\end{subfigure}\\
\vspace{0.5cm}
\begin{subfigure}{.5\textwidth}
  \centering
  \includegraphics[scale=0.6]{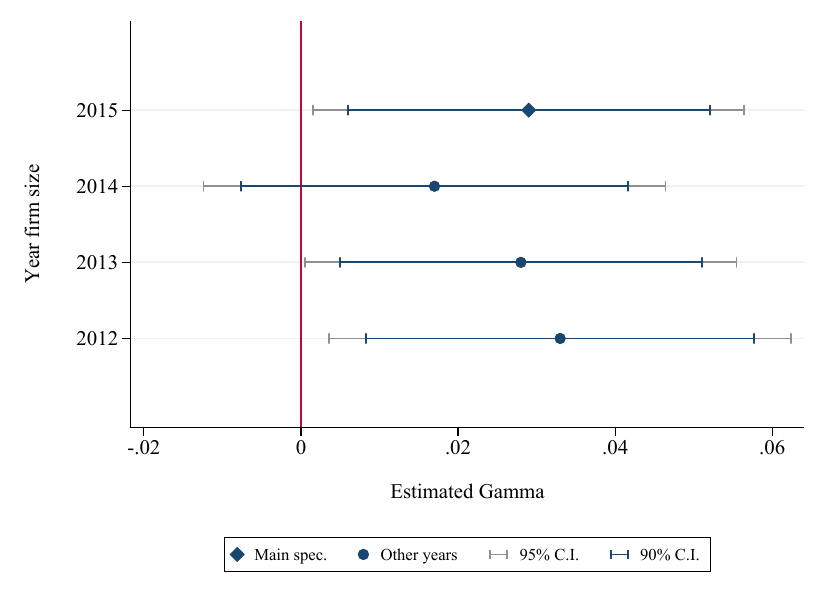}
  \caption{Changing year treatment status}
\end{subfigure}%
\begin{subfigure}{.5\textwidth}
  \centering
 \includegraphics[scale=0.6]{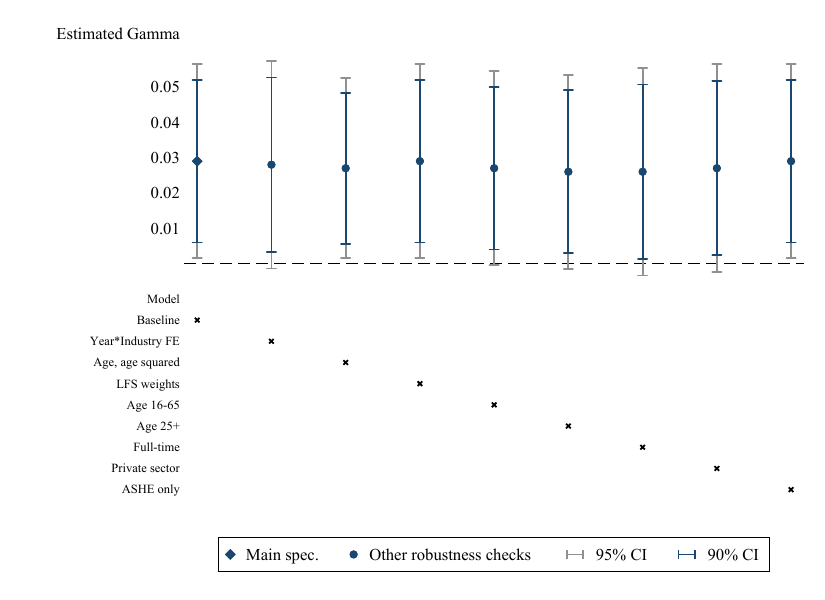}
  \caption{Other robustness checks}
\end{subfigure}\\
  \vspace{0.5cm}
\begin{tablenotes}
\item \textit{Source:} ASHE, 2013--2019.
\item\textit{Notes:} These graphs present a series of robustness checks on the impact of the policy on the gender pay gap. Detailed results are presented in Appendix Tables \ref{placebo regressions}-\ref{other robustness checks}.
\end{tablenotes}
\end{figure}

\begin{table}[htbp]
\def\sym#1{\ifmmode^{#1}\else\(^{#1}\)\fi}
\caption{Public gender equality indicators}\label{public_indicators}
\begin{center}
\begin{threeparttable}
\begin{footnotesize}
\begin{tabular}{l*{2}{c}}
\toprule
                    &\multicolumn{1}{c}{2017/18}&\multicolumn{1}{c}{2018/19}\\
                    &\multicolumn{1}{c}{(1)}&\multicolumn{1}{c}{(2)}\\
            
\midrule
Gender median hourly pay gap (\%) & 11.79 & 11.88  \\
  & (15.84) & (15.51)  \\
   \addlinespace
Gender mean hourly pay gap (\%) & 14.33 & 14.19   \\
  & (14.91) & (14.21)   \\
  \addlinespace
Gender median bonus gap (\%) & -21.72 & -0.86   \\
 & (1,399.04) & (270.51)    \\
  \addlinespace
Gender mean bonus gap (\%) & 7.66 & 15.44   \\
 & (833.06) & (200.70)   \\
  \addlinespace
  \% men receiving bonus & 35.39 & 35.72   \\
 & (36.33) & (36.68)    \\
  \addlinespace
 \% women receiving bonus & 33.92 & 34.40   \\
 & (36.01) & (36.38)    \\
\addlinespace
  \% women lower quartile & 53.67 & 53.88  \\
 & (24.13) & (24.11)    \\
 \addlinespace
 \% women lower-middle quartile & 49.49 & 49.82   \\
 & (26.09) & (26.19)   \\
\addlinespace
\% women upper-middle quartile & 45.15 & 45.62   \\
 & (26.22) & (26.32)   \\
\addlinespace
\% women top quartile & 39.20 & 39.75   \\
 & (24.41) & (24.48)    \\
\addlinespace
Observations & 10,557 & 10,812   \\  
\bottomrule
\end{tabular}
\end{footnotesize}
\begin{tablenotes}
\item {\footnotesize \textit{Source:} UK Government Equalities Office (GEO).}
\item {\footnotesize \textit{Notes:} This table reports mean and standard deviation of gender equality indicators published by targeted firms, separately by year of publication.}
\end{tablenotes}
\end{threeparttable}
\end{center}
\end{table}
\clearpage
\newpage

\begin{table}[htbp]
\def\sym#1{\ifmmode^{#1}\else\(^{#1}\)\fi}
\caption{ASHE Summary statistics - pre-policy period}\label{summary_stats}
\begin{threeparttable}
\begin{footnotesize}
\begin{tabular}{l*{4}{c}}
\toprule
                    &\multicolumn{1}{c}{Treated men}&\multicolumn{1}{c}{Control men}&\multicolumn{1}{c}{Treated women}&\multicolumn{1}{c}{Control women}\\
 &\multicolumn{1}{c}{(1)}&\multicolumn{1}{c}{(2)}&\multicolumn{1}{c}{(3)} &\multicolumn{1}{c}{(4)}\\
\midrule
Hourly pay (\textsterling) & 15.93 & 15.59 & 13.36 & 13.40\\
 & (14.24) & (11.68) & (8.87) & (10.70)\\
 \addlinespace
Weekly pay (\textsterling) & 581.51 & 569.38 & 414.60 & 411.69\\
 & (533.28) & (429.76) & (307.32) & (316.98)\\
 \addlinespace
Weekly hours & 36.41 & 36.67 & 30.69 & 30.49\\
  & (8.53) & (8.50) & (10.53) & (10.69)\\
  \addlinespace
Receiving allowances/bonuses & 0.29 & 0.29 & 0.19 & 0.18\\
 & (0.45) & (0.46) & (0.39) & (0.38)\\
 \addlinespace
Allowance/bonus amount (\textsterling) & 26.03 & 26.08 & 10.07 & 9.47\\
 & (102.33) & (114.65) & (38.71) & (42.15)\\
 \addlinespace
Promotion & 0.03 & 0.03 & 0.04 & 0.04\\
 & (0.17) & (0.18) & (0.18) & (0.18)\\
 \addlinespace
Managerial occupation & 0.10 & 0.10 & 0.07 & 0.06\\
 & (0.30) & (0.30) & (0.26) & (0.24)\\
 \addlinespace
Highest-paid occupations & 0.42 & 0.42 & 0.40 & 0.37\\
 & (0.49) & (0.49) & (0.49) & (0.48)\\
\addlinespace
Tenure in months & 87.40 & 86.28 & 74.37 & 72.10\\
 & (98.47) & (96.65) & (80.85) & (80.34)\\
 \addlinespace
Leaving firm in t+1 & 0.36 & 0.35 & 0.37 & 0.35\\
 & (0.48) & (0.48) & (0.48) & (0.48)\\
 \addlinespace
Private sector & 0.91 & 0.92 & 0.80 & 0.78\\
 & (0.29) & (0.27) & (0.40) & (0.41)\\
 \addlinespace
Covered by collective agreement & 0.28 & 0.27 & 0.32 & 0.34\\
 & (0.45) & (0.44) & (0.47) & (0.47)\\

\addlinespace
Observations & 6,916 & 8,677 & 5,870 & 7,711\\
\bottomrule
\end{tabular}
\end{footnotesize}
\begin{tablenotes}
\item{\footnotesize \textit{Source:} ASHE, 2013--2017.}
\item {\footnotesize \textit{Notes:} This table reports mean and standard deviation of the main variables used in the analysis, separately for men and women, and treatment and control group, before the implementation of the mandate. }
\end{tablenotes}
\end{threeparttable}
\end{table}
\clearpage
\newpage

\begin{table}[htbp]
\def\sym#1{\ifmmode^{#1}\else\(^{#1}\)\fi}
\caption{Impact on pay outcomes}\label{pay outcomes}
\begin{threeparttable}
\begin{footnotesize}
\begin{tabular}{l*{4}{c}}
\toprule
                   
                    &\multicolumn{1}{c}{Log hourly}&\multicolumn{1}{c}{Log hourly}&\multicolumn{1}{c}{Allowances}&\multicolumn{1}{c}{Promotion}\\
&\multicolumn{1}{c}{pay}&\multicolumn{1}{c}{basic pay}&\multicolumn{1}{c}{\& bonuses}&\\
                    &\multicolumn{1}{c}{(1)}&\multicolumn{1}{c}{(2)}&\multicolumn{1}{c}{(3)}&\multicolumn{1}{c}{(4)}\\
\midrule
Treated firm*post   &      -0.026\sym{***} &      -0.024\sym{***}&      -0.025    &      -0.002     \\
                    &     (0.008)    &     (0.009)         &     (0.024)     &     (0.008)     \\
\addlinespace
Treated firm*post*fem&       0.029\sym{**}  &       0.031\sym{**} &       0.002   &   0.011    \\
                    &     (0.014)         &     (0.014)         &     (0.032)    &     (0.011)      \\
\addlinespace
Observations        &       29,226         &       29,226         &       29,226     &       29,226     \\
Adjusted \(R^{2}\)  &       0.909         &       0.911         &       0.622     &       0.003     \\
P-value Women Coeff &       0.788          &       0.578         &       0.306      &       0.281   \\
Men's pre-policy mean&        15.93          &        15.25         &        0.68     &        0.03     \\
Women's pre-policy mean&        13.36         &        13.02         &        0.34    &        0.04     \\
\bottomrule
\end{tabular}
\end{footnotesize}
\begin{tablenotes}
\item{\footnotesize \textit{Source:} ASHE, 2013--2019.}
\item {\footnotesize \textit{Notes:} This table reports the impact of pay transparency on pay outcomes, obtained from the estimation of regression \ref{tripledid}. Each column refers to a different outcome, as specified at the top of it. The estimation sample comprises men  and women working in firms that have between 200 and 300 employees. All regressions include firm*individual fixed effects, gender*year fixed effects, and region-specific time shocks. A treated firm is defined as having at least 250 employees in 2015. The post dummy is equal to one from 2018 onward. Heteroskedasticity-robust standard errors clustered at firm level in parentheses. The p-value at the bottom of the table refers to the t-test on the sum of the two reported coefficients, corresponding to the effect of the policy on female employees. The pre-policy mean represents the mean of the outcome variable for the treated group between 2013 and 2017.}
\item{\footnotesize *** p$<$0.01, ** p$<$0.05, * p$<$0.1.}
\end{tablenotes}
\end{threeparttable}
\end{table}
\clearpage
\newpage

\begin{table}[htbp]\centering
\def\sym#1{\ifmmode^{#1}\else\(^{#1}\)\fi}
\caption{Impact on wage-posting decision}\label{bgt other outcomes}
\begin{threeparttable}
\begin{footnotesize}
\begin{tabular}{l*{4}{c}}
\toprule
                    
                    &\multicolumn{1}{c}{Wage}&\multicolumn{1}{c}{Log annual}&\multicolumn{1}{c}{Wage}&\multicolumn{1}{c}{Interval}\\
                     &\multicolumn{1}{c}{posted}&\multicolumn{1}{c}{wage}&\multicolumn{1}{c}{ interval}&\multicolumn{1}{c}{dispersion}\\
                    &\multicolumn{1}{c}{(1)}&\multicolumn{1}{c}{(2)}&\multicolumn{1}{c}{(3)}&\multicolumn{1}{c}{(4)}\\
\midrule
Treated firm*post &       0.043\sym{*}  &       0.040         &       0.036         &      -0.046\sym{*}  \\
                    &     (0.022)         &     (0.029)         &     (0.028)         &     (0.026)         \\
\addlinespace
Observations        &       97,467         &       43,752         &       43,752         &       19,342         \\
Adjusted \(R^{2}\)  &       0.470         &       0.584         &       0.414         &       0.382         \\
Pre-policy mean     &        0.46         &        26,116         &        0.51         &        1.25         \\
\bottomrule
\end{tabular}
\end{footnotesize}
\begin{tablenotes}
\item{\footnotesize \textit{Source:} BGT, 2015--2019.}
\item {\footnotesize \textit{Notes:} This table reports the impact of pay transparency on firms' decision to post wage information in job vacancies and on the characteristics of posted wages, obtained from the estimation of regression \ref{did BGT}. The estimation sample comprises BGT firms that have between 200 and 300 employees. In Column 2, it is restricted to vacancies with wage information. In Column 4, it is further restricted to vacancies posting a wage interval. All regressions include firm fixed effects and 5-digit SIC-specific time shocks. A treated firm is defined as having at least 250 employees in 2015. The post dummy is equal to one from the second quarter of 2018 onward. Heteroskedasticity-robust standard errors clustered at firm level in parentheses. The pre-policy mean represents the mean of the outcome variable for the treated group between 2015 and 2017.}
\item{\footnotesize *** p$<$0.01, ** p$<$0.05, * p$<$0.1.}
\end{tablenotes}
\end{threeparttable}
\end{table}
\clearpage
\newpage

\newpage
\appendix


\pagenumbering{arabic}
\renewcommand*{\thepage}{A-\arabic{page}}

\setcounter{equation}{0}
\renewcommand{\theequation}{A.\arabic{equation}}
\setcounter{footnote}{0}
\renewcommand{\thefootnote}{A.\arabic{footnote}}
\setcounter{table}{0}
\renewcommand{\thetable}{\Alph{section}\arabic{table}}
\setcounter{figure}{0}
\renewcommand{\thefigure}{\Alph{section}\arabic{figure}}

\begin{center}
{\Large Appendix}
\end{center}


\clearpage

\section{Further results and robustness checks}\label{detailed robustness checks}

\begin{figure}[H]
\caption{Firm size distribution}\label{density}
\centering
\includegraphics[scale=0.5]{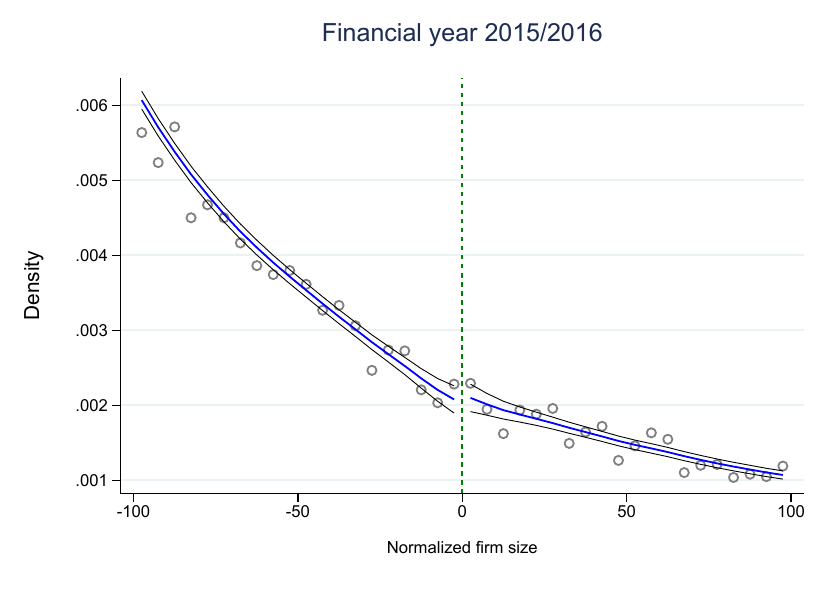}
\includegraphics[scale=0.5]{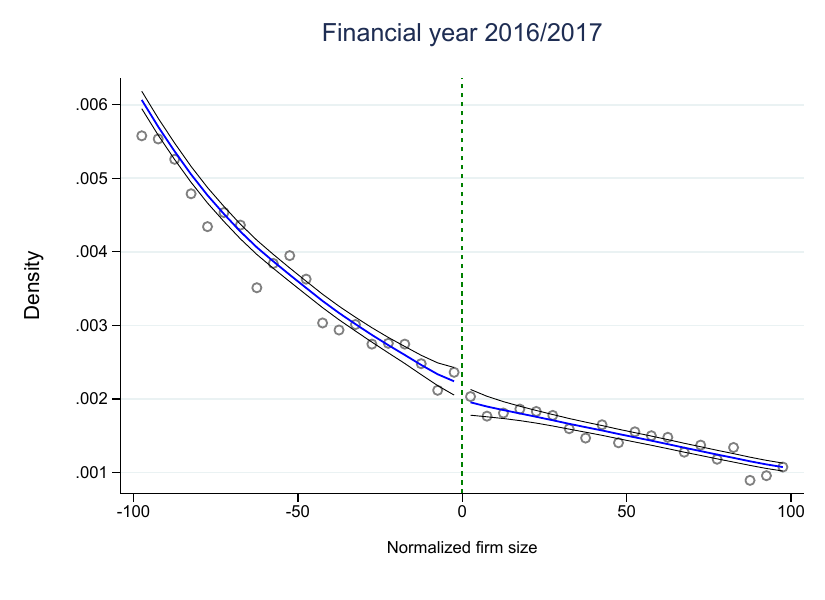}
\includegraphics[scale=0.5]{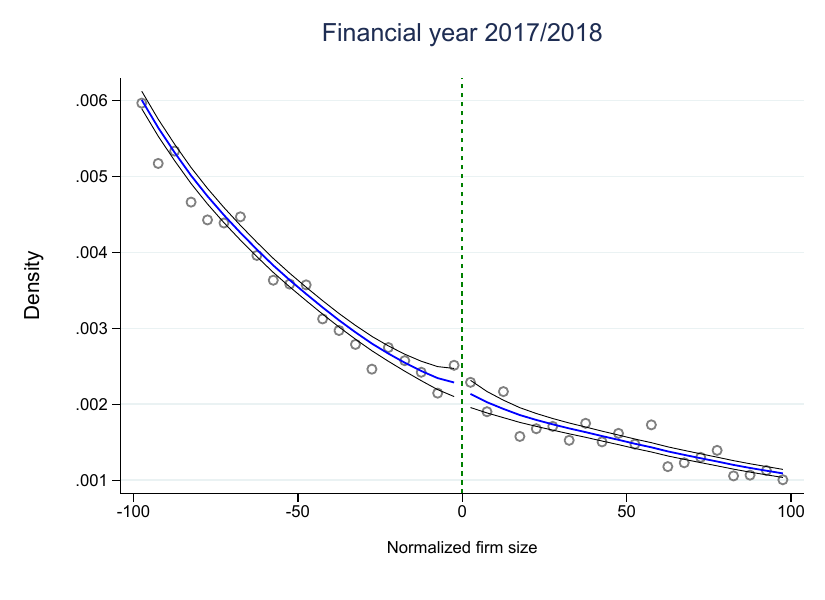}
\\

\floatfoot{{\textit{Source:} BSD, 2016--2018.\\
\textit{Note:} These graphs show the distribution of firms around the 250-employee cutoff in each year since the announcement of the policy. In each figure, the sample includes firms with +/100 employees from the threshold, grouped in 20 bins. Each dot represents the share of firms with a number of employees comprised in the corresponding bin. }}
\end{figure}
\clearpage
\newpage

\begin{figure}[H]
\caption{Event studies - other pay outcomes}\label{event studies other pay outcomes} 
\centering

    \vspace{0.2cm}
 \captionsetup[subfigure]{justification=centering}
\begin{subfigure}[a]{\textwidth}
\centerline{\includegraphics[scale=0.42]{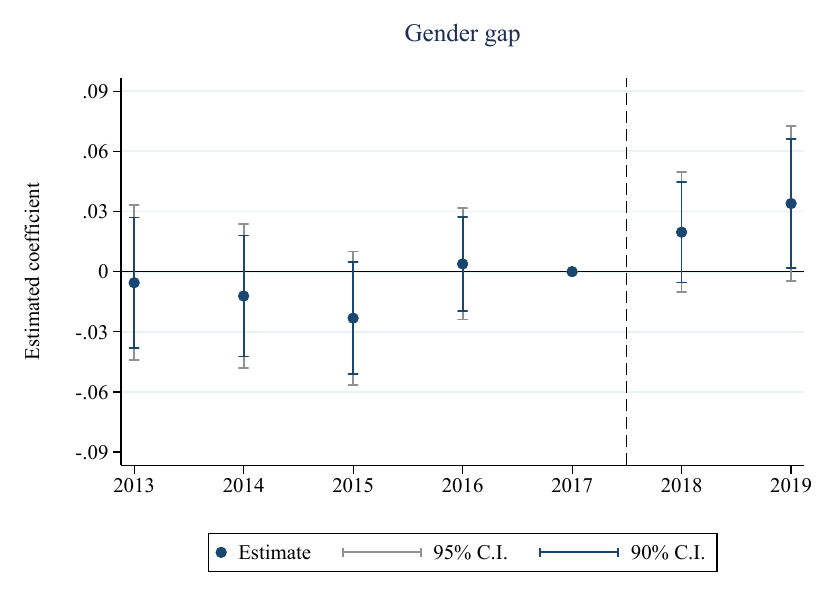} \hspace{0.3cm}\includegraphics[scale=0.42]{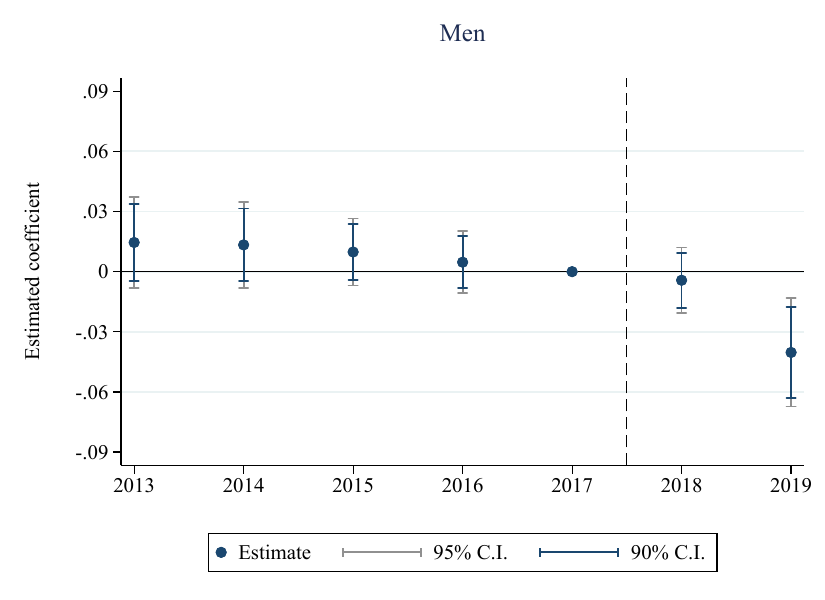}  \hspace{0.3cm} \includegraphics[scale=0.42]{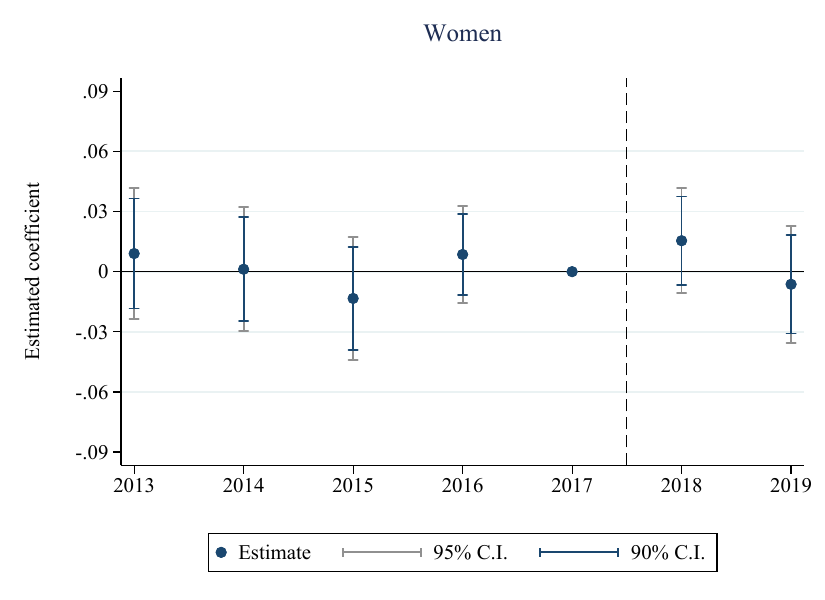}}
\caption{Log hourly basic pay}
\end{subfigure}\\
    \vspace{0.5cm}
\captionsetup[subfigure]{justification=centering}
\begin{subfigure}[b]{\textwidth}
\centerline{\includegraphics[scale=0.42]{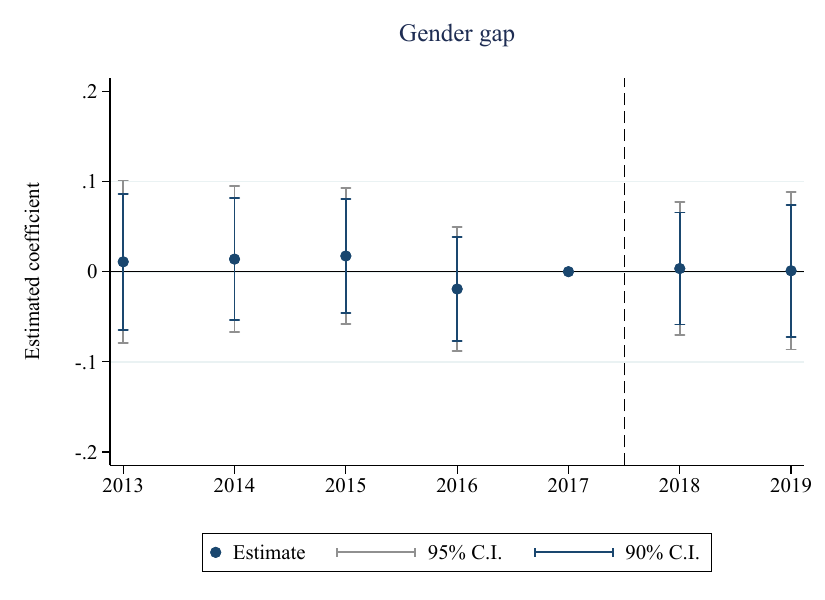} \hspace{0.3cm}\includegraphics[scale=0.42]{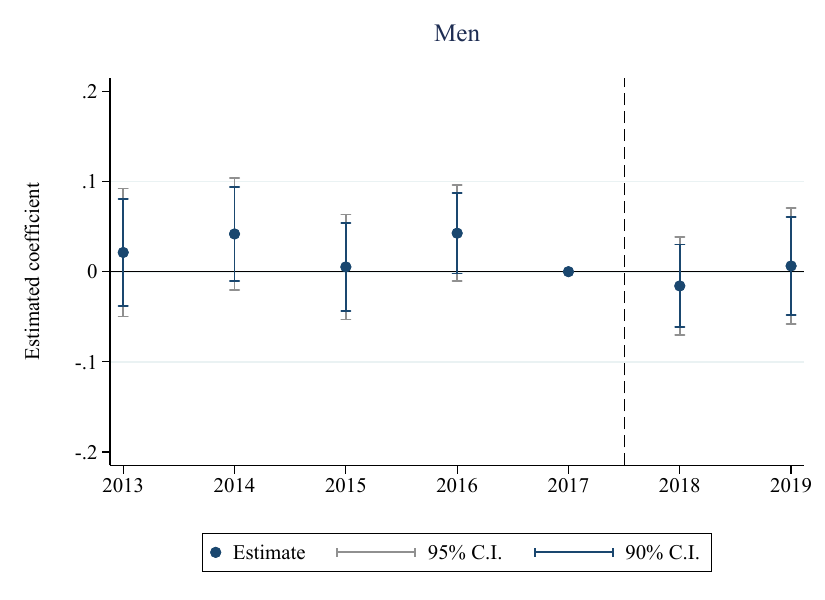}  \hspace{0.3cm} \includegraphics[scale=0.42]{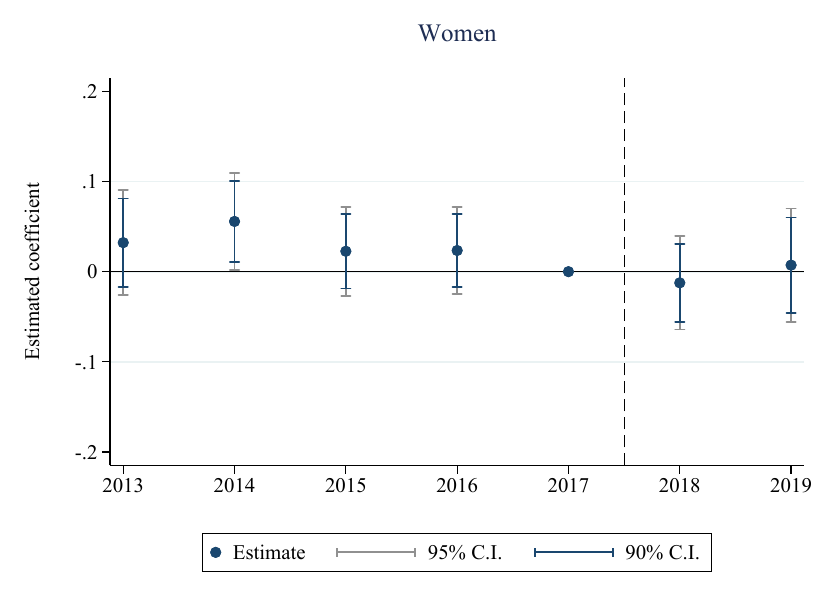}}
\caption{Allowances and bonuses}
\end{subfigure}\\
   \vspace{0.5cm}
\captionsetup[subfigure]{justification=centering}
\begin{subfigure}[c]{\textwidth}
\centerline{\includegraphics[scale=0.42]{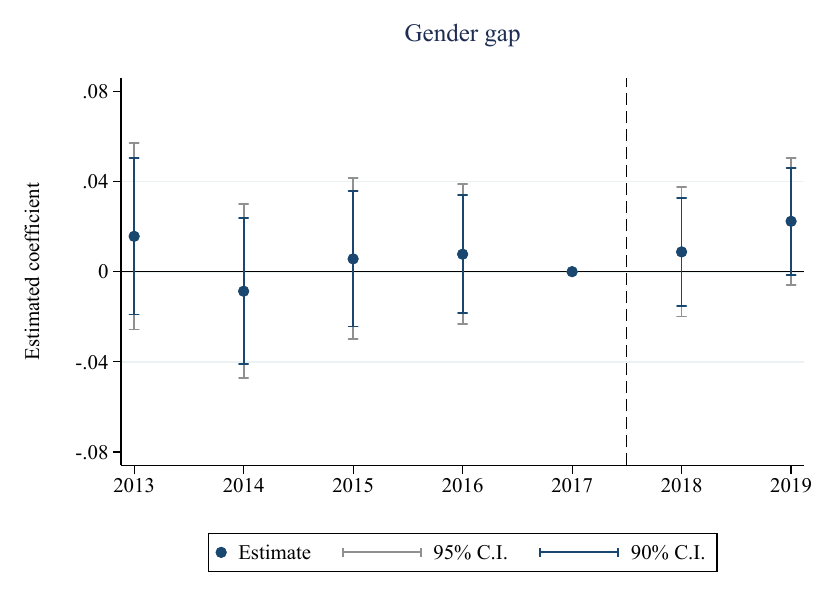} \hspace{0.3cm}\includegraphics[scale=0.42]{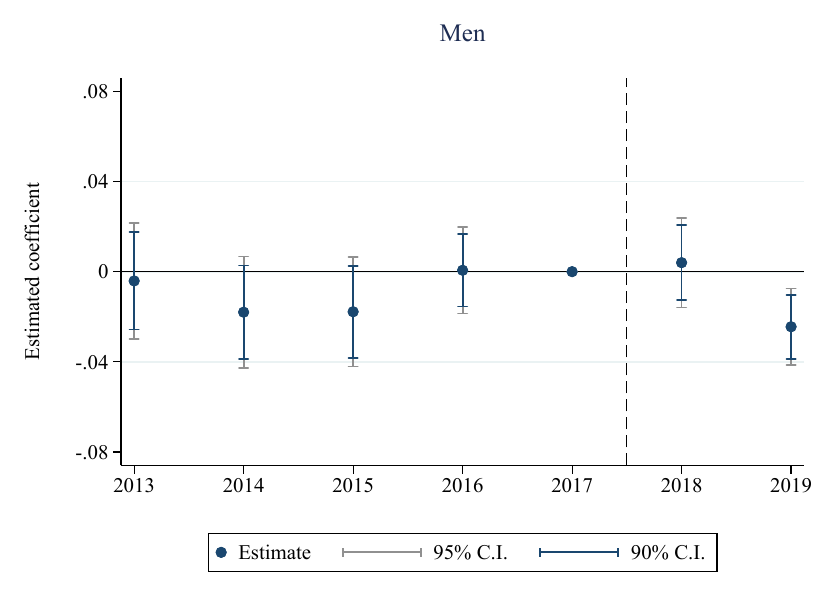}   \hspace{0.3cm} \includegraphics[scale=0.42]{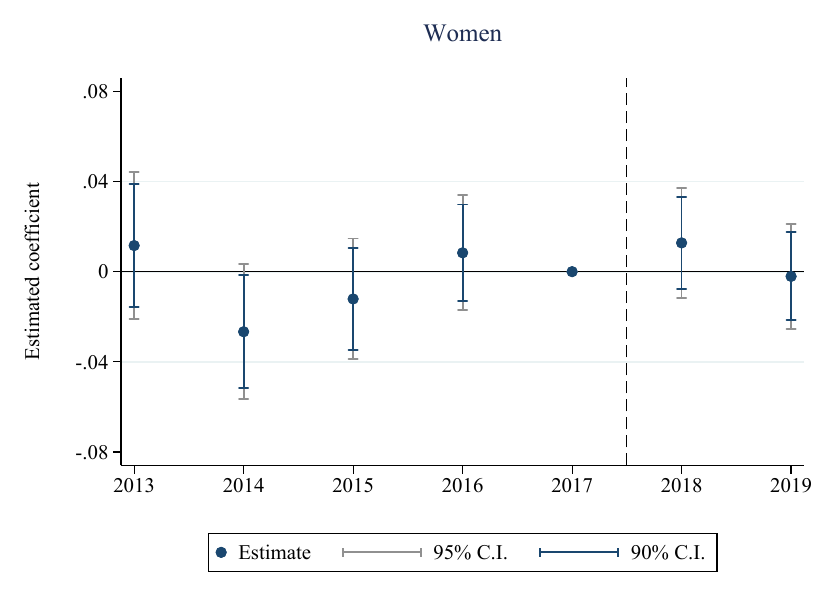}}
\caption{Promotion}
\end{subfigure}\\
    \vspace{0.5cm}
\floatfoot{{\textit{Source:} ASHE, 2013--2019.\\
\textit{Notes:} These graphs present the estimates of the leads and lags of the policy on different pay outcomes. These results are obtained from the estimation of regression \ref{event_study_reg}. In each graph, the estimation sample includes workers employed in firms with 200 to 300 employees. The graphs also report 90 and 95 percent confidence intervals associated with firm-level clustered standard errors. The dash vertical line indicates the month when the mandate is approved, i.e., February 2017.}}
\end{figure}
\clearpage
\newpage

\begin{figure}[H]
\caption{Unconditional trends in men's basic pay}\label{trends men basic pay} 
\centering

    \vspace{0.2cm}
 \captionsetup[subfigure]{justification=centering}
\begin{subfigure}[a]{\textwidth}
\centerline{\includegraphics[scale=0.6]{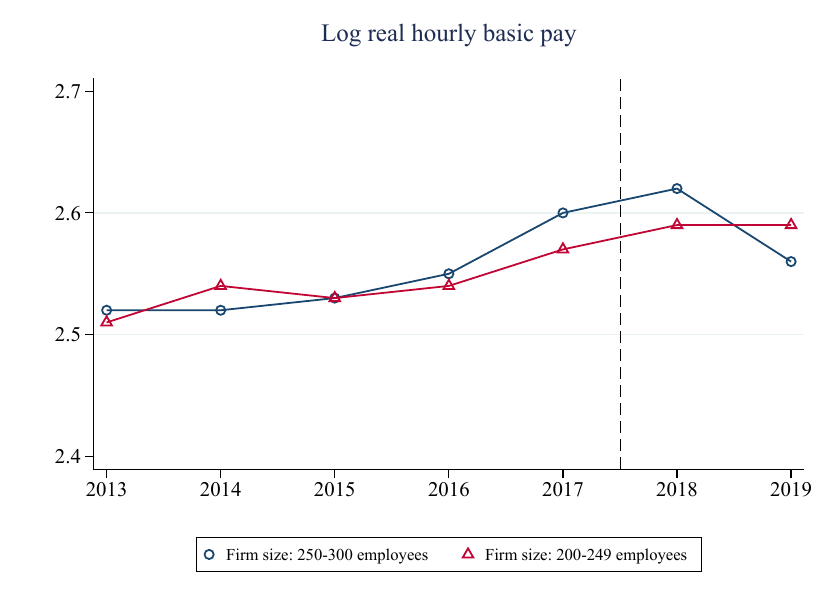} \hspace{0.3cm}\includegraphics[scale=0.6]{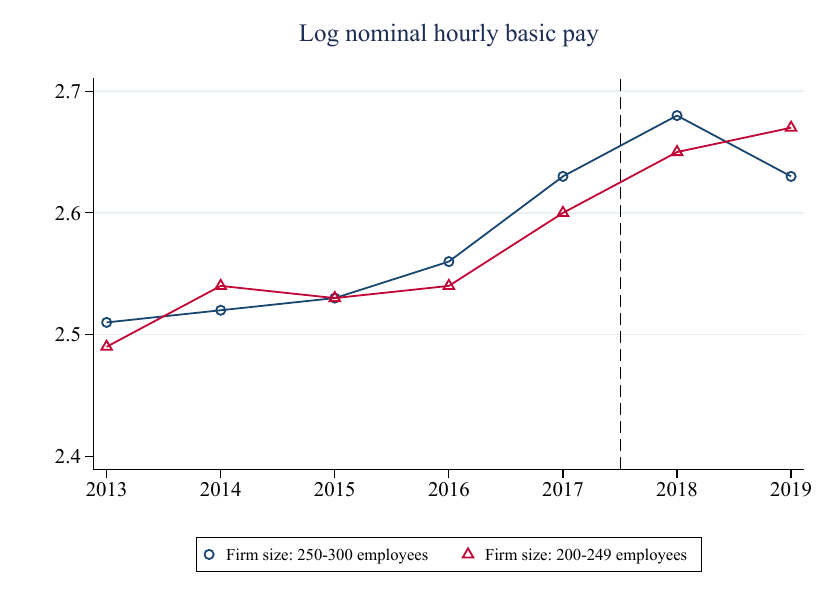}  }
\caption{Men's real vs. nominal pay}
\end{subfigure}\\
    \vspace{0.5cm}
\captionsetup[subfigure]{justification=centering}
\begin{subfigure}[b]{\textwidth}
\centerline{\includegraphics[scale=0.6]{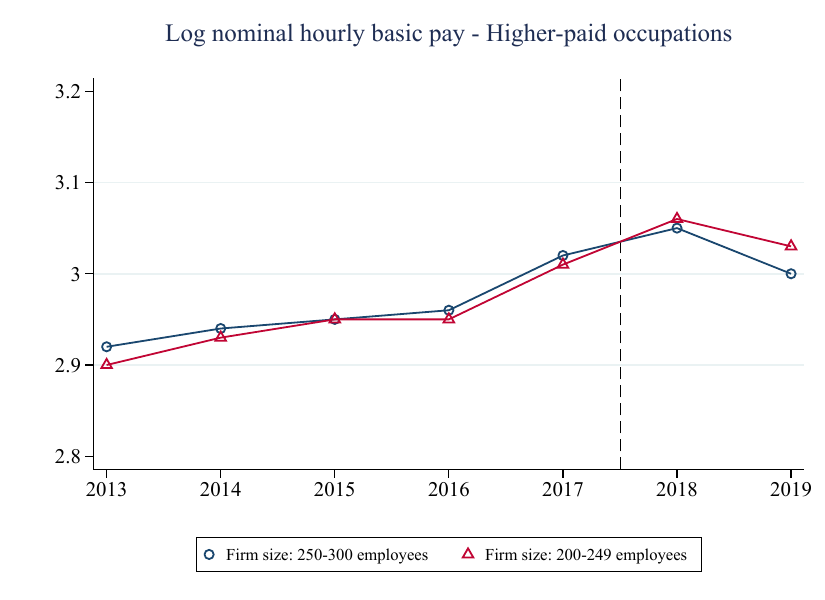} \hspace{0.3cm}\includegraphics[scale=0.6]{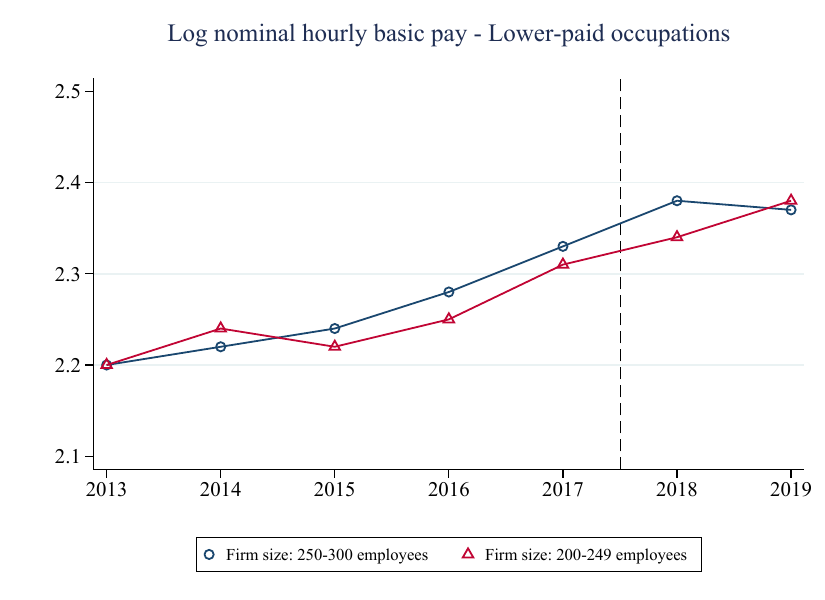}}
\caption{Men's nominal pay - higher vs. lower paid occupations}
\end{subfigure}\\
    \vspace{0.5cm}
\floatfoot{{\textit{Source:} ASHE, 2013--2019.\\
\textit{Notes:} The graphs in Panel A present the unconditional trends in men's log real and nominal basic pay, separately for treatment and control groups. The graphs in Panel B further compare trends in nominal basic pay across higher and lower-paid occupations. Higher paid occupations include managerial, professional, and technical occupations (1-digit SOC 1-3), while lower paid occupations, i.e. administrative, skilled-trades, caring and leisure, sales and customer service, plant and machine operative, and elementary occupations (1-digit SOC 4-9). In each graph, the blue line represents the treatment group, individuals working in firms with 250-300 employees, and the red line the control group, individuals working in firms with 200-249 employees. The dash vertical line indicates the month when the mandate is approved, i.e., February 2017.}}
\end{figure}
\clearpage
\newpage

\begin{figure}[H]
\caption{Unconditional trends in women's basic pay}\label{trends women basic pay} 
\centering

    \vspace{0.2cm}
 \captionsetup[subfigure]{justification=centering}
\begin{subfigure}[a]{\textwidth}
\centerline{\includegraphics[scale=0.6]{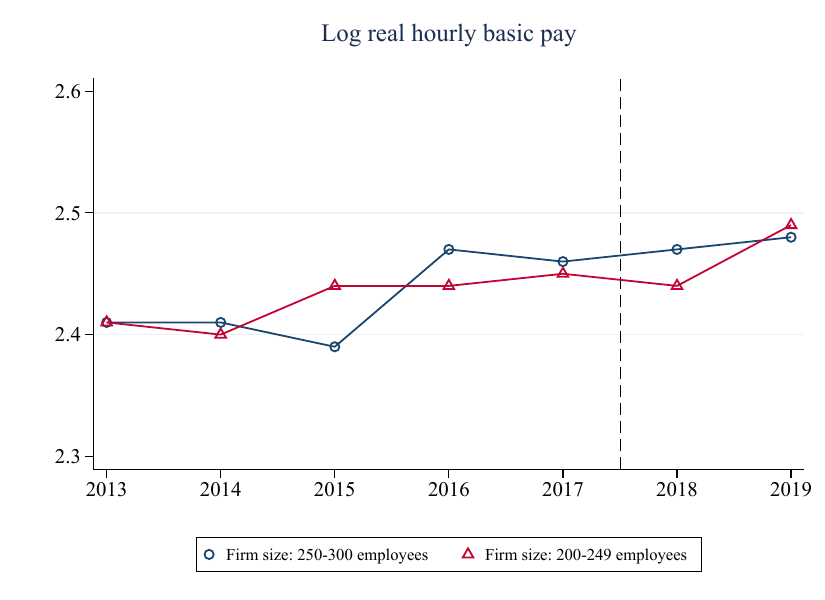} \hspace{0.3cm}\includegraphics[scale=0.6]{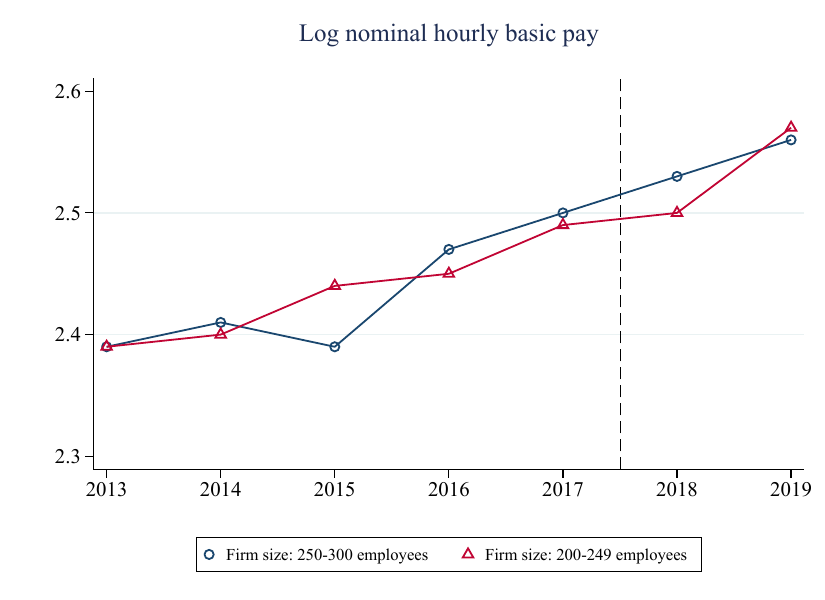}}
\caption{Women's real vs. nominal pay}
\end{subfigure}\\
    \vspace{0.5cm}
\captionsetup[subfigure]{justification=centering}
\begin{subfigure}[b]{\textwidth}
\centerline{\includegraphics[scale=0.6]{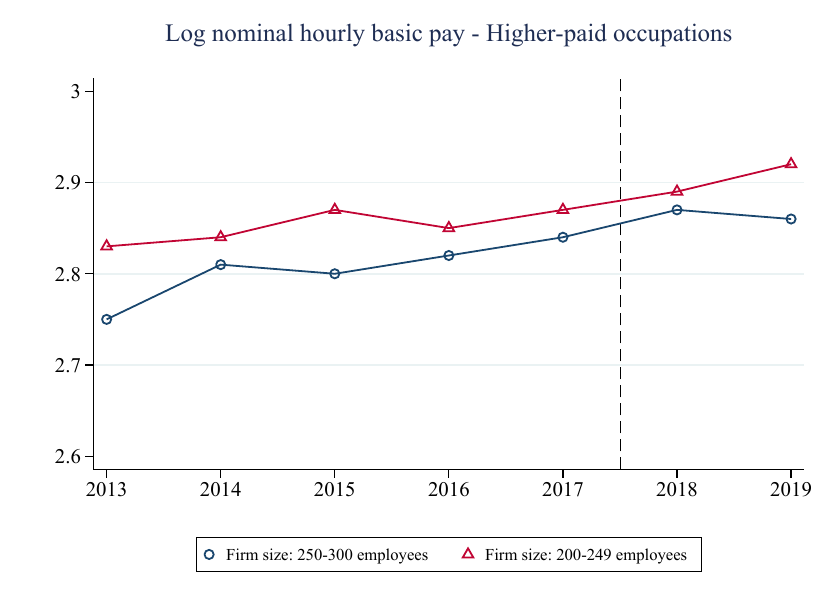} \hspace{0.3cm}\includegraphics[scale=0.6]{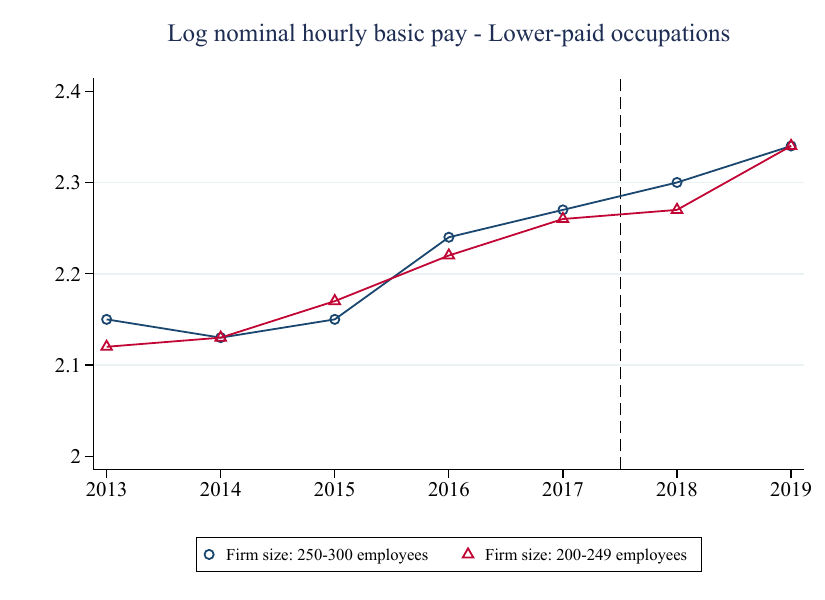}}
\caption{Women's nominal pay - higher vs. lower paid occupations}
\end{subfigure}\\
    \vspace{0.5cm}
\floatfoot{{\textit{Source:} ASHE, 2013--2019.\\
\textit{Notes:} The graphs in Panel A present the unconditional trends in women's log real and nominal basic pay, separately for treatment and control groups. The graphs in Panel B further compare trends in nominal basic pay across higher and lower-paid occupations. Higher paid occupations include managerial, professional, and technical occupations (1-digit SOC 1-3), while lower paid occupations, i.e. administrative, skilled-trades, caring and leisure, sales and customer service, plant and machine operative, and elementary occupations (1-digit SOC 4-9). In each graph, the blue line represents the treatment group, individuals working in firms with 250-300 employees, and the red line the control group, individuals working in firms with 200-249 employees. The dash vertical line indicates the month when the mandate is approved, i.e., February 2017.}}
\end{figure}
\clearpage
\newpage

\begin{figure}[H]
\caption{Estimation sample versus entire ASHE}\label{comp_samples}

\centering
    \vspace{0.2cm}
\captionsetup[subfigure]{justification=centering}

\begin{subfigure}[a]{\textwidth}
\centerline{\includegraphics[scale=0.43]{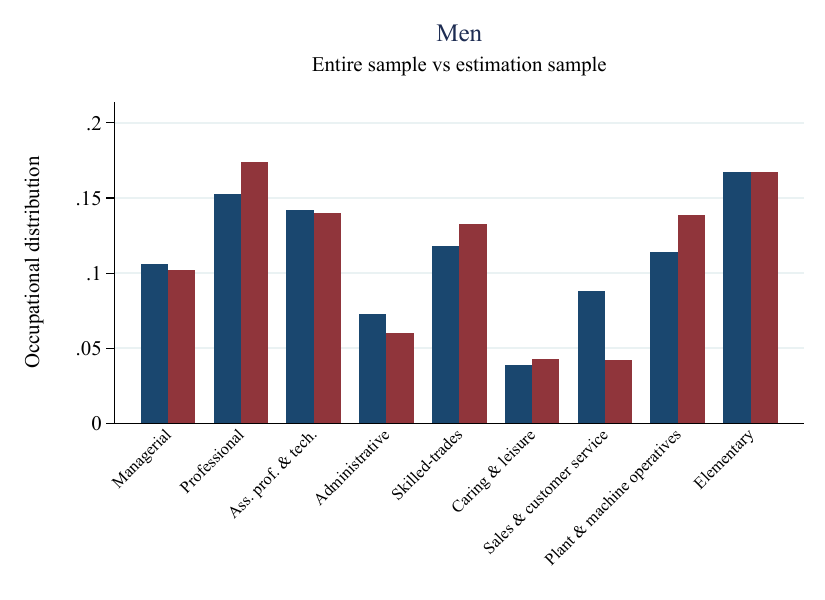}}
\centerline{\includegraphics[scale=0.43]{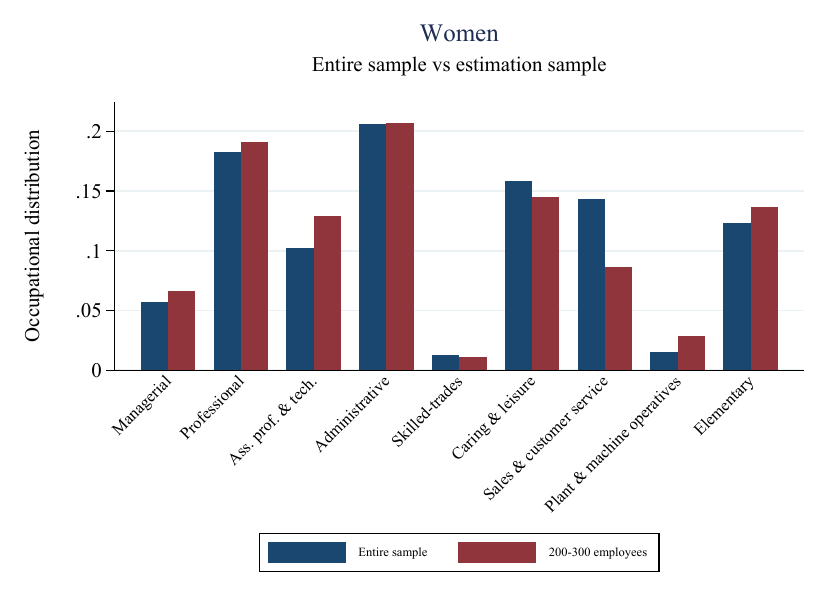}}
\caption{Occupational distribution}
\end{subfigure}\\
    \vspace{0.5cm}
\begin{subfigure}[b]{\textwidth}
\centerline{\includegraphics[scale=0.43]{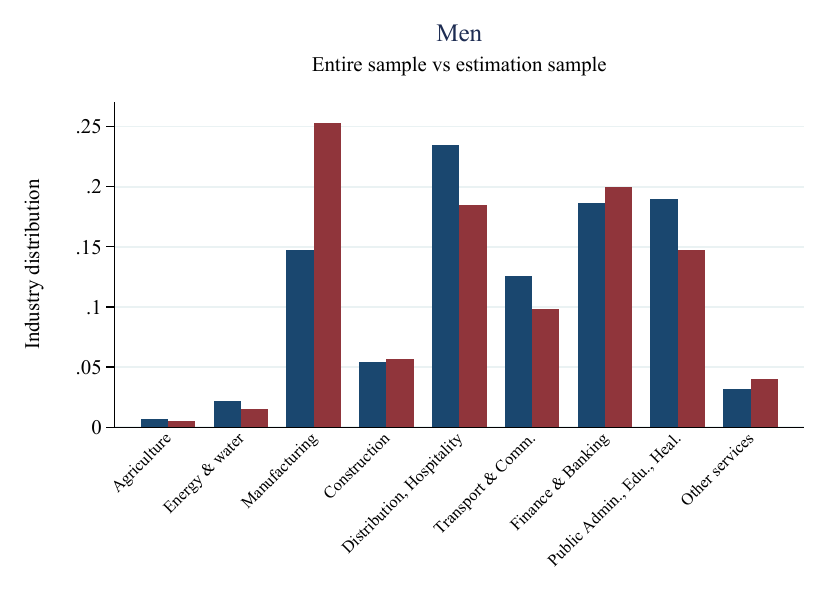}}
\centerline{\includegraphics[scale=0.43]{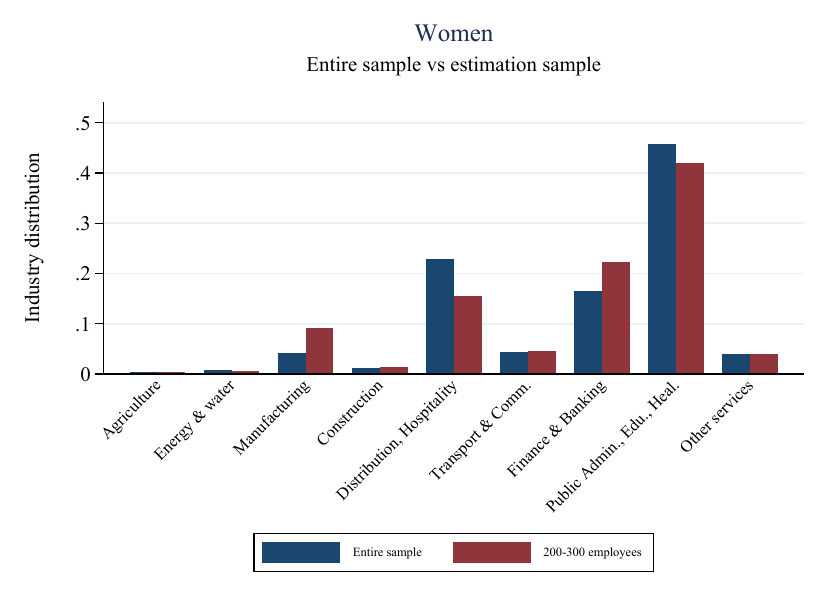}}
\caption{Industry distribution}
\end{subfigure}\\
    \vspace{0.2cm}
    
\floatfoot{\textit{Source:} ASHE, 2013--2019.\\
	\textit{Note:} These figures compare the occupational and industry distributions of men and women in the estimation sample and in the entire population of ASHE, over the period of analysis. }
\end{figure}
\clearpage
\newpage

\begin{table}[htbp]
\def\sym#1{\ifmmode^{#1}\else\(^{#1}\)\fi}
\caption{Impact on log hourly pay - extending treatment period}\label{including 2020-21}
\begin{threeparttable}
\begin{footnotesize}
\begin{tabular}{l*{3}{c}}
\toprule

                    &\multicolumn{1}{c}{2013-19}&\multicolumn{1}{c}{2013-20}&\multicolumn{1}{c}{2013-21}\\
                    &\multicolumn{1}{c}{(1)}&\multicolumn{1}{c}{(2)}&\multicolumn{1}{c}{(3)}\\
\midrule
Treated firm*post   &      -0.026\sym{***}&      -0.028\sym{***}&      -0.029\sym{***}\\
                    &     (0.008)         &     (0.009)         &     (0.009)         \\
\addlinespace
Treated firm*post*fem&       0.029\sym{**} &       0.031\sym{**} &       0.030\sym{**} \\
                    &     (0.014)         &     (0.014)         &     (0.013)         \\
\addlinespace
Observations        &       29,226         &       33,421         &       35,123         \\
Adjusted \(R^{2}\)  &       0.909         &       0.897         &       0.894         \\
P-value Women Coeff &       0.788         &       0.785         &       0.909         \\
Men's pre-policy mean&        2.58         &        2.58         &        2.58         \\
Women's pre-policy mean&        2.45         &        2.45         &        2.45         \\
\bottomrule
\end{tabular}
\end{footnotesize}
\begin{tablenotes}
\item{\footnotesize \textit{Source:} ASHE, 2013--2021.}
\item {\footnotesize \textit{Notes:} This table reports the impact of pay transparency on employees' hourly pay, obtained from the estimation of regression \ref{tripledid}. Each column refers to a different estimation period, as specified at the top of it. The estimation sample comprises men  and women working in firms that have between 200 and 300 employees. All regressions include firm*individual fixed effects, gender*year fixed effects, and region-specific time shocks. A treated firm is defined as having at least 250 employees in 2015. The post dummy is equal to one from 2018 onward. Heteroskedasticity-robust standard errors clustered at firm level in parentheses. The p-value at the bottom of the table refers to the t-test on the sum of the two reported coefficients, corresponding to the effect of the policy on female employees. The pre-policy mean represents the mean of the outcome variable for the treated group between 2013 and 2017.}
\item{\footnotesize *** p$<$0.01, ** p$<$0.05, * p$<$0.1.}
\end{tablenotes}
\end{threeparttable}
\end{table}
\clearpage
\newpage

\begin{table}[htbp]
\def\sym#1{\ifmmode^{#1}\else\(^{#1}\)\fi}
\caption{Impact on job mobility}\label{job mobility}
\begin{threeparttable}
\begin{footnotesize}
\begin{tabular}{l*{3}{c}}
\toprule
                   
                    &\multicolumn{1}{c}{Tenure}&\multicolumn{1}{c}{Leaving the firm}&\multicolumn{1}{c}{Managerial}\\
&\multicolumn{1}{c}{in months}&\multicolumn{1}{c}{in t+1}&\multicolumn{1}{c}{occupation}\\
                    &\multicolumn{1}{c}{(1)}&\multicolumn{1}{c}{(2)}&\multicolumn{1}{c}{(3)}\\
\midrule
Treated firm*post        &      -0.542         &      -0.002      &      -0.001    \\
                         &     (0.736)         &     (0.024)     &     (0.008)    \\
\addlinespace
Treated firm*post*fem    &       0.580         &       0.021     &       0.006    \\
                         &     (1.008)         &     (0.034)       &     (0.011)    \\
\addlinespace
Observations             &       29,226         &       28,714         &       29,226         \\
Adjusted \(R^{2}\)       &       0.989         &       0.111       &       0.898    \\
P-value Women Coeff      &       0.957         &       0.497      &       0.466    \\
Men's pre-policy mean    &       87.40         &        0.36       &        0.10     \\
Women's pre-policy mean  &       74.37         &        0.37      &        0.07    \\
\bottomrule
\end{tabular}
\end{footnotesize}
\begin{tablenotes}
\item{\footnotesize \textit{Source:} ASHE, 2013--2019.}
\item {\footnotesize \textit{Notes:} This table reports the impact of pay transparency on job mobility, obtained from the estimation of regression \ref{tripledid}. Each column refers to a different outcome, as specified at the top of it. The estimation sample comprises men  and women working in firms that have between 200 and 300 employees. All regressions include firm*individual fixed effects, gender*year fixed effects, and region-specific time shocks. A treated firm is defined as having at least 250 employees in 2015. The post dummy is equal to one from 2018 onward. Heteroskedasticity-robust standard errors clustered at firm level in parentheses. The p-value at the bottom of the table refers to the t-test on the sum of the two reported coefficients, corresponding to the effect of the policy on female employees. The pre-policy mean represents the mean of the outcome variable for the treated group between 2013 and 2017.}
\item{\footnotesize *** p$<$0.01, ** p$<$0.05, * p$<$0.1.}
\end{tablenotes}
\end{threeparttable}
\end{table}
\clearpage
\newpage

\begin{table}[htbp]
\def\sym#1{\ifmmode^{#1}\else\(^{#1}\)\fi}
\caption{Impact on pay and hours worked}\label{hours vs. pay}
\begin{threeparttable}
\begin{footnotesize}
\begin{tabular}{l*{4}{c}}
\toprule
                   
        &\multicolumn{1}{c}{Log hourly}&\multicolumn{1}{c}{Log weekly}&\multicolumn{1}{c}{Weekly}&\multicolumn{1}{c}{Part-time}\\
&\multicolumn{1}{c}{pay}&\multicolumn{1}{c}{ pay}&\multicolumn{1}{c}{hours}&\\
                    &\multicolumn{1}{c}{(1)}&\multicolumn{1}{c}{(2)}&\multicolumn{1}{c}{(3)}&\multicolumn{1}{c}{(4)}\\
\midrule
Treated firm*post   &      -0.026\sym{***} &      -0.019\sym{*}&      0.084    &      -0.001     \\
                    &     (0.008)    &     (0.010)         &     (0.009)     &     (0.009)     \\
\addlinespace
Treated firm*post*fem&       0.029\sym{**}  &       0.010 &       -0.461   &   0.016    \\
                    &     (0.014)         &     (0.018)         &     (0.371)    &     (0.016)      \\
\addlinespace
Observations        &       29,226         &       29,226         &       29,226     &       29,226     \\
Adjusted \(R^{2}\)  &       0.909         &       0.913        &       0.798     &       0.752     \\
P-value Women Coeff &       0.788          &       0.532         &       0.242     &       0.340   \\
Men's pre-policy mean&        15.93          &        581.51         &       36.41     &        0.10     \\
Women's pre-policy mean&        13.36         &        414.60         &        30.69    &        0.34     \\
\bottomrule
\end{tabular}
\end{footnotesize}
\begin{tablenotes}
\item{\footnotesize \textit{Source:} ASHE, 2013--2019.}
\item {\footnotesize \textit{Notes:} This table reports the impact of pay transparency on pay outcomes and hours worked, obtained from the estimation of regression \ref{tripledid}. Each column refers to a different outcome, as specified at the top of it. The estimation sample comprises men  and women working in firms that have between 200 and 300 employees. All regressions include firm*individual fixed effects, gender*year fixed effects, and region-specific time shocks. A treated firm is defined as having at least 250 employees in 2015. The post dummy is equal to one from 2018 onward. Heteroskedasticity-robust standard errors clustered at firm level in parentheses. The p-value at the bottom of the table refers to the t-test on the sum of the two reported coefficients, corresponding to the effect of the policy on female employees. The pre-policy mean represents the mean of the outcome variable for the treated group between 2013 and 2017.}
\item{\footnotesize *** p$<$0.01, ** p$<$0.05, * p$<$0.1.}
\end{tablenotes}
\end{threeparttable}
\end{table}
\clearpage
\newpage

\begin{table}[htbp]
\def\sym#1{\ifmmode^{#1}\else\(^{#1}\)\fi}
\caption{Impact on pay outcomes by occupation}\label{results by occupation}
\begin{threeparttable}
\begin{footnotesize}
\begin{tabular}{l*{6}{c}}
\toprule
                     
                    &\multicolumn{3}{c}{Log hourly pay}&\multicolumn{3}{c}{Promotion}\\
                    \cmidrule(l){2-4} \cmidrule(l){5-7}
                    &\multicolumn{1}{c}{Entire}&\multicolumn{1}{c}{Lower}&\multicolumn{1}{c}{Higher}&\multicolumn{1}{c}{Entire}&\multicolumn{1}{c}{Lower}&\multicolumn{1}{c}{Higher}\\
                     &\multicolumn{1}{c}{sample}&\multicolumn{1}{c}{paid}&\multicolumn{1}{c}{paid}&\multicolumn{1}{c}{sample}&\multicolumn{1}{c}{paid}&\multicolumn{1}{c}{paid}\\
                    &\multicolumn{1}{c}{(1)}&\multicolumn{1}{c}{(2)}&\multicolumn{1}{c}{(3)} &\multicolumn{1}{c}{(4)}&\multicolumn{1}{c}{(5)}&\multicolumn{1}{c}{(6)}\\
\midrule
Treated firm*post   &      -0.026\sym{***}&    -0.020\sym{*}&    -0.030\sym{**}  &      -0.002&    -0.002&    -0.002   \\
                    &     (0.008)         &     (0.010)         &     (0.014)    &     (0.008)         &     (0.010)         &     (0.012)   \\
\addlinespace
Treated firm*post*fem&       0.029\sym{**} &   0.017  &       0.046\sym{**} &  0.011 &   0.006  &       0.013   \\
                    &     (0.014)       &     (0.018)         &     (0.022)      &     (0.011)       &     (0.014)         &     (0.019)      \\
\addlinespace
Observations        &       29,226       &      16,623         &       12,082      &       29,226       &      16,623         &       12,082     \\
Adjusted \(R^{2}\)  &       0.909    &       0.776     &       0.904    &       0.003    &       0.025    &       -0.016            \\
P-value Women Coeff &       0.788      &       0.868     &       0.405    &       0.281     &       0.701     &       0.540    \\
P-value High vs. Low M &             & 0.554 &       &             & 0.968 &          \\
P-value High vs. Low W &             & 0.443 &     &             & 0.747 &         \\
Men's pre-policy mean&        15.93         &        10.51        &        23.53   &        0.03         &        0.03        &         0.03        \\
Women's pre-policy mean&        13.36      &        9.73        &        18.87   &         0.04      &        0.03         &        0.05      \\
\bottomrule
\end{tabular}
\end{footnotesize}
\begin{tablenotes}
\item{\footnotesize \textit{Source:} ASHE, 2013--2019.}
\item {\footnotesize \textit{Notes:} This table compares the impact of pay transparency on employees' hourly pay and the probability of getting promoted across occupations, by estimating regression \ref{tripledid} by subgroup. Column 1 (4) reports the estimate on log hourly pay (promotion) for the entire sample, employees working in firms that have between 200 and 300 employees. Columns 2 (5) and 3 (5) compare the impact across lower and higher-paid occupations, where this grouping is based on the ranking of pre-policy 1-digit SOC-specific median wages. All regressions include firm*individual fixed effects, gender*year fixed effects, and region-specific time shocks. A treated firm is defined as having at least 250 employees in 2015. The post dummy is equal to one from 2018 onward. Heteroskedasticity-robust standard errors clustered at firm level in parentheses. The `` P-value Women Coeff'' refers to the t-test on the sum of the two reported coefficients, corresponding to the effect of the policy on female employees. The ``P-value High vs. Low M (W)'' refers to the t-test on the equality of effects on men's (women's) pay across occupations. The pre-policy mean represents the mean of the outcome variable for the treated group and subgroup considered between 2013 and 2017.}
\item{\footnotesize *** p$<$0.01, ** p$<$0.05, * p$<$0.1.}
\end{tablenotes}
\end{threeparttable}
\end{table}
\clearpage
\newpage

\begin{table}[htbp]\centering
\def\sym#1{\ifmmode^{#1}\else\(^{#1}\)\fi}
\caption{{ Impact on log hourly pay - placebo regressions}}\label{placebo regressions}
\begin{threeparttable}
\begin{footnotesize}
\begin{tabular}{l*{9}{c}}
\toprule
                    
                    &\multicolumn{1}{c}{50}&\multicolumn{1}{c}{100}&\multicolumn{1}{c}{150}&\multicolumn{1}{c}{200}&\multicolumn{1}{c}{{\bf 250}}&\multicolumn{1}{c}{300}&\multicolumn{1}{c}{350}&\multicolumn{1}{c}{400}&\multicolumn{1}{c}{450}\\
                    &\multicolumn{1}{c}{(1)}&\multicolumn{1}{c}{(2)}&\multicolumn{1}{c}{(3)}&\multicolumn{1}{c}{(4)}&\multicolumn{1}{c}{(5)}&\multicolumn{1}{c}{(6)}&\multicolumn{1}{c}{(7)}&\multicolumn{1}{c}{(8)}&\multicolumn{1}{c}{(9)}\\
\midrule
Treated firm*post&      -0.005         &         -0.009              &         -0.003             &      -0.006                &                     -0.026\sym{***} &           -0.011          &           -0.013            &         0.001             &          -0.012            \\
                    &     (0.004)         &         (0.006)             &     (0.007)                   &     (0.008)                 &                    (0.008)  &             (0.010)        &          (0.010)           &         (0.012)            &        (0.016)              \\
\addlinespace
Treated firm*post*fem&      -0.005         &         0.005              &        -0.002              &             0.006         &                     0.029\sym{**}  &          0.005            &             0.017          &           0.015           &         0.028             \\
                    &     (0.006)         &          (0.009)           &        (0.010)             &          (0.013)             &                    (0.014) &         (0.015)            &           (0.018)            &         (0.018)            &           (0.021)           \\

\addlinespace
Observations        &      243,795         &       85,204         &       52,438         &       39,429         &       29,226         &       22,964         &       19,406         &       15,994         &       13,616         \\
Adjusted \(R^{2}\)  &       0.866         &       0.899         &       0.905         &       0.908         &       0.909         &       0.912         &       0.912         &       0.908         &       0.909         \\
P-value Women Coeff &       0.044         &       0.548         &       0.543         &       0.995         &       0.788         &       0.638         &       0.783         &       0.239         &       0.301         \\
Men's pre-policy mean&       14.86         &       15.36         &       15.80         &       15.71         &       15.93         &       15.66         &       16.03         &       16.04         &       16.05         \\
Women's pre-policy mean&       11.99         &       12.79         &       12.87         &       13.49         &       13.36         &       13.53         &       13.04         &       13.50         &       13.06         \\
\bottomrule
\end{tabular}
\end{footnotesize}
\begin{tablenotes}
\item{\footnotesize \textit{Source:} ASHE, 2013--2019.}
\item {\footnotesize \textit{Notes:} This table reports the impact of placebo policies on log hourly pay, obtained from the estimation of regression \ref{tripledid}. In each regression, the estimation sample comprises employees working in firms that have +/- 50 employees from the threshold $c$ specified at the top of each column.  All regressions include firm*individual fixed effects, gender*year fixed effects, and region-specific time shocks. A treated firm is defined as having at least 250 employees in 2015. The post dummy is equal to one from 2018 onward. Heteroskedasticity-robust standard errors clustered at firm level in parentheses. The p-value at the bottom of the table refers to the t-test on the sum of the two reported coefficients, corresponding to the effect of the policy on female employees. The pre-policy mean represents the mean of the outcome variable for the treated group between 2013 and 2017.}
\item{\footnotesize *** p$<$0.01, ** p$<$0.05, * p$<$0.1.}
\end{tablenotes}
\end{threeparttable}
\end{table}
\clearpage
\newpage

\begin{landscape}
\begin{table}[htbp]\centering
\def\sym#1{\ifmmode^{#1}\else\(^{#1}\)\fi}
\caption{{ Impact on log hourly pay - different bandwidths}}\label{changing bandwidth}
\begin{threeparttable}
\begin{footnotesize}
\begin{tabular}{l*{8}{c}}
\toprule
                    
                    &\multicolumn{1}{c}{30}&\multicolumn{1}{c}{40}&\multicolumn{1}{c}{ {\bf 50}}&\multicolumn{1}{c}{60}&\multicolumn{1}{c}{70}&\multicolumn{1}{c}{80}&\multicolumn{1}{c}{90}&\multicolumn{1}{c}{100}\\
                    &\multicolumn{1}{c}{(1)}&\multicolumn{1}{c}{(2)}&\multicolumn{1}{c}{(3)}&\multicolumn{1}{c}{(4)}&\multicolumn{1}{c}{(5)}&\multicolumn{1}{c}{(6)}&\multicolumn{1}{c}{(7)}&\multicolumn{1}{c}{(8)}\\
\midrule
Treated firm*post   &      -0.020\sym{*}  &      -0.020\sym{**} &      -0.026\sym{***}&      -0.022\sym{***}&      -0.023\sym{***}&      -0.024\sym{***}&      -0.019\sym{***}&      -0.018\sym{***}\\
                    &     (0.011)         &     (0.010)         &     (0.008)         &     (0.008)         &     (0.007)         &     (0.007)         &     (0.006)         &     (0.006)         \\
\addlinespace
Treated firm*post*fem&       0.038\sym{**} &       0.030\sym{*}  &       0.029\sym{**} &       0.022\sym{*}  &       0.019\sym{*}  &       0.023\sym{**} &       0.019\sym{*}  &       0.015         \\
                    &     (0.019)         &     (0.016)         &     (0.014)         &     (0.012)         &     (0.011)         &     (0.011)         &     (0.010)         &     (0.009)         \\
\addlinespace
Observations        &       15,867         &       22,533         &       29,226         &       36,201         &       43,478         &       50,693         &       58,291         &       66,553         \\
Adjusted \(R^{2}\)  &       0.909         &       0.911         &       0.909         &       0.907         &       0.909         &       0.908         &       0.908         &       0.908         \\
P-value Women Coeff &       0.249         &       0.444         &       0.788         &       0.979         &       0.662         &       0.952         &       0.961         &       0.629         \\
Men's pre-policy mean&        16.08         &        16.13        &        15.93        &        15.78        &        15.87         &        15.88        &        15.83         &        15.83         \\
Women's pre-policy mean&        13.37        &        13.37        &        13.36         &        13.37         &        13.46        &        13.45        &        13.43         &        13.40         \\
\bottomrule
\end{tabular}
\end{footnotesize}
\begin{tablenotes}
\item{\footnotesize \textit{Source:} ASHE, 2013--2019.}
\item {\footnotesize \textit{Notes:} This table reports the impact of pay transparency on log hourly pay, obtained from the estimation of regression \ref{tripledid}. In each regression, the estimation sample comprises individuals working in firms that have +/- $h$ employees from the 250-employee threshold, where $h$ is indicated at the top of each column.  All regressions include firm*individual fixed effects, gender*year fixed effects, and region-specific time shocks. A treated firm is defined as having at least 250 employees in 2015. The post dummy is equal to one from 2018 onward. Heteroskedasticity-robust standard errors clustered at firm level in parentheses. The p-value at the bottom of the table refers to the t-test on the sum of the two reported coefficients, corresponding to the effect of the policy on female employees. The pre-policy mean represents the mean of the outcome variable for the treated group between 2013 and 2017.}
\item{\footnotesize *** p$<$0.01, ** p$<$0.05, * p$<$0.1.}
\end{tablenotes}
\end{threeparttable}
\end{table} 
\end{landscape}
\clearpage
\newpage

\begin{table}[htbp]\centering
\def\sym#1{\ifmmode^{#1}\else\(^{#1}\)\fi}
\caption{{Impact on log hourly pay - changing year to define treatment status}}\label{treatment year}
\begin{threeparttable}
\begin{footnotesize}
\begin{tabular}{l*{4}{c}}
\toprule
                    
                    &\multicolumn{1}{c}{Main}&\multicolumn{1}{c}{Firm size}&\multicolumn{1}{c}{Firm size}&\multicolumn{1}{c}{Firm size}\\
                    &\multicolumn{1}{c}{spec}&\multicolumn{1}{c}{2014}&\multicolumn{1}{c}{2013}&\multicolumn{1}{c}{2012}\\
                    &\multicolumn{1}{c}{(1)}&\multicolumn{1}{c}{(2)}&\multicolumn{1}{c}{(3)}&\multicolumn{1}{c}{(4)}\\
\midrule
Treated firm*post   &      -0.026\sym{***} &       -0.020\sym{**}  &            -0.019\sym{**}    &   -0.032\sym{***}                                                           \\
                    &     (0.008)     &         (0.009)           &            (0.008)                    &           (0.009)                      \\
\addlinespace
Treated firm*post*fem&       0.029\sym{**}  &     0.017    &           0.028\sym{*}                           &         0.033\sym{**}                                      \\
                    &     (0.014)    &      (0.015)                            &          (0.014)                   &         (0.015)                       \\
\addlinespace
Observations        &       29,226     &       29,030    &       28,765  &       28,378                                          \\
Adjusted \(R^{2}\)  &       0.909      &       0.908   &       0.908      &       0.908                                    \\
P-value Women Coeff &       0.788     &       0.804   &       0.462      &       0.938                                       \\
Men's pre-policy mean&       15.93    &       15.80    &       16.09     &       16.21                                    \\
Women's pre-policy mean&       13.36     &       13.44    &       13.40      &       13.25                                   \\
\bottomrule
\end{tabular}
\end{footnotesize}
\begin{tablenotes}
\item{\footnotesize \textit{Source:} ASHE, 2013--2019.}
\item {\footnotesize \textit{Notes:} This table reports the impact of pay transparency on log hourly pay, obtained from the estimation of regression \ref{tripledid}. In each regression, the estimation sample comprises men and women working in firms that have between 200 and 300 employees.  All regressions include firm*individual fixed effects, gender*year fixed effects, and region-specific time shocks. A treated firm is defined as having at least 250 employees in 2015 or in the year indicated on top of each column. The post dummy is equal to one from 2018 onward. Heteroskedasticity-robust standard errors clustered at firm level in parentheses. The p-value at the bottom of the table refers to the t-test on the sum of the two reported coefficients, corresponding to the effect of the policy on female employees. The pre-policy mean represents the mean of the outcome variable for the treated group between 2013 and 2017.}
\item{\footnotesize *** p$<$0.01, ** p$<$0.05, * p$<$0.1.}
\end{tablenotes}
\end{threeparttable}
\end{table} 
\clearpage
\newpage

\begin{landscape}
\begin{table}[htbp]\centering
\def\sym#1{\ifmmode^{#1}\else\(^{#1}\)\fi}
\caption{{ Impact on log hourly pay - other robustness checks}}\label{other robustness checks}
\scalebox{0.9}{
\begin{threeparttable}
\begin{footnotesize}
\begin{tabular}{l*{9}{c}}
\toprule
                    
                    &\multicolumn{1}{c}{Main}&\multicolumn{1}{c}{1-digit SIC}&\multicolumn{1}{c}{Age}&\multicolumn{1}{c}{LFS}&\multicolumn{1}{c}{25}&\multicolumn{1}{c}{16-65}&\multicolumn{1}{c}{Private}&\multicolumn{1}{c}{Full-time}&\multicolumn{1}{c}{ASHE}\\
                    &\multicolumn{1}{c}{spec}&\multicolumn{1}{c}{FE}&\multicolumn{1}{c}{controls}&\multicolumn{1}{c}{weights}&\multicolumn{1}{c}{+}&&\multicolumn{1}{c}{sector}&&\multicolumn{1}{c}{only}\\
                    &\multicolumn{1}{c}{(1)}&\multicolumn{1}{c}{(2)}&\multicolumn{1}{c}{(3)}&\multicolumn{1}{c}{(4)}&\multicolumn{1}{c}{(5)}&\multicolumn{1}{c}{(6)}&\multicolumn{1}{c}{(7)}&\multicolumn{1}{c}{(8)}&\multicolumn{1}{c}{(9)}\\
\midrule
Treated firm*post   &      -0.026\sym{***}&      -0.026\sym{***}&      -0.020\sym{**} &      -0.026\sym{***}&      -0.019\sym{**} &      -0.026\sym{***}&      -0.028\sym{***}&      -0.021\sym{***}&         -0.025\sym{***}            \\
                    &     (0.008)         &     (0.009)         &     (0.008)         &     (0.009)         &     (0.008)         &     (0.009)         &     (0.009)         &     (0.008)         &         (0.009)                \\
\addlinespace
Treated firm*post*fem&       0.029\sym{**} &       0.028\sym{**} &       0.027\sym{**} &       0.029\sym{**} &       0.026\sym{*}  &       0.027\sym{*}  &       0.027\sym{*}  &       0.026\sym{*}  &        0.029\sym{**}             \\
                    &     (0.014)         &     (0.014)         &     (0.013)         &     (0.014)         &     (0.014)         &     (0.014)         &     (0.015)         &     (0.015)         &               (0.014)                \\
\addlinespace
Observations        &       29,226         &       29,226         &       29,220         &       29,226         &       26,695         &       28,561         &       24,984         &       23,084         &       25,845                  \\
Adjusted \(R^{2}\)  &       0.909         &       0.909         &       0.912         &       0.915         &       0.916         &       0.908         &       0.910         &       0.935         &       0.912                 \\
Firm*Ind FE         &  \checkmark         &  \checkmark         &  \checkmark         &  \checkmark         &  \checkmark         &  \checkmark         &  \checkmark         &  \checkmark         &  \checkmark                 \\
Year*Reg FE     &  \checkmark         &                     &  \checkmark         &  \checkmark         &  \checkmark         &  \checkmark         &  \checkmark         &  \checkmark         &  \checkmark                 \\
Year*SIC1 FE    &                     &  \checkmark         &                     &                     &                     &                     &                     &                     &                                          \\
P-value Women Coeff &       0.788         &       0.868         &       0.526         &       0.761         &       0.576         &       0.916         &       0.991         &       0.713         &       0.731                  \\
Men's pre-policy mean&       15.93         &       15.93         &       15.93         &       17.07         &       16.83         &       15.97         &       15.88         &       16.48         &       16.03                 \\
Women's pre-policy mean&       13.36         &       13.36         &       13.36         &       13.88         &       14.05         &       13.39         &       13.03         &       14.05         &       13.36        \\
\bottomrule
\end{tabular}
\end{footnotesize}
\begin{tablenotes}
\item{\footnotesize \textit{Source:} ASHE, 2013--2019.}
\item {\footnotesize \textit{Notes:} This table reports a series of robustness checks on the impact of pay transparency on log hourly pay, obtained from the estimation of regression \ref{tripledid}. In each regression, the estimation sample comprises men and women working in firms that have between 200 and 300 employees. A treated firm is defined as having at least 250 employees in 2015. The post dummy is equal to one from 2018 onward. All regressions include firm*individual fixed effects, gender*year fixed effects, and region-specific time shocks -- with the exception of Column 2 that controls for 1-digit SIC-specific time shocks. Heteroskedasticity- robust standard errors clustered at firm level in parentheses. The p-value at the bottom of the table refers to the t-test on the sum of the two reported coefficients, corresponding to the effect of the policy on female employees. The pre-policy mean represents the mean of the outcome variable for the treated group between 2013 and 2017.}
\item{\footnotesize *** p$<$0.01, ** p$<$0.05, * p$<$0.1.}
\end{tablenotes}
\end{threeparttable}
}
\end{table} 
\end{landscape}
\clearpage
\newpage

\section{Mechanisms}\label{appendix mechanisms}

\setcounter{table}{0}
\setcounter{figure}{0}

\subsection{Performance comparisons}\label{performance comparisons}

\begin{table}[htbp]\centering
\def\sym#1{\ifmmode^{#1}\else\(^{#1}\)\fi}
\caption{2018 Gender pay gap and changes in employees' pay}\label{behavioral response table}
\begin{threeparttable}
\begin{footnotesize}
\begin{tabular}{l*{3}{c}}
\toprule
                    
                    &\multicolumn{1}{c}{\% $\Delta$}&\multicolumn{1}{c}{\% $\Delta$}&\multicolumn{1}{c}{\% $\Delta$}\\
                     &\multicolumn{1}{c}{Gender pay gap}&\multicolumn{1}{c}{Men's pay}&\multicolumn{1}{c}{Women's pay}\\
                    &\multicolumn{1}{c}{(1)}&\multicolumn{1}{c}{(2)} &\multicolumn{1}{c}{(3)}\\
\midrule
2018 Gender pay gap &    -1.665\sym{***}    &   -0.057\sym{*} &    0.050\sym{*}     \\
                           &     (0.111)  &     (0.032)    &     (0.030)                \\
\addlinespace
Observations        &  9,599     &        4,163  &        3,890            \\

\bottomrule
\end{tabular}
\end{footnotesize}
\begin{tablenotes}
\item {\footnotesize \textit{Source:} UK Government Equalities Office (GEO), ASHE 2018-2019.}
\item {\footnotesize \textit{Notes:} This table shows the correlation between the publicly available 2018 gender median hourly pay gap and 2018-19 changes in, respectively, the firm's gender pay gap, men's and women's median hourly pay. All regressions control for 5-digit SIC fixed effects. The sample in Column 1 includes 9,599 firms that publish gender equality indicators in both 2018 and 2019. Columns 2 and 3 include the subgroup of these firms that are also present in ASHE in 2018 and 2019 with either male or female employees (respectively 4,163 and 3,890 firms). In all regressions, outliers (the bottom and top 1 percent) in the distribution of the y variable are excluded. Standard errors are clustered at the level of 5-digit SIC.}
\end{tablenotes}
\end{threeparttable}
\end{table}
\clearpage
\newpage

\subsection{YouGov data and firms' reputation}\label{appendix yougov}

\begin{figure}[H]
\caption{YouGov vs. GEO sample}\label{yougov distributions}
\centering
    \vspace{0.2cm}
\captionsetup[subfigure]{justification=centering}
\begin{subfigure}[a]{\textwidth}
\centerline{\includegraphics[scale=0.6]{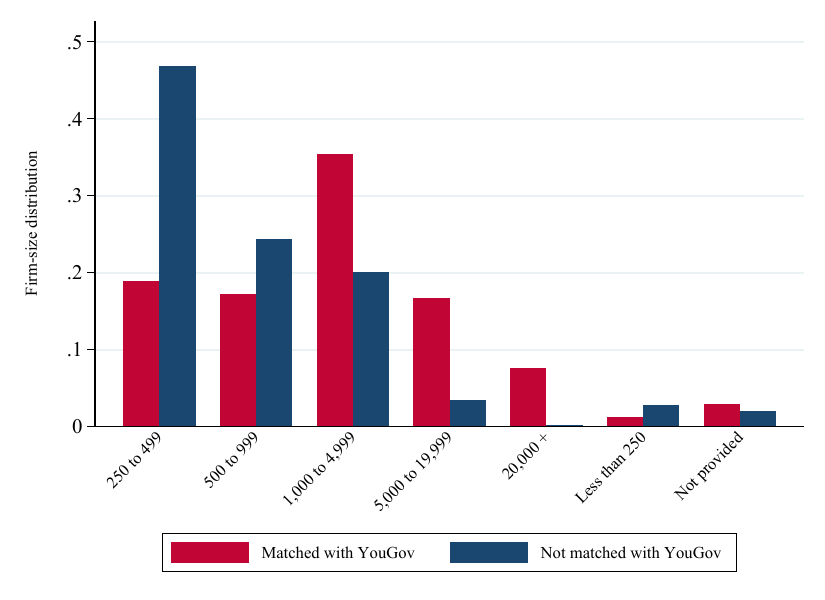}}
\caption{Firm-size distribution}
\end{subfigure}\\
    \vspace{0.5cm}
\begin{subfigure}[b]{\textwidth}
\centerline{\includegraphics[scale=0.6]{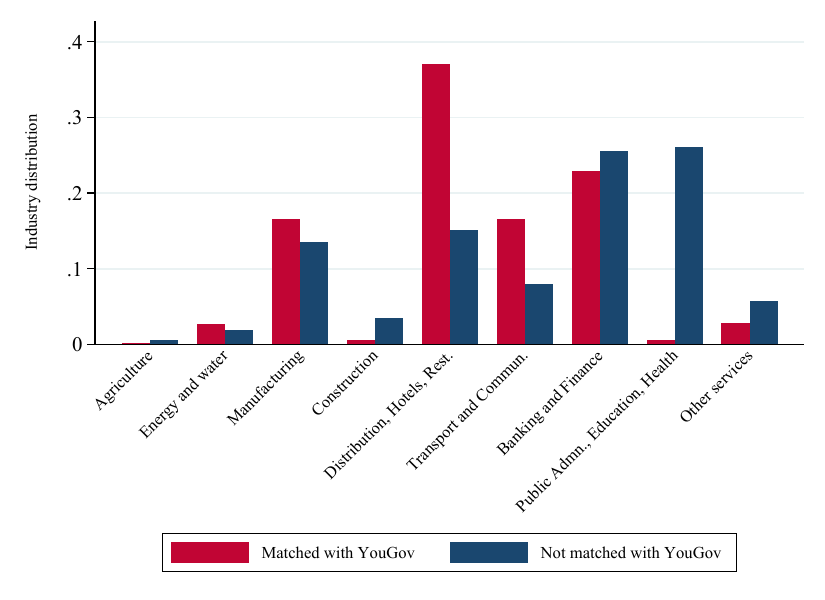}}
\caption{Industry distribution}
\end{subfigure}\\
    \vspace{0.2cm}
\floatfoot{\textit{Source:} UK Government Equality Office (GEO), YouGov 2018-2019.\\
	\textit{Note:} These figures compare the firm-size and industry distributions in the YouGov-GEO matched sample and in the entire GEO sample.}
\end{figure}
\clearpage
\newpage

\begin{table}[htbp]
\def\sym#1{\ifmmode^{#1}\else\(^{#1}\)\fi}
\caption{Gender equality performance and presence in YouGov}\label{selection_gpg_yougov}
\scalebox{0.9}{
\begin{threeparttable}
\begin{footnotesize}
\begin{tabular}{l*{8}{c}}
\toprule
                   &\multicolumn{4}{c}{2017/18}&\multicolumn{4}{c}{2018/19}\\
                  \cmidrule(l){2-5} \cmidrule(l){6-9} &\multicolumn{1}{c}{Entire}&\multicolumn{2}{c}{Matched with YouGov}&\multicolumn{1}{c}{P-value}&\multicolumn{1}{c}{Entire}&\multicolumn{2}{c}{Matched with YouGov}&\multicolumn{1}{c}{P-value}\\
                    &\multicolumn{1}{c}{sample}&\multicolumn{1}{c}{Yes}&\multicolumn{1}{c}{No}&\multicolumn{1}{c}{difference}&\multicolumn{1}{c}{sample}&\multicolumn{1}{c}{Yes}&\multicolumn{1}{c}{No}&\multicolumn{1}{c}{difference}\\
    
                    &\multicolumn{1}{c}{(1)}&\multicolumn{1}{c}{(2)}&\multicolumn{1}{c}{(3)}&\multicolumn{1}{c}{(4)}&\multicolumn{1}{c}{(5)}&\multicolumn{1}{c}{(6)}&\multicolumn{1}{c}{(7)}&\multicolumn{1}{c}{(8)}\\
                    
\midrule
Gender pay gap (\%) & 11.79 & 12.92 & 11.73 & 0.09 & 11.88 & 12.37 & 11.85 & 0.46\\
 & (15.84) & (13.80) & (15.94) &  & (15.51) & (13.38) & (15.61) & \\
\addlinespace

Observations & 10,557 & 540 & 10,017 &  & 10,812 & 527 & 10,285 & \\
\bottomrule
\end{tabular}
\end{footnotesize}
\begin{tablenotes}
\item{\footnotesize \textit{Source:} GEO, YouGov 2018--2019.}
\item {\footnotesize \textit{Notes:} This table explores potential selection patterns of GEO firms matched with YouGov. Column 1 (5) reports the gender median hourly pay gap for all GEO firms in 2017/2018 (2018/2019); Column 2 (6) refers to firms matched with YouGov; Column 3 (7) refers to firms that we do not find in YouGov; Column 4 (8) reports the p-value of the difference in the sample means of these two groups. }
\item{\footnotesize *** p$<$0.01, ** p$<$0.05, * p$<$0.1.}
\end{tablenotes}
\end{threeparttable}}
\end{table}
\clearpage
\newpage

\begin{table}[htbp]\centering
\def\sym#1{\ifmmode^{#1}\else\(^{#1}\)\fi}
\caption{Gender pay gap and placement in YouGov Rankings}\label{Corr YouGov GPG}
\begin{threeparttable}
\begin{footnotesize}
\begin{tabular}{l*{2}{c}}
\toprule
                    
                    &\multicolumn{1}{c}{Women's Rankings}&\multicolumn{1}{c}{Workforce Rankings}\\
                    &\multicolumn{1}{c}{(1)}&\multicolumn{1}{c}{(2)}\\
\midrule
Gender pay gap      &      -0.672\sym{*}  &      -0.794\sym{**} \\
                    &     (0.378)         &     (0.373)         \\
\addlinespace
Observations        &        1,830         &        1,836         \\
Year FE             &  \checkmark         &  \checkmark         \\
GEO firm FE         &  \checkmark         &  \checkmark         \\
\bottomrule
\end{tabular}
\end{footnotesize}
\begin{tablenotes}
\item{\footnotesize \textit{Source:} UK Government Equalities Office (GEO), YouGov, 2018--2019.}
\item {\footnotesize \textit{Notes:} This table shows the correlation between firms’ gender median hourly pay gap and, respectively, firms' placement in  YouGov Women’s Rankings (Column 1), and YouGov Workforce Rankings (Column 2). The gender pay gap is expressed relative to men’s pay. Firms' placement in YouGov Rankings is measured such that a smaller number indicates a lower position in the ranking. In each column, the sample includes YouGov Rankings’ firms that either publish directly or have a parent company that publishes gender equality indicators in at least one year. Each year, data for YouGov Women's rankings are missing for 3 firms, compared to the list of employers included in YouGov Workforce Rankings. Standard errors are clustered at the level of the GEO company.}
\item{\footnotesize *** p$<$0.01, ** p$<$0.05, * p$<$0.1.}
\end{tablenotes}
\end{threeparttable}
\end{table}
\clearpage
\newpage

\subsection{Firms' response to public scrutiny}\label{public scrutiny}

\begin{table}[htbp]
\def\sym#1{\ifmmode^{#1}\else\(^{#1}\)\fi}
\caption{Impact on log hourly pay by industry}\label{results by industry}
\begin{threeparttable}
\begin{footnotesize}
\begin{tabular}{l*{3}{c}}
\toprule

                    &\multicolumn{1}{c}{Entire sample}&\multicolumn{1}{c}{Low exposure}&\multicolumn{1}{c}{High exposure}\\
                    &\multicolumn{1}{c}{(1)}&\multicolumn{1}{c}{(2)}&\multicolumn{1}{c}{(3)}\\
\midrule
Treated firm*post   &      -0.026\sym{***}&    -0.016\sym{*}&    -0.052\sym{***}   \\
                    &     (0.008)         &     (0.010)         &     (0.017)  \\
\addlinespace
Treated firm*post*fem&       0.029\sym{**} &   0.030\sym{*}  &       0.050\sym{*}   \\
                    &     (0.014)       &     (0.016)         &     (0.026)      \\
\addlinespace
Observations        &       29,226       &      18,900         &       10,189         \\
Adjusted \(R^{2}\)  &       0.909    &       0.902      &       0.918         \\
P-value Women Coeff &       0.788       &       0.322       &       0.903      \\
P-value High vs. Low M &             & 0.056 &         \\
P-value High vs. Low W &             & 0.527 &        \\
Men's pre-policy mean&        15.93         &        15.56        &        16.39       \\
Women's pre-policy mean&        13.36      &        13.30         &        13.46        \\
\bottomrule
\end{tabular}
\end{footnotesize}
\begin{tablenotes}
\item{\footnotesize \textit{Source:} ASHE, 2013--2019.}
\item {\footnotesize \textit{Notes:} This table compares the impact of pay transparency on employees' hourly pay across industries, by estimating regression \ref{tripledid} by subgroup. The first column reports the estimate for the entire sample, employees working in firms that have between 200 and 300 employees. Columns 2 and 3 compare the impact across industries that are less or more exposed to publicly scrutiny, based on their presence in YouGov surveys (see Section \ref{mechanisms} for the definition of these two groups). All regressions include firm*individual fixed effects, gender*year fixed effects, and region-specific time shocks. A treated firm is defined as having at least 250 employees in 2015. The post dummy is equal to one from 2018 onward. Heteroskedasticity-robust standard errors clustered at firm level in parentheses. The `` P-value Women Coeff'' refers to the t-test on the sum of the two reported coefficients, corresponding to the effect of the policy on female employees. The ``P-value High vs. Low M (W)'' refers to the t-test on the equality of effects on men's (women's) pay across industries. The pre-policy mean represents the mean of the outcome variable for the treated group and subgroup considered between 2013 and 2017. }
\item{\footnotesize *** p$<$0.01, ** p$<$0.05, * p$<$0.1.}
\end{tablenotes}
\end{threeparttable}
\end{table}
\clearpage
\newpage

\section{Firms' profits}\label{impact on profits} 

\setcounter{table}{0}
\setcounter{figure}{0}
\setcounter{footnote}{0}

Our analysis shows that pay transparency pushes firms to reduce men's pay growth. A natural question to ask is whether this reduction of the wage bill translates into higher profits. Answering this question is important to assess the overall welfare implications of pay transparency policy. Moreover, the answer to this question is not obvious, as many studies outside of the gender literature show that pay transparency policies have a priori ambiguous effects on workers' productivity (\citealt{card2012inequality}, \citealt{breza2018morale}, \citealt{dube2019fairness}, \citealt{cullen2018much}.). On the one hand, learning about pay inequality in the workplace may decrease the job satisfaction and productivity of lower paid employees, while the higher-paid may feel threatened by any attempt of the employer to mitigate inequality. On the other hand, if firms respond to the policy by improving gender equality, this could boost the productivity of those workers who care about working in a fair environment.

While the data available do not allow us to rigorously estimate the impact of the policy on labour productivity, in Table \ref{impact firm level outcomes}, we study its implications for firms' wage bill and profits by estimating the following difference-in-differences model on the sample of firms with +/- 50 employees from the 250-employee cutoff:
\begin{equation}\label{reg ABS}
Y_{jt}=\alpha_{j}+\theta_{t}+\beta(TreatedFirm_{j}*Post_{t})+Z_{jt}'\delta+u_{jt}, \tag{C.1} 
\end{equation} 

\noindent where $Y_{jt}$ is either labor costs or profits of firm $j$ in year $t$; $\alpha_{j}$ and $\theta_{t}$ are firm and year fixed effects respectively, and $Z_{jt}$ includes region-specific time shocks. As in the ASHE analysis, $TreatedFirm_{j}$ is a dummy equal to one if a firm has at least 250 employees in 2015, and $Post_{t}$ is a dummy equal to one from the second quarter of 2018 onward. Standard errors are clustered at the firm level. 

For this part of the analysis, we use the Annual Business Survey (ABS),\footnote{Office for National Statistics. (2021). Annual Business Survey, 2005-2019: Secure Access. [data collection]. 15th Edition. UK Data Service. SN: 7451, DOI: 10.5255/UKDA-SN-7451-15.}, an annual survey of businesses covering the production, construction, distribution, and service industries, which represent about two-thirds of the UK economy in terms of gross value added (GVA). We measure labor costs using the log of wage costs, and consider the inverse hyperbolic transformation of GVA to account for negative profits. All monetary values are deflated using the ONS 2015 CPI Index.

\begin{figure}[H]
\caption{Event studies - firm-level outcomes}\label{event studies firm-level outcomes}
\centering

    \vspace{0.2cm}
 \captionsetup[subfigure]{justification=centering}
\begin{subfigure}[a]{\textwidth}
\centerline{\includegraphics[scale=0.6]{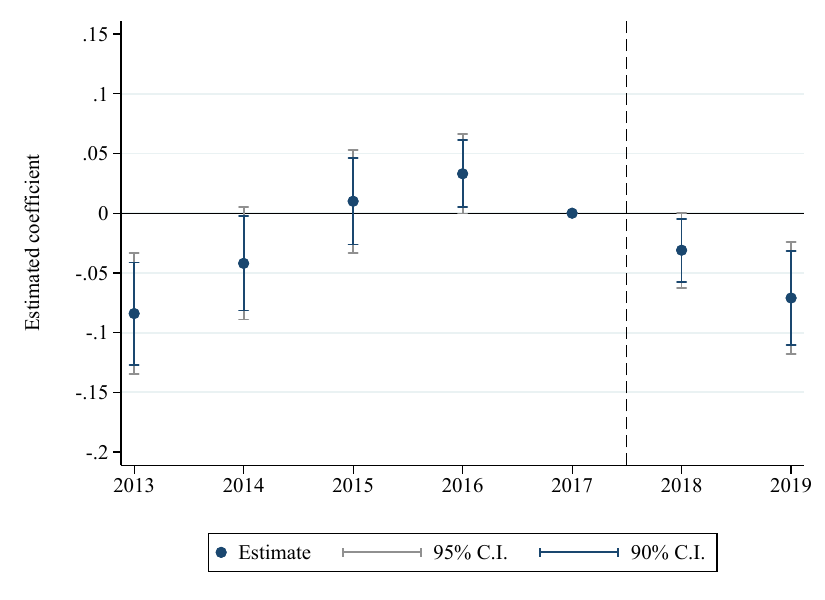}}
\caption{Wage costs}
\end{subfigure}\\
    \vspace{0.5cm}
\captionsetup[subfigure]{justification=centering}
\begin{subfigure}[b]{\textwidth}
\centerline{\includegraphics[scale=0.6]{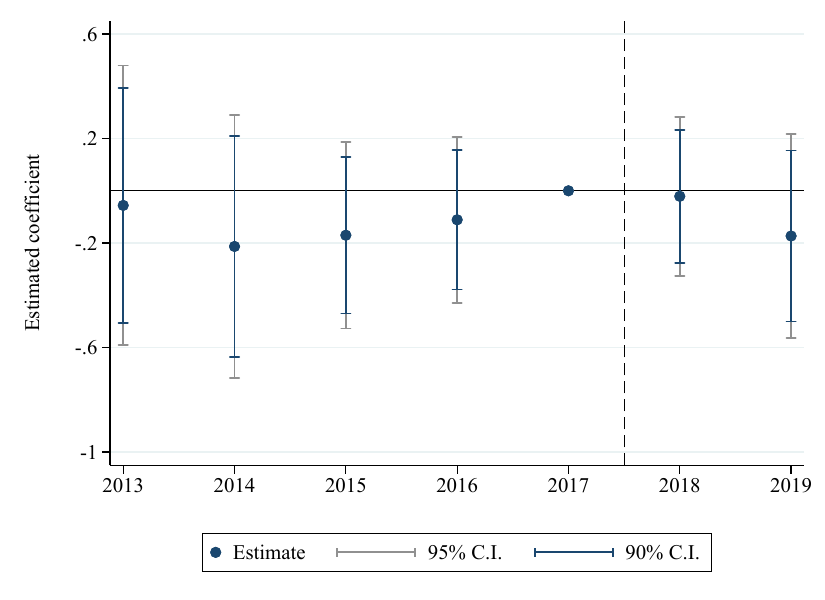}}
\caption{Profits}
\end{subfigure}\\
    \vspace{0.5cm}
\floatfoot{{\textit{Source:} ABS, 2013--2019.\\
\textit{Notes:} These graphs present the estimates of the leads and lags of the policy on firm-level outcomes. These results are obtained from the estimation of the dynamic version of regression \ref{reg ABS}. In each graph, the estimation sample includes firms with 200 to 300 employees. The graphs also report 90 and 95 percent confidence intervals associated with firm-level clustered standard errors. The dash vertical line indicates the month when the mandate is approved, i.e., February 2017.}}
\end{figure}

Table \ref{impact firm level outcomes} offers two insights. First, in line with the negative effect that we find on men's pay, the policy leads to a significant reduction in firms' wage bill.\footnote{Admittedly, point estimates are larger than what a 2.6 percent reduction in the pay of male employees could imply. As shown in Figure \ref{event studies firm-level outcomes}, these estimates may in part capture the effect of pre-policy differential trends between treated and control firms.} Second and despite this, the policy has no impact on firms' profits, which points to a negative effect on employees' productivity. The studies cited above show that pay transparency reduces workers’ effort only when these cannot assess what determines pay differentials across colleagues (\citealt{card2012inequality}, \citealt{dube2019fairness}). In contrast, when workers can clearly perceive that their higher-paid peers are more productive than themselves, learning about pay disparity has no detrimental impact on their effort (\citealt{breza2018morale}, \citealt{cullen2018much}). As discussed in Section \ref{institutional_setting}, the fact that the UK transparency policy imposes firms to publish only their raw gender pay gap, rather than pay gaps within profession and hierarchy position, does not allow to distinguish what part of the gap is due to occupation or experience differences across genders, and what part could instead depend on factors such as explicit discrimination, subtle discrimination, or implicit biases. In turn, this opacity of the transparency indicators may contribute to explain the likely detrimental effect on workers' productivity. 

\clearpage
\newpage

\begin{table}[htbp]
\def\sym#1{\ifmmode^{#1}\else\(^{#1}\)\fi}
\caption{Impact on firm-level outcomes}\label{impact firm level outcomes}
\begin{threeparttable}
\begin{footnotesize}
\begin{tabular}{l*{2}{c}}
\toprule
                    
                    &\multicolumn{1}{c}{Wage costs}&\multicolumn{1}{c}{Profits}\\
                    &\multicolumn{1}{c}{(1)}&\multicolumn{1}{c}{(2)}\\
\midrule
Treated firm*post   &      -0.050\sym{***} &      -0.002         \\
                    &     (0.015)   &     (0.137)                \\
\addlinespace
Observations        &       10,397         &       10,397         \\
Adjusted \(R^{2}\)  &       0.858    &       0.695               \\
Pre-policy mean     &        8.78   &        8.86                \\
\bottomrule
\end{tabular}
\end{footnotesize}
\begin{tablenotes}
\item{\footnotesize \textit{Source:} ABS 2013--2019.}
\item {\footnotesize \textit{Notes:} This table reports the impact of pay transparency on firms' profits and labor costs, obtained from the estimation of regression \ref{reg ABS}. Each column refers to a different outcome, as specified at the top of it. The estimation sample comprises firms that have between 200 and 300 employees. All regressions include year and firm fixed effects, and region-specific time shocks. A treated firm is defined as having at least 250 employees in 2015. The post dummy is equal to one from 2018 onward. Heteroskedasticity-robust standard errors clustered at firm level in parentheses. The pre-policy mean represents the mean of the outcome variable for the treated group between 2013 and 2017.}
\item{\footnotesize *** p$<$0.01, ** p$<$0.05, * p$<$0.1.}
\end{tablenotes}
\end{threeparttable}
\end{table}
\clearpage
\newpage

\section{Burning Glass Technologies data}\label{appendix BGT}

\setcounter{table}{0}
\setcounter{figure}{0}

\subsection{Name matching algorithm}\label{name matching algorithm}

We merge two different firm-level data sets, A and B, through the only common identifier available: firm name. We first collapse all firm names in each data set down to a unique set of firm names using standard text cleaning procedures; this includes dropping any exact duplicates. We then use firm names from one of the datasets, A, to define a vector space using all character-level 1--4-grams (with a maximum of 30,000 features) to create a matrix with dimensions number of entries in A times number of text features. This is achieved using Python's scikit-learn's \citep{scikit-learn} TF-IDF Vectorizer, so that frequently appearing 1--4 character grams are down-weighted. As the final stage of preparation for matching, the cleaned firm names in B are expressed in the vector space defined by the cleaned firm names from A. 

To  perform the matching, we use cosine similarity. Note that this involves taking the inner vector product of every firm name in A with every firm name in B so is computationally intensive. To facilitate this, we use the sparse\_dot\_topn package, developed by ING Bank, to perform parallel computation of the closest matches across A and B.

The result is an array of scores of the firm name matches between A and B that we are then able to use at different thresholds according to how close a match we prefer, with unity reflecting a perfect match in the vector space, and 0 reflecting two firm names that are entirely orthogonal in the vector space.

\subsection{GEO firms in BGT}\label{GPG matched with BGT}

In order to explore correlations between firms' hiring practices and gender equality indicators, we match GEO firms with BGT using the name-matching strategy explained above. We retain only employers with a match score of one, for a total of 6,852 GEO firms, of which 5,107 publish gender equality indicators in both 2017/18 and 2018/19 and have non-missing SIC and SOC codes. Table \ref{selectivity_gpg_bgt} explores selection patterns of the matched sample. While GEO firms matched with BGT have, on average, a larger and statistically different gender pay gap in both years than firms that do not match with BGT, the percentage of women in the top quartile of the firm wage distribution is not statistically different across the two groups.

\subsection{Difference-in-differences sample}\label{Did sample BGT}

To implement the difference-in-differences strategy, we need to identify a treatment and a control group in BGT, and for this we need information on firms' number of employees. We now describe in detail how we create this sample, putting particular care in explaining why we need to exclude some firms at each stage, and what impact this could have in terms of sample selection. 

\begin{figure}[H]
\caption{Representativity of FAME sample}\label{repr_FAME}
\includegraphics[scale=1]{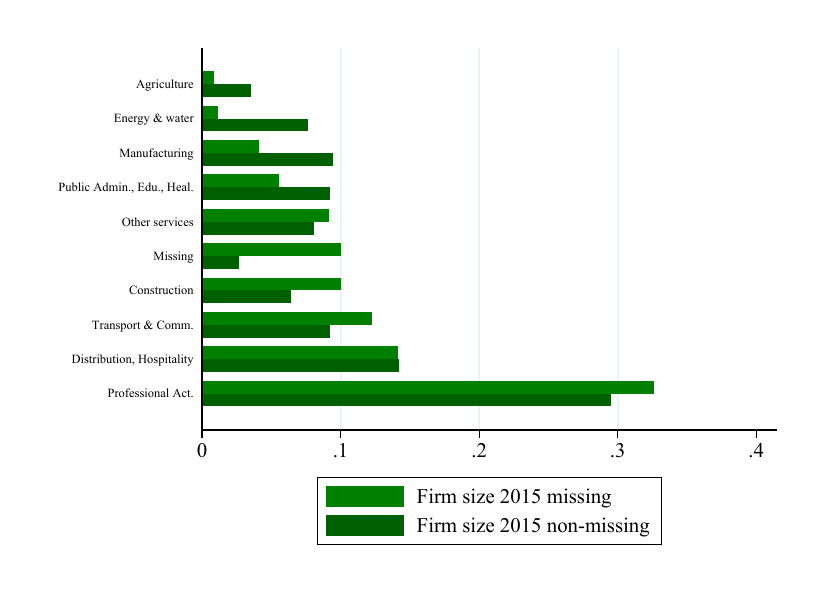}\\
	\floatfoot{\footnotesize{\textit{Source:} FAME, 2015.\\
	\textit{Note:} This figure compares the industry distribution of FAME firms with missing and non-missing size information in 2015, the year used to to define firms' treatment status.}}
\end{figure}

\paragraph{FAME.} We start from FAME, the UK version of Amadeus, covering all UK-registered firms. For around 30 percent of them we have information on the number of employees for at least one year in the pre-policy period. To address selectivity concerns at this stage, in Figure \ref{repr_FAME} we compare the industry distribution for firms with and without information on the number of employees in 2015, the year used to define the treatment status. Reassuringly, while firms with missing information on employees' numbers also tend to have missing information on industry, the rest of the distribution appears similar. At this stage, we retain all FAME firms with a number of employees between 200 and 300 in at least one year and non-missing firm size in 2015, with a resulting sample of 9,771 firms.

\paragraph{FAME and BGT.} We then merge this sample of firms with BGT using the name-matching algorithm described above and only retain firms with a match score equal to one, for a total of 5,140 FAME firms, or 53 percent of the FAME sample. Table \ref{selectivity_fame_bgt} explores selection patterns at this stage by comparing the 2015 number of employees for FAME firms that do or do not match with a BGT employer. While the average firm size is statistically different across these two groups, the difference does not appear very large, with firms matching with BGT having on average 231 employees in 2015, and firms that do not match having on average 240 employees in 2015.

\paragraph{Final sample restrictions.} We make three last restrictions to create the difference-in-differences sample. First, we keep only vacancies with non-missing SIC and SOC codes. Second, we restrict our sample to full-time vacancies, representing around 90 percent of the sample. We do this because, together with the wage posting decision, we also estimate the impact of the policy on wage offers. As vacancies for part-time jobs report either the full-time equivalent salary or the part-time salary but we cannot distinguish these two types of vacancies, we exclude the vacancies that we identify as offering a part-time job. Third, in the main analysis, we exclude vacancies for the financial years 2013 and 2014, as the data for this period are known to be of lower quality (\citealt{adams2020flex}). The final sample includes 97,467 vacancies of 2,556 firms, over the fiscal years 2015--2019.
\clearpage
 \newpage

\begin{figure}[H]
\caption{Wage posting and equality indicators - conditional correlations}\label{bgt gpg correlations}
\includegraphics[scale=0.8]{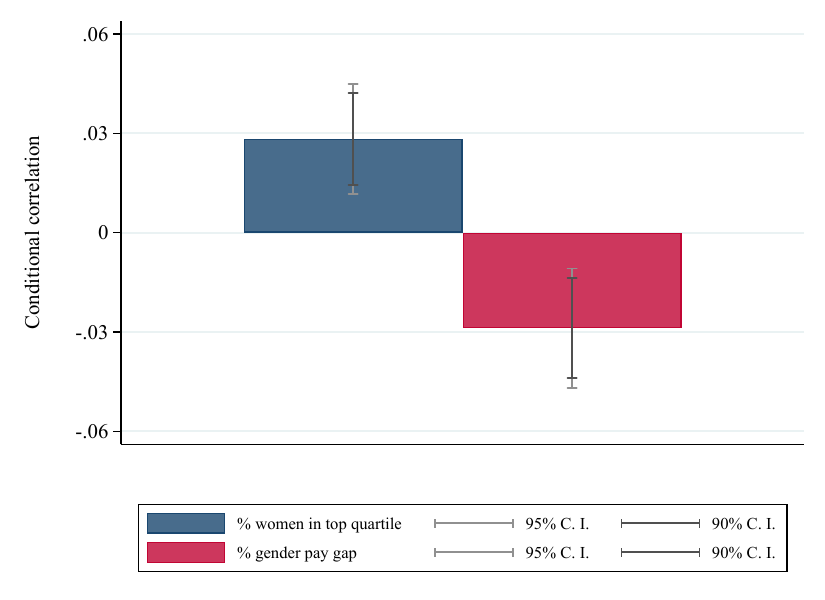}\\
	\floatfoot{
	\textit{Source:} BGT 2015--2019. GEO 2018--2019. \\
	\textit{Note:} The bar graph reports estimated coefficients from regressions of gender equality indicators (averaged across 2017/18 and 2018/19) on the average percentage of vacancies posting wage information over the period 2015--2019, the occupational composition of firms' vacancies, firms' size bands, and 5-digit SIC fixed effects. The graph also displays 90 and 95 percent confidence intervals associated with heteroskedasticity-robust standard errors. The sample includes firms publishing gender equality indicators both in 2017/18 and in 2018/19, with non-missing registration numbers and SIC codes, and matched with BGT with a match score of 1 (See Appendix \ref{name matching algorithm} for a description of the name-matching procedure). N. observations = 5,107.}
\end{figure}
\clearpage
\newpage

\begin{figure}[H]
\caption{Industry distribution in BGT and ONS Vacancy Survey}\label{bgt vs ons}
\centering
\includegraphics[scale=1]{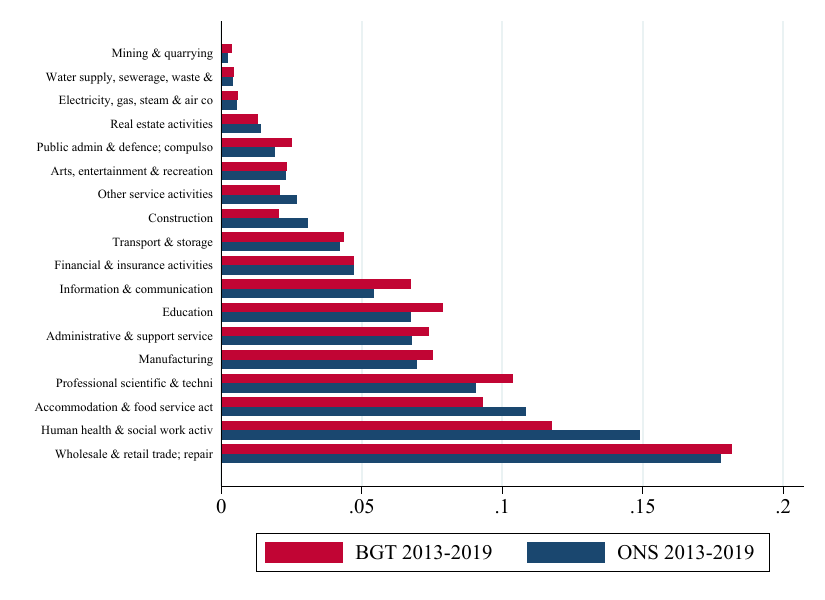}
\\
\floatfoot{{\textit{Source:} BGT, ONS Vacancy Survey, 2013--2019. \\
\textit{Note}: This figure compares the industry distribution in the stock of BGT vacancies with non-missing employer name and in the ONS Vacancy Survey.}}
\end{figure}
\clearpage
 \newpage
 
\clearpage
\newpage

\begin{figure}[H]
\caption{Wage posting by industry and occupation}\label{BGT distributions wage posting}
\centering

    \vspace{0.2cm}
 \captionsetup[subfigure]{justification=centering}
\begin{subfigure}[a]{\textwidth}
\centerline{\includegraphics[scale=0.6]{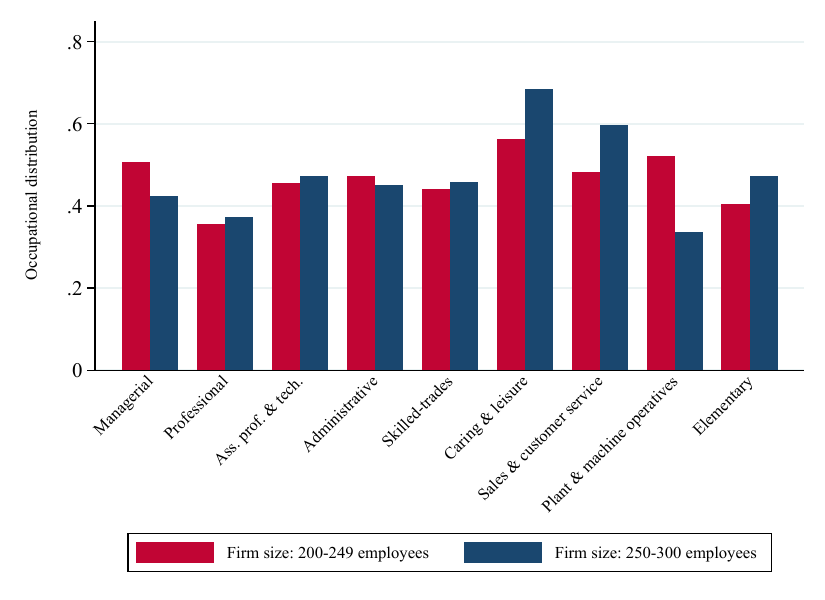}}
\caption{Occupational distribution}
\end{subfigure}\\
    \vspace{0.5cm}
\captionsetup[subfigure]{justification=centering}
\begin{subfigure}[b]{\textwidth}
\centerline{\includegraphics[scale=0.6]{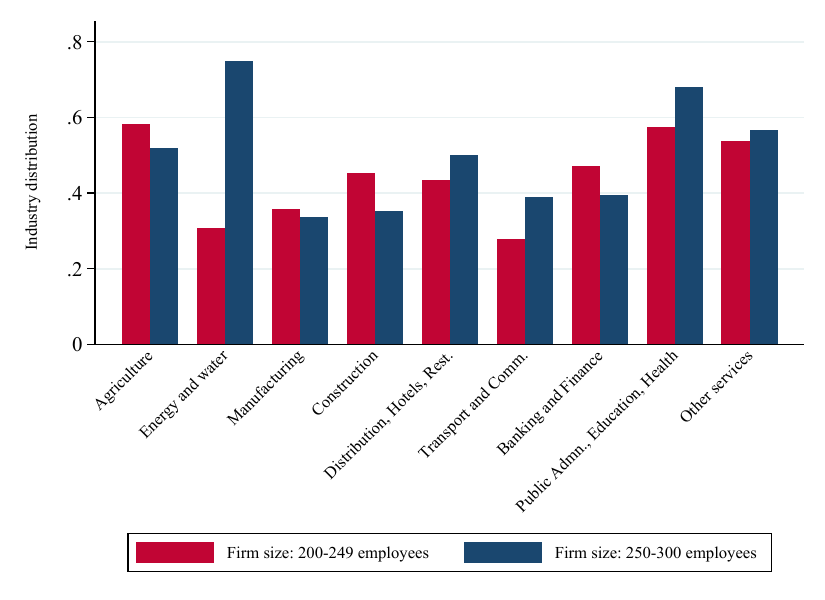}}
\caption{Industry distribution}
\end{subfigure}\\
    \vspace{0.5cm}
\floatfoot{{\textit{Source:} BGT, 2015--2017.\\
\textit{Notes:} These graphs present the occupational and industry distribution of wage posting, separately for treated and control firms, in the pre-policy period. In each graph, the blue bars refer to the treatment group, firms with 250-300 employees, and the red bars to the control group, firms with 200-249 employees.}}
\end{figure}
\clearpage
\newpage

\begin{figure}[H]
\caption{Event studies  - firms' hiring practices}\label{event study wage posting}
\centering
    \vspace{0.2cm}
\captionsetup[subfigure]{justification=centering}

\begin{subfigure}{.5\textwidth}
  \centering
  \includegraphics[scale=0.45]{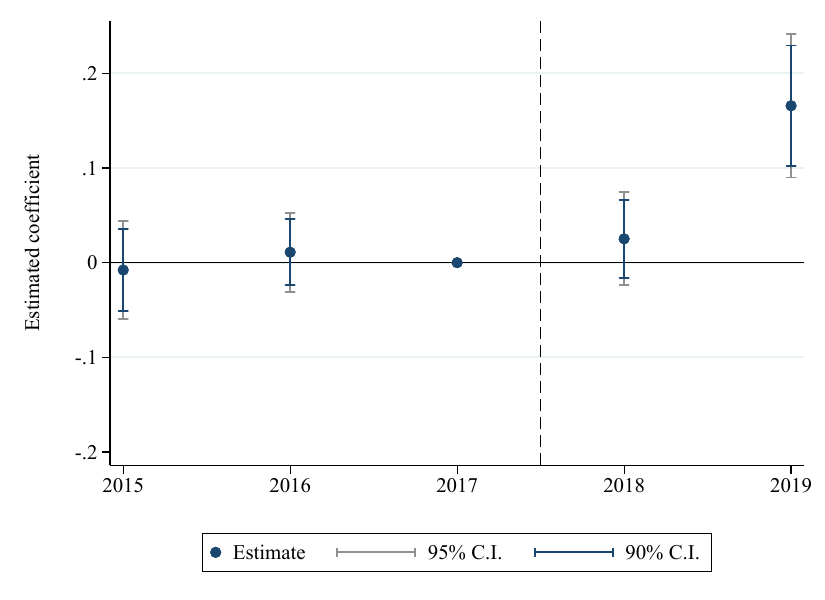}
\caption{Wage posting}
\end{subfigure}%
\begin{subfigure}{.5\textwidth}
  \centering
  \includegraphics[scale=0.45]{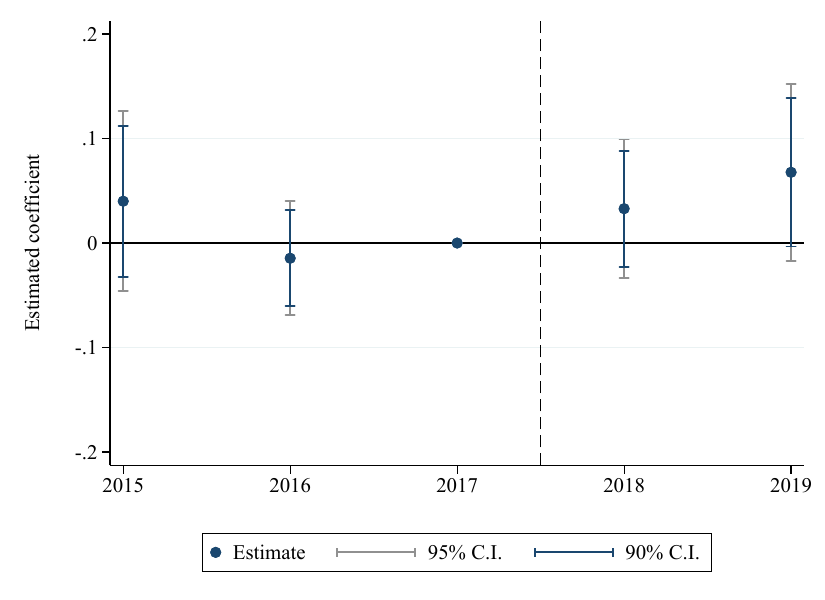}
\caption{Log annual wage}
\end{subfigure}\\
    \vspace{0.5cm}
\begin{subfigure}{.5\textwidth}
  \centering
  \includegraphics[scale=0.45]{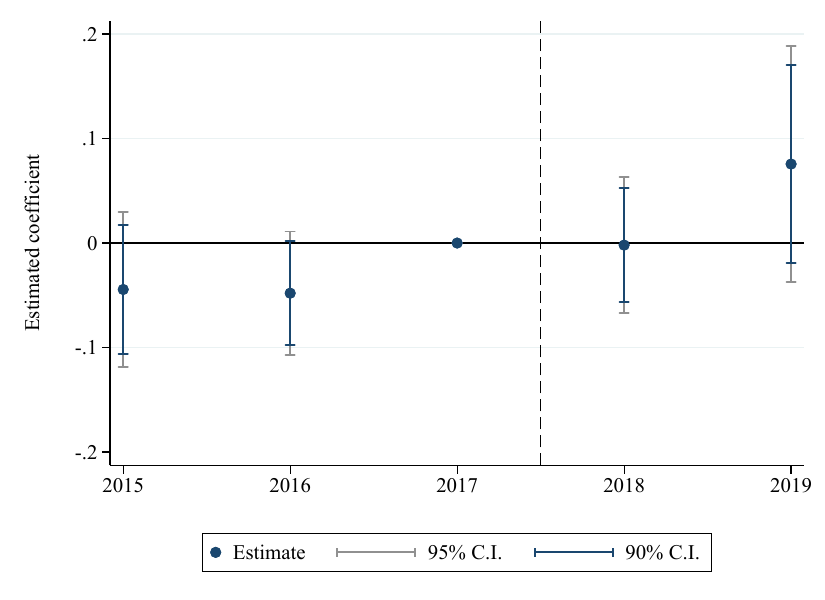}
\caption{Wage interval}
\end{subfigure}%
\begin{subfigure}{.5\textwidth}
  \centering
  \includegraphics[scale=0.45]{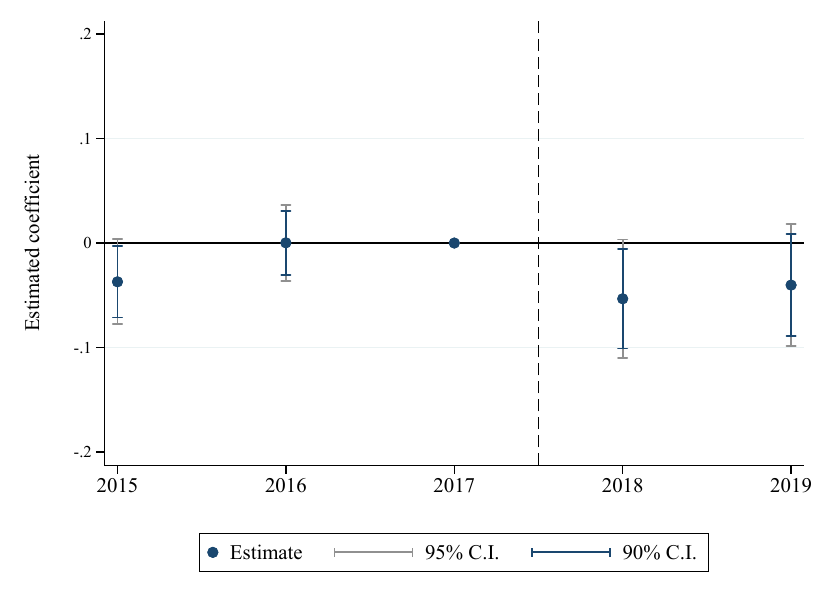}
\caption{Interval width}
\end{subfigure}\\   
    \vspace{0.2cm}
    
	\floatfoot{
	\textit{Source:} BGT, FAME 2015--2019. \\
	\textit{Note:} This graph presents the estimates of the leads and lags of the policy on firms' wage-posting decision and the characteristics of posted wages. These results are obtained from the estimation of the dynamic version of regression \ref{did BGT}. The estimation sample includes firms with 200-300 employees, between the financial years 2015 and 2019. The graph also reports 90 and 95 percent confidence intervals associated with firm-level clustered standard errors. The dash  vertical  line  indicates  the  month  when  the  mandate  is  approved,  i.e.,  February  2017.}
\end{figure}
\clearpage
\newpage

\begin{table}[htbp]
\def\sym#1{\ifmmode^{#1}\else\(^{#1}\)\fi}
\caption{ {Gender equality performance and presence in BGT}}\label{selectivity_gpg_bgt}
\scalebox{0.9}{
\begin{threeparttable}
\begin{footnotesize}
\begin{tabular}{l*{8}{c}}
\toprule
                   &\multicolumn{4}{c}{2017/18}&\multicolumn{4}{c}{2018/19}\\
                  \cmidrule(l){2-5} \cmidrule(l){6-9} &\multicolumn{1}{c}{Entire}&\multicolumn{2}{c}{Matched with BGT}&\multicolumn{1}{c}{P-value}&\multicolumn{1}{c}{Entire}&\multicolumn{2}{c}{Matched with BGT}&\multicolumn{1}{c}{P-value}\\
                    &\multicolumn{1}{c}{sample}&\multicolumn{1}{c}{Yes}&\multicolumn{1}{c}{No}&\multicolumn{1}{c}{difference}&\multicolumn{1}{c}{sample}&\multicolumn{1}{c}{Yes}&\multicolumn{1}{c}{No}&\multicolumn{1}{c}{difference}\\
    
                    &\multicolumn{1}{c}{(1)}&\multicolumn{1}{c}{(2)}&\multicolumn{1}{c}{(3)}&\multicolumn{1}{c}{(4)}&\multicolumn{1}{c}{(5)}&\multicolumn{1}{c}{(6)}&\multicolumn{1}{c}{(7)}&\multicolumn{1}{c}{(8)}\\
                    
\midrule
Gender pay gap (\%) & 11.79 & 12.11 & 11.31 & 0.01 & 11.88 & 12.26 & 11.30 & 0.00\\
 & (15.84) & (15.84) & (15.81) &  & (15.51) & (15.73) & (15.15) & \\
\addlinespace
Women in top quartile (\%) & 39.20 & 39.10 & 39.35 & 0.60 & 39.75 & 39.70 & 39.84 & 0.77\\
 & (24.41) & (24.90) & (23.65) &  & (24.48) & (24.94) & (23.78) & \\
\addlinespace
Observations & 10,557 & 6,323 & 4,234 &  & 10,812 & 6,514 & 4,298 & \\
\bottomrule
\end{tabular}
\end{footnotesize}
\begin{tablenotes}
\item{\footnotesize \textit{Source:} BGT 2013--2019, GEO 2018--2019.}
\item {\footnotesize \textit{Notes:} This table explores potential selection patterns of GEO firms matched with BGT. Column 1 (5) reports the gender median hourly pay gap and percentage of women in the top quartile of the firm wage distribution for all GEO firms in 2017/18 (2018/19); Column 2 (6) refers to firms that match perfectly with a BGT employer; Column 3 (7) refers to firms that do not match perfectly with a BGT employer; Column 4 (8) reports the p-value of the difference in the sample means of these two groups. }
\item{\footnotesize *** p$<$0.01, ** p$<$0.05, * p$<$0.1.}
\end{tablenotes}
\end{threeparttable}}
\end{table}
\clearpage
\newpage

\begin{table}[htbp]
\def\sym#1{\ifmmode^{#1}\else\(^{#1}\)\fi}
\caption{FAME firm size and presence in BGT}\label{selectivity_fame_bgt}
\begin{threeparttable}
\begin{footnotesize}
\begin{tabular}{l*{4}{c}}
\toprule
                    &\multicolumn{1}{c}{Entire}&\multicolumn{2}{c}{Matched with BGT}&\multicolumn{1}{c}{P-value}\\
                    &\multicolumn{1}{c}{sample}&\multicolumn{1}{c}{Yes}&\multicolumn{1}{c}{No} &\multicolumn{1}{c}{Difference}\\
                  &\multicolumn{1}{c}{(1)}&\multicolumn{1}{c}{(2)}&\multicolumn{1}{c}{(3)}\\
\midrule
Number of employees &  236  & 231  &  240 & 0.05\\
 & (230) & (164) & (286) & \\
 \addlinespace
Observations &  9,771 &  5,140 & 4,631 & \\

\bottomrule
\end{tabular}
\end{footnotesize}
\begin{tablenotes}
\item{\footnotesize \textit{Source:} BGT, FAME, 2013--2019.}
\item {\footnotesize \textit{Notes:} This table explores potential selection patterns of FAME firms matched with BGT. Column 1 reports the 2015 number of employees of all FAME firms with a number of employees between 200 and 300 in at least one year and non-missing firm size in 2015; Columns 2 and 3 reports the number of employees in 2015 for FAME that do or do not match perfectly with a BGT employer. Finally, Column 4 reports the p-value of the difference in the sample means of these two groups. }
\end{tablenotes}
\end{threeparttable}
\end{table}
\clearpage
\newpage

\begin{table}[htbp]
\def\sym#1{\ifmmode^{#1}\else\(^{#1}\)\fi}
\caption{BGT Summary statistics - pre-policy period}\label{bgt summary_stats}
\begin{threeparttable}
\begin{footnotesize}
\begin{tabular}{l*{2}{c}}
\toprule
                    &\multicolumn{1}{c}{Treated}&\multicolumn{1}{c}{Control}\\
 &\multicolumn{1}{c}{(1)}&\multicolumn{1}{c}{(2)}\\
 
 \midrule
Full-time & 0.92 & 0.90\\
 & (0.27) & (0.30)\\
 \addlinespace
Observations & 25,145 & 45,563\\
\midrule
{\bf Full-time vacancies} & & \\
\midrule
Higher-paid occupations & 0.61 & 0.56\\
 & (0.49) & (0.50)\\
 \addlinespace
Posting wage & 0.46 & 0.45\\
 & (0.50) & (0.50)\\
 \addlinespace
Observations & 23,211 & 41,125\\
\midrule
{\bf Vacancies with wage information} & & \\
\midrule
Wage interval & 0.51 & 0.41\\
 & (0.50) & (0.49)\\
 \addlinespace
 Interval dispersion & 1.25 & 1.23\\
 & (0.23) & (0.20)\\
 \addlinespace
Annual wage offered & 26,145 & 29,395\\
 & (14,492) & (19,466)\\
 \addlinespace
Observations & 10,735 & 18,400\\
\bottomrule
\end{tabular}
\end{footnotesize}
\begin{tablenotes}
\item{\footnotesize \textit{Source:} BGT, 2015--2017.}
\item {\footnotesize \textit{Notes:} This table reports mean and standard deviation of the main variables used in the analysis of BGT data, separately for treatment and control groups, before the implementation of the mandate.}
\end{tablenotes}
\end{threeparttable}
\end{table}
\clearpage
\newpage


\begin{table}[htbp]\centering
\def\sym#1{\ifmmode^{#1}\else\(^{#1}\)\fi}
\caption{Impact on wage-posting decision - longer pre-period}\label{bgt outcomes from 2013}
\begin{threeparttable}
\begin{footnotesize}
\begin{tabular}{l*{4}{c}}
\toprule
                    
                    &\multicolumn{1}{c}{Wage}&\multicolumn{1}{c}{Log annual}&\multicolumn{1}{c}{Wage}&\multicolumn{1}{c}{Interval}\\
                     &\multicolumn{1}{c}{posted}&\multicolumn{1}{c}{wage}&\multicolumn{1}{c}{ interval}&\multicolumn{1}{c}{dispersion}\\
                    &\multicolumn{1}{c}{(1)}&\multicolumn{1}{c}{(2)}&\multicolumn{1}{c}{(3)}&\multicolumn{1}{c}{(4)}\\
\midrule
{\bf Panel A: 2015--2019} & & & & \\
Treated firm*post &       0.043\sym{*}  &       0.040         &       0.036         &      -0.046\sym{*}  \\
                    &     (0.022)         &     (0.029)         &     (0.028)         &     (0.026)         \\
\addlinespace
Observations        &       97,467         &       43,752         &       43,752         &       19,342         \\
Adjusted \(R^{2}\)  &       0.470         &       0.584         &       0.414         &       0.382         \\
Pre-policy mean     &        0.46         &        26,116         &        0.51         &        1.25         \\
\midrule
{\bf Panel B: 2013--2019} & & & & \\
Treated firm*post &       0.039\sym{*}  &       0.027         &       0.016         &      -0.042\sym{*}  \\
                    &     (0.022)         &     (0.028)         &     (0.031)         &     (0.025)         \\
\addlinespace
Observations        &      125,479         &       56,933         &       56,933         &       23,912         \\
Adjusted \(R^{2}\)  &       0.474         &       0.565         &       0.437         &       0.378         \\
Pre-policy mean     &        0.49         &        26,315       &        0.44         &        1.26         \\
\bottomrule
\end{tabular}
\end{footnotesize}
\begin{tablenotes}
\item{\footnotesize \textit{Source:} BGT, 2013--2019.}
\item {\footnotesize \textit{Notes:} This table compares the impact of pay transparency on wage posting and wage characteristics when using the main sample (Panel A) and when including the fiscal years 2013 and 2014. Both sets of results are obtained from the estimation of regression \ref{did BGT}. In both panels, the estimation sample comprises BGT firms that have between 200 and 300 employees. In Column 2, it is restricted to vacancies with wage information. In Column 4, it is further restricted to vacancies posting a wage interval. All regressions include firm fixed effects and 5-digit SIC-specific time shocks. A treated firm is defined as having at least 250 employees in 2015. The post dummy is equal to one from the second quarter of 2018 onward. Heteroskedasticity-robust standard errors clustered at firm level in parentheses. The pre-policy mean represents the mean of the outcome variable for the treated group between 2015 (2013) and 2017.}
\item{\footnotesize *** p$<$0.01, ** p$<$0.05, * p$<$0.1.}
\end{tablenotes}
\end{threeparttable}
\end{table}
\clearpage
\newpage

\begin{table}[htbp]\centering
\def\sym#1{\ifmmode^{#1}\else\(^{#1}\)\fi}
\caption{Impact on wage-posting decision - controlling for SOC FE}\label{bgt outcomes with soc fe}
\begin{threeparttable}
\begin{footnotesize}
\begin{tabular}{l*{4}{c}}
\toprule
                    
                    &\multicolumn{1}{c}{Wage}&\multicolumn{1}{c}{Log annual}&\multicolumn{1}{c}{Wage}&\multicolumn{1}{c}{Interval}\\
                     &\multicolumn{1}{c}{posted}&\multicolumn{1}{c}{wage}&\multicolumn{1}{c}{ interval}&\multicolumn{1}{c}{dispersion}\\
                    &\multicolumn{1}{c}{(1)}&\multicolumn{1}{c}{(2)}&\multicolumn{1}{c}{(3)}&\multicolumn{1}{c}{(4)}\\
\midrule
{\bf Panel A: Main specification} & & & & \\
Treated firm*post &       0.043\sym{*}  &       0.040         &       0.036         &      -0.046\sym{*}  \\
                    &     (0.022)         &     (0.029)         &     (0.028)         &     (0.026)         \\
\addlinespace
Observations        &       97,467         &       43,752         &       43,752         &       19,342         \\
Adjusted \(R^{2}\)  &       0.470         &       0.584         &       0.414         &       0.382         \\
Pre-policy mean     &        0.46         &        26,116         &        0.51         &        1.25         \\
\midrule
{\bf Panel B: Adding 4-digit SOC FE} & & & & \\
Treated firm*post &       0.044\sym{**} &       0.040         &       0.038         &      -0.039         \\
                    &     (0.021)         &     (0.027)         &     (0.028)         &     (0.024)         \\
\addlinespace
Observations        &       97,458         &       43,725         &       43,725         &       19,310         \\
Adjusted \(R^{2}\)  &       0.484         &       0.681         &       0.429         &       0.414         \\
Pre-policy mean     &        0.46         &        26,116         &        0.51         &        1.25         \\
\bottomrule
\end{tabular}
\end{footnotesize}
\begin{tablenotes}
\item{\footnotesize \textit{Source:} BGT, 2015--2019.}
\item {\footnotesize \textit{Notes:} This table compares the impact of pay transparency on wage posting and wage characteristics in the main specification and in a model that controls for 4-digit SOC fixed effects. Both sets of results are obtained from the estimation of regression \ref{did BGT}. In both panels, the estimation sample comprises BGT firms that have between 200 and 300 employees. In Column 2, it is restricted to vacancies with wage information. In Column 4, it is further restricted to vacancies posting a wage interval. All regressions include firm fixed effects and 5-digit SIC-specific time shocks. A treated firm is defined as having at least 250 employees in 2015. The post dummy is equal to one from the second quarter of 2018 onward. Heteroskedasticity-robust standard errors clustered at firm level in parentheses. The pre-policy mean represents the mean of the outcome variable for the treated group between 2015 and 2017.}
\item{\footnotesize *** p$<$0.01, ** p$<$0.05, * p$<$0.1.}
\end{tablenotes}
\end{threeparttable}
\end{table}
\clearpage
\newpage

\begin{landscape}
\begin{table}[htbp]\centering
\def\sym#1{\ifmmode^{#1}\else\(^{#1}\)\fi}
\caption{Impact on number of vacancies and occupational distribution }\label{bgt vacs}
\begin{threeparttable}
\begin{footnotesize}
\begin{tabular}{l*{10}{c}}
\toprule
                    
                    &\multicolumn{1}{c}{Log number}&\multicolumn{1}{c}{SOC1}&\multicolumn{1}{c}{SOC2}&\multicolumn{1}{c}{SOC3}&\multicolumn{1}{c}{SOC4}&\multicolumn{1}{c}{SOC5}&\multicolumn{1}{c}{SOC6}&\multicolumn{1}{c}{SOC7}&\multicolumn{1}{c}{SOC8}&\multicolumn{1}{c}{SOC9}\\
                    &\multicolumn{1}{c}{vacancies}&\multicolumn{1}{c}{share}&\multicolumn{1}{c}{share}&\multicolumn{1}{c}{share}&\multicolumn{1}{c}{share}&\multicolumn{1}{c}{share}&\multicolumn{1}{c}{share}&\multicolumn{1}{c}{share}&\multicolumn{1}{c}{share}&\multicolumn{1}{c}{share}\\
                    &\multicolumn{1}{c}{(1)}&\multicolumn{1}{c}{(2)}&\multicolumn{1}{c}{(3)}&\multicolumn{1}{c}{(4)}&\multicolumn{1}{c}{(5)}&\multicolumn{1}{c}{(6)}&\multicolumn{1}{c}{(7)}&\multicolumn{1}{c}{(8)}&\multicolumn{1}{c}{(9)}&\multicolumn{1}{c}{(10)}\\
\midrule
Treated firm*post&       0.028         &       0.001         &       0.011         &      -0.011         &      -0.005         &       0.004         &      -0.001         &       0.005         &      -0.002         &      -0.001         \\
                    &     (0.071)         &     (0.015)         &     (0.018)         &     (0.018)         &     (0.014)         &     (0.011)         &     (0.010)         &     (0.013)         &     (0.008)         &     (0.009)         \\
\addlinespace
Observations        &       11,397         &       11,397         &       11,397         &       11,397         &       11,397         &       11,397         &       11,397         &       11,397         &       11,397         &       11,397         \\
Adjusted \(R^{2}\)  &       0.534         &       0.188         &       0.392         &       0.219         &       0.204         &       0.380         &       0.464         &       0.356         &       0.477         &       0.358         \\
Pre-policy mean     &        1.20         &        0.12         &        0.22         &        0.20         &        0.11         &        0.08         &        0.05         &        0.12         &        0.04         &        0.05         \\
\bottomrule
\end{tabular}
\end{footnotesize}
\begin{tablenotes}
\item{\footnotesize \textit{Source:} BGT, 2015--2019.}
\item {\footnotesize \textit{Notes:} This table reports the impact of pay transparency on firms' number of vacancies and the occupational distribution of vacancies. The results are obtained from the estimation of regression \ref{did BGT} at firm level. The estimation sample comprises BGT firms that have between 200 and 300 employees. All regressions include firm fixed effects and 5-digit SIC-specific time shocks. A treated firm is defined as having at least 250 employees in 2015. The post dummy is equal to one from the second quarter of 2018 onward. Heteroskedasticity-robust standard errors clustered at firm level in parentheses. The pre-policy mean represents the mean of the outcome variable for the treated group between 2015 and 2017.}
\item{\footnotesize *** p$<$0.01, ** p$<$0.05, * p$<$0.1.}
\end{tablenotes}
\end{threeparttable}
\end{table}
\clearpage
\newpage
\end{landscape}

\begin{table}[htbp]
\def\sym#1{\ifmmode^{#1}\else\(^{#1}\)\fi}
\caption{Impact on pay outcomes by tenure in the firm}\label{results by tenure}
\begin{threeparttable}
\begin{footnotesize}
\begin{tabular}{l*{3}{c}}
\toprule
                    &\multicolumn{1}{c}{Entire}&\multicolumn{1}{c}{$\leq$ \text{2 years}}&\multicolumn{1}{c}{More than 2 years}\\
                     &\multicolumn{1}{c}{sample}&\multicolumn{1}{c}{tenure}&\multicolumn{1}{c}{tenure}\\
                    &\multicolumn{1}{c}{(1)}&\multicolumn{1}{c}{(2)}&\multicolumn{1}{c}{(3)}\\
\midrule
Treated firm*post   &      -0.026\sym{***}&    -0.018 &    -0.019\sym{**}   \\
                    &     (0.008)         &     (0.024)         &     (0.008)   \\
\addlinespace
Treated firm*post*fem&       0.029\sym{**} &   0.067  &       0.017 \\
                    &     (0.014)       &     (0.045)         &     (0.015)     \\
\addlinespace
Observations        &       29,226       &      5,002         &       21,084      \\
Adjusted \(R^{2}\)  &       0.909    &       0.900     &       0.916 \\
P-value Women Coeff &       0.788       &       0.198     &       0.839    \\
P-value High vs. Low M &             & 0.747 &      \\
P-value High vs. Low W &             & 0.200 &     \\
Men's pre-policy mean&        15.93         &        12.59        &        17.46   \\
Women's pre-policy mean&        13.36      &        11.37        &        14.39 \\
\bottomrule
\end{tabular}
\end{footnotesize}
\begin{tablenotes}
\item{\footnotesize \textit{Source:} ASHE, 2013--2019.}
\item {\footnotesize \textit{Notes:} This table compares the impact of pay transparency on employees' hourly pay across workers with more or at most 2 years of tenure in the firm, by estimating regression \ref{tripledid} by subgroup. Column 1 reports the estimate on log hourly pay for the entire sample, employees working in firms that have between 200 and 300 employees. Columns 2 and 3 compare the impact across tenure groups. All regressions include firm*individual fixed effects, gender*year fixed effects, and region-specific time shocks. A treated firm is defined as having at least 250 employees in 2015. The post dummy is equal to one from 2018 onward. Heteroskedasticity-robust standard errors clustered at firm level in parentheses. The `` P-value Women Coeff'' refers to the t-test on the sum of the two reported coefficients, corresponding to the effect of the policy on female employees. The ``P-value High vs. Low M (W)'' refers to the t-test on the equality of effects on men's (women's) pay across tenure groups. The pre-policy mean represents the mean of the outcome variable for the treated group and subgroup considered between 2013 and 2017.}
\item{\footnotesize *** p$<$0.01, ** p$<$0.05, * p$<$0.1.}
\end{tablenotes}
\end{threeparttable}
\end{table}
\clearpage
\newpage

\end{document}